%% file: BGSP.tex
\documentclass[aps,prd,nofootinbib,showkeys,preprint,floatfix]{revtex4-1}

\pdfoutput=1

\makeatletter

\usepackage{graphicx}
\usepackage{graphics}
\usepackage[mathscr]{eucal}
\usepackage{amssymb}
\usepackage{amsmath}
\usepackage{xspace}
\usepackage{listings}
\usepackage{ulem}
\usepackage{color}
\usepackage{multirow}
\usepackage{enumerate}

\newcommand{\SARAH}{{\tt SARAH}\xspace}
\newcommand{\SPheno}{{\tt SPheno}\xspace}
\newcommand{\HiggsBounds}{{\tt HiggsBounds}\xspace}

\def\hE{\hat{E}}
\def\hEt{\hat{\tilde{E}}}
\def\nn{\nonumber}
\def\ov{\overline}

\def\C{\mathcal}

\def\tr{\mathrm{tr}}

\def\beq{\begin{equation}}
\def\eeq{\end{equation}}
\def\bal{\begin{align}}
\def\eal{\end{align}}

\def\L{\mathcal{L}}

\def\twovec[#1,#2]{\left( \begin{array}{c} #1 \\ #2\end{array}\right)}
\def\threevec[#1,#2,#3]{\left( \begin{array}{c} #1 \\ #2 \\ #3 \end{array}\right)}
\def\twomatarix[#1,#2][#3,#4]{\left( \begin{array}{cc} #1 & #2 \\ #3 & #4 \end{array}
\right)}
\def\threematrix[#1,#2,#3][#4,#5,#6][#7,#8,#9]{\left( \begin{array}{ccc} #1 & #2 &#3 \\
#4 & #5 & #6\\#7&#8&#9\end{array} \right)}
\def\threediag[#1,#2,#3]{\left( \begin{array}{ccc} #1 & 0 & 0\\
0 & #2 & 0\\0& 0&#3\end{array} \right)}

\def\fdir{Figs/}

\let\oldincludegraphics\includegraphics
\renewcommand{\includegraphics}[2][]{\oldincludegraphics[#1]{\fdir #2}}

\newcommand{\AddrLPTHE}{%
1-- Sorbonne Universit\'es, UPMC Univ Paris 06, UMR 7589, LPTHE, F-75005, Paris, France \\
2-- CNRS, UMR 7589, LPTHE, F-75005, Paris, France 
}

\newcommand{\AddrWur}{%
Institut f\"ur Theoretische Physik und Astronomie,
Universit\"at W\"urzburg\\
Am Hubland,
97074 Wuerzburg, Germany}

\newcommand{\AddrBonn}{%
Bethe Center for Theoretical Physics \& Physikalisches Institut der Universit\"at Bonn\\Nu{\ss}allee 12, 53115 Bonn, Germany}

\begin{document}

\title{The Constrained Minimal Dirac Gaugino Supersymmetric Standard Model}

\author{K.\ Benakli} \email{kbenakli@lpthe.jussieu.fr}\affiliation{\AddrLPTHE}

\author{M.\ Goodsell} \email{goodsell@lpthe.jussieu.fr}\affiliation{\AddrLPTHE}

\author{W.\ Porod} \email{porod@physik.uni-wuerzburg.de}\affiliation{\AddrWur}

\author{F.\ Staub}\email{fnstaub@th.physik.uni-bonn.de}\affiliation{\AddrBonn}



\preprint{Bonn-TH-2014-06}
\begin{abstract}
We examine the possibilities for constructing models with Dirac gaugino masses and unification of gauge couplings. We identify one promising model, and discuss to what extent it can have a ``natural SUSY'' spectrum. We then determine the low-energy constraints upon it, and propose a constrained set of boundary conditions at the unification scale. We describe the implementation of these boundary conditions in the  spectrum-generator generator \SARAH and we perform a first exploration of the parameter space, specifically searching for points where the spectrum is relatively light. It is shown that the pattern of the masses of SUSY states is very different compared to any expectations from the constrained MSSM. 
\end{abstract}

\maketitle

\section{Introduction}

Supersymmetry (SUSY) is still the best-motivated framework for physics beyond the Standard Model. The main reasons for studying it remain that it provides a solution to the hierarchy problem; it appears to be required for string theory; it provides candidates for dark matter; and in the simplest implementation predicts unification of the gauge couplings at a high scale $M_{GUT} \simeq 2 \times 10^{16}$ GeV.  
However, since the latest searches from the LHC combined with the measured value of the Higgs mass have placed stringent 
exclusion bounds on the parameter space of its simpler manifestations, either simplified models or various variants of the 
Minimal Supersymmetric Standard Model (MSSM), it is now time to seriously consider \emph{non-minimal} realisations. One 
increasingly interesting example is to add Dirac masses to gauginos 
\cite{fayet,Polchinski:1982an,Hall:1990hq,fnw,Nelson:2002ca,Antoniadis:2005em,Antoniadis:2006uj,kpw,%
Amigo:2008rc,Plehn:2008ae,Benakli:2008pg,Belanger:2009wf,Benakli:2009mk,Choi:2009ue,Benakli:2010gi,Choi:2010gc,%
Carpenter:2010as,Kribs:2010md,Abel:2011dc,Davies:2011mp,Benakli:2011vb,Benakli:2011kz,Kalinowski:2011zz,Frugiuele:2011mh,%
Itoyama:2011zi,Rehermann:2011ax,Bertuzzo:2012su,Davies:2012vu,Argurio:2012cd,Fok:2012fb,Argurio:2012bi,Frugiuele:2012pe,%
Frugiuele:2012kp,Benakli:2012cy,Itoyama:2013sn,Chakraborty:2013gea,Csaki:2013fla,Itoyama:2013vxa,Beauchesne:2014pra,%
Benakli:2014daa,%
Bertuzzo:2014bwa}. 
This is particularly well-motivated, both from the top-down -- permitting simpler SUSY-breaking models due to preserving an R-
symmetry; having a possible relation to $N=2$ SUSY in the gauge sector; or arising in string models -- and from the bottom up: 
they allow increased naturalness; they can help enhance the Higgs mass without the danger of charge or colour breaking 
minima \cite{Camargo-Molina:2013sta,Blinov:2013fta,Chowdhury:2013dka}; and can weaken both LHC search bounds \cite{Heikinheimo:2011fk,Kribs:2012gx,Alves:2013wra} and flavour constraints \cite{kpw, Fok:2012me, Dudas:2013gga}. 

Giving Dirac masses to gauginos requires an extra adjoint chiral superfield for each gauge group. 
This changes the running of the gauge couplings, and, if only these fields are added to the MSSM, we lose the ``prediction'' of unification of gauge couplings. Of course, this is not necessarily a problem -- the apparent unification in the MSSM could be accidental. However, since the apparent unification in the MSSM can be taken as a motivation for supersymmetry, in this work we reconsider the consequences when we take unification of the gauge couplings in Dirac gaugino models seriously.

\begin{center}
\begin{figure}
\includegraphics[width=0.8\textwidth]{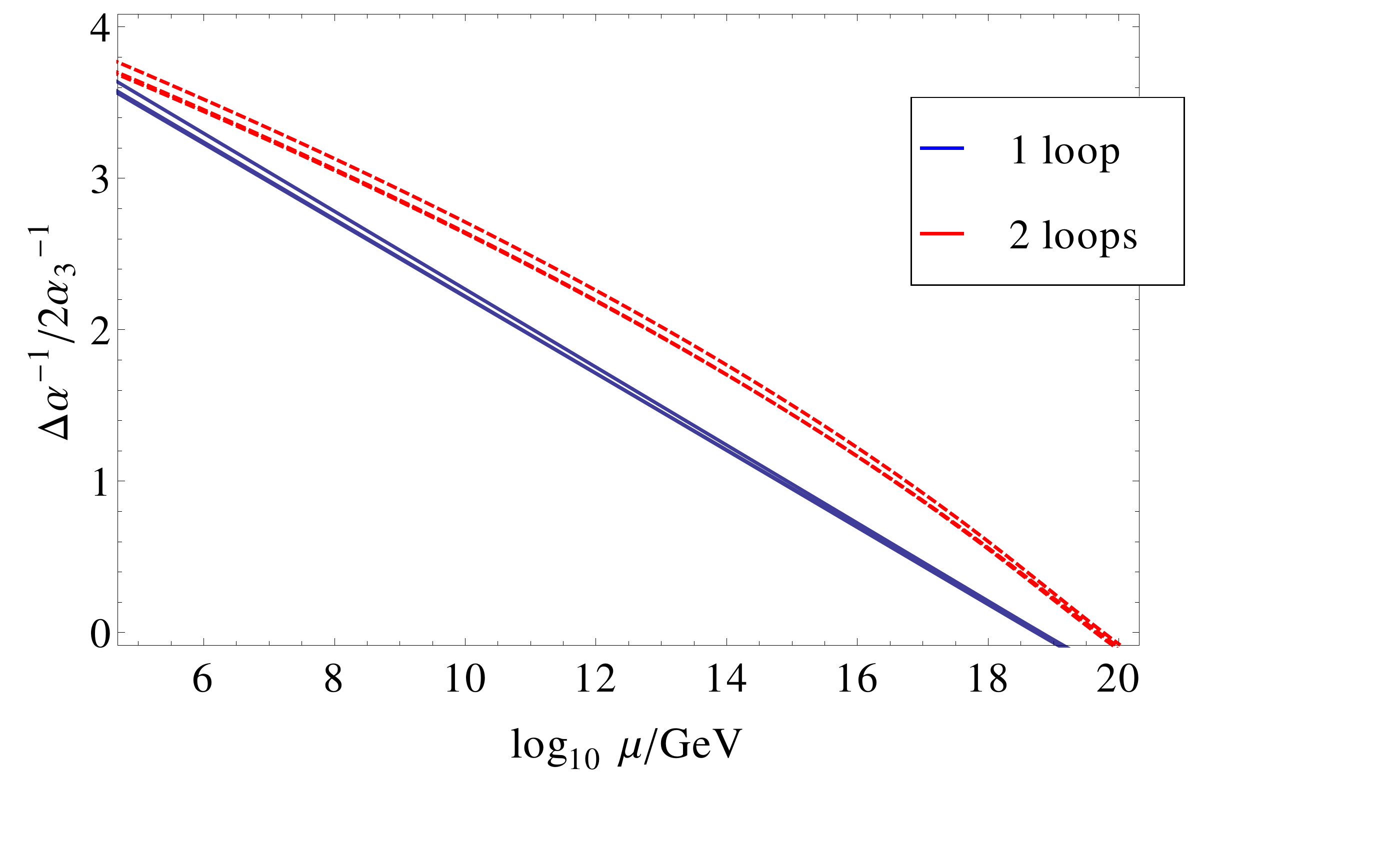} \vspace{-1cm}
\caption{F-theory unification: plotted is the logarithm of the renormalisation scale $\mu$ versus $(5 g_1^{-2} - 3 g_2^{-2} - 2 g_3^{-2})/2 g_3^{-1}$, which is a measure of the relative deviation from ``unification'' which should be close to zero and certainly less than unity for good agreement. We show one-loop RGE evolution with a blue solid lines and two-loop evolution with red dashed lines; there are three closely spaced curves for both corresponding to $\tan \beta =3,10,40$. Unfortunately, while the ``unification'' is marginal at one loop, at two loops it takes place well above the Planck scale.}
\label{FIG:FTheory}\end{figure}\end{center}

\begin{center}
\begin{figure}
\includegraphics[width=0.8\textwidth]{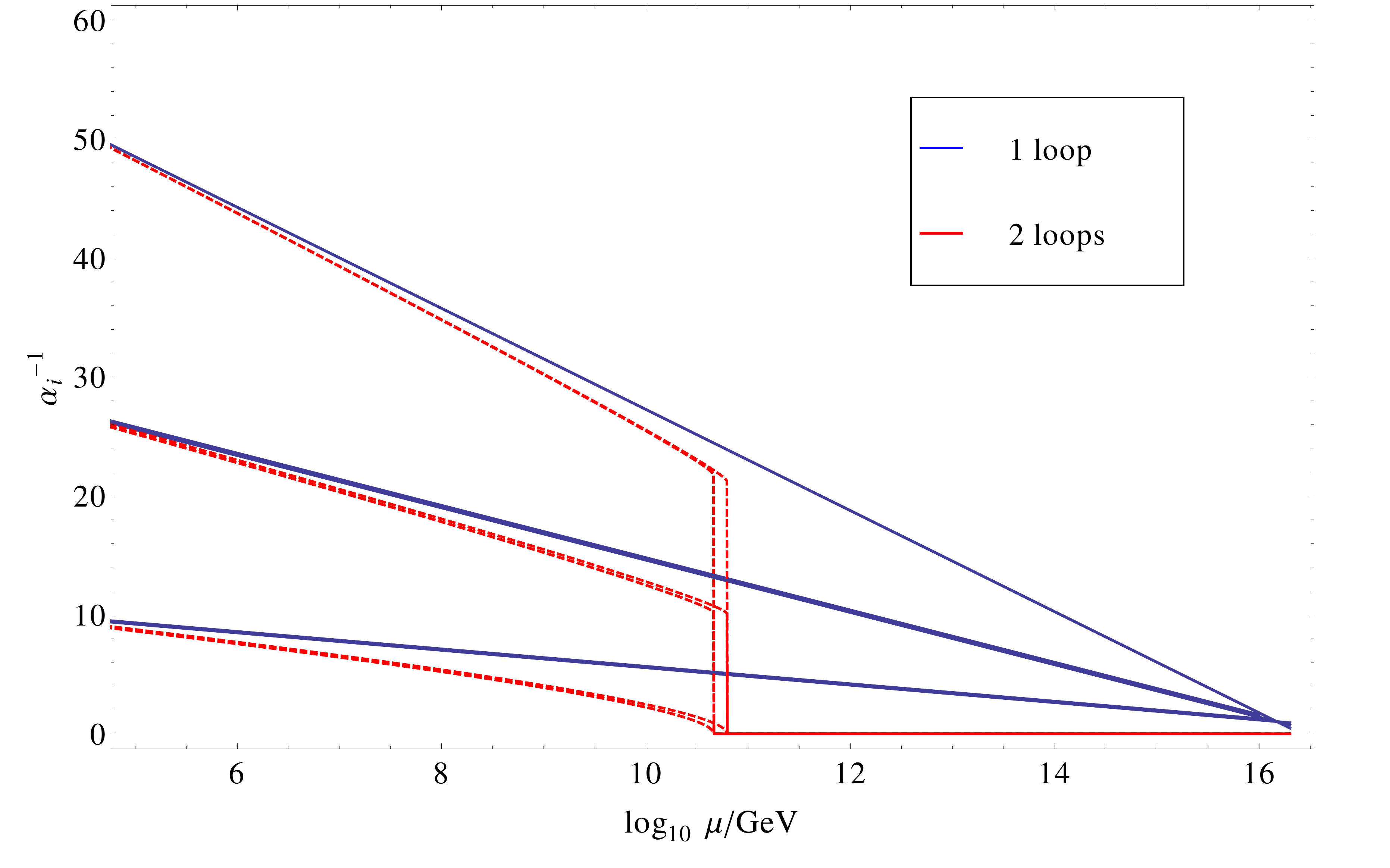}\vspace{-0.5cm}
\caption{$SU(5)$ bachelor unification: plotted is  the logarithm of the renormalisation scale $\mu$ versus $\alpha_i^{-1}$ for $i=1,2,3$ being the inverses of the GUT-normalised fine structure constants. Again we show one-loop RGE evolution with a blue solid lines and two-loop evolution with red dashed lines; there are three closely spaced curves for both corresponding to $\tan \beta =3,10,40$. It can be seen that, although perturbative unification is (just) possible at one loop, at two loops the couplings diverge at an intermediate scale.}\label{FIG:SU5UNI}\end{figure}
\end{center}

\begin{figure}
\includegraphics[width=0.8\textwidth]{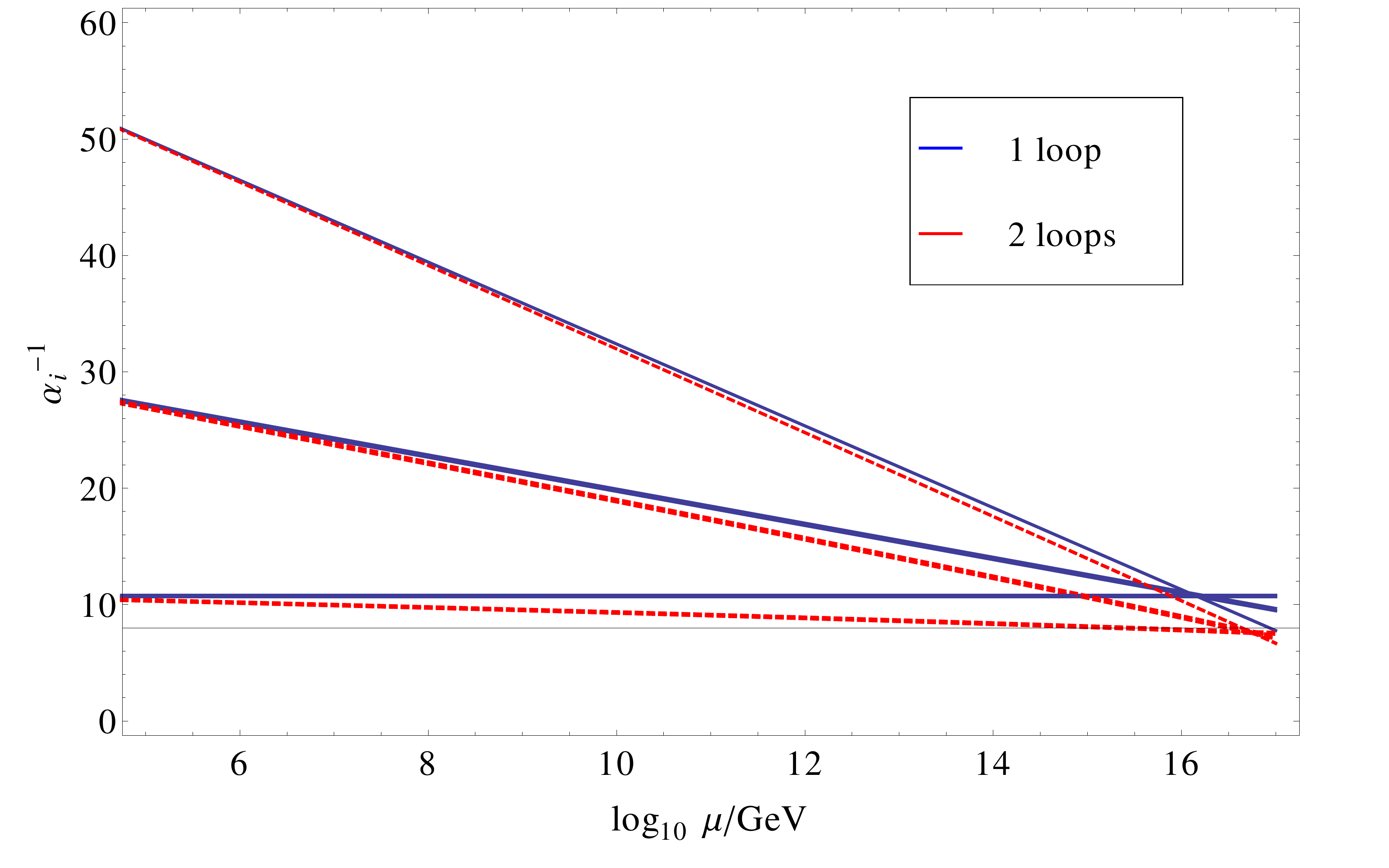} \vspace{-0.5cm}
\caption{As figure \ref{FIG:SU5UNI} but for (SU($3$))$^3$ matter content. Perturbative unification is achieved to a good accuracy both at one and two loops.}\label{FIG:SU33UNI}\end{figure}

As the most obvious shortcoming of a minimal extension of the MSSM by Dirac gaugino masses, several approaches to this problem have been suggested, which we now reconsider:
\begin{enumerate}
\item Suppose that there are extra fields with masses intermediate between the electroweak/supersymmetry and unification scales which do not fall into complete Grand Unified Theory (GUT) multiplets. Although the intermediate-scale masses may appear tuned (in that they cannot be arbitrary), the new states could restore unification without affecting the low-energy phenomenology -- allowing us to justify studying the minimal Dirac gaugino model at low energies -- and could even play a role as messengers of gauge mediation \cite{Benakli:2010gi}.  
\item We could change our definition of unification. Specifically, in string theory models such as in F-theory GUTs \cite{Donagi:2008ca,Beasley:2008dc,Beasley:2008kw,Blumenhagen:2008aw,Mayrhofer:2013ara}, although there is an underlying unified structure there is no actual unification of the gauge couplings, instead merely a weaker condition:\footnote{This condition could hold with different coefficients in different models, but we give here the one relevant for F-theory $SU(5)$ GUTS.}
$$ 5 \alpha_1^{-1} - 3 \alpha_2^{-1} - 2 \alpha_3^{-1} = 0 $$
where $\alpha_i$ are the structure constants for the three gauge couplings, with $\alpha_1$ having $SU(5)$ normalisation. To satisfy this condition in Dirac gaugino models we must still add some additional states, but there is a certain minimal choice that we can make: add one vector-like pair of right-handed electrons \cite{Davies:2012vu}. We plot the running of the gauge couplings for this scenario in figure \ref{FIG:FTheory}:  we find that, although the condition can be easily satisfied (for the minimal choice of extra states) at one loop, at two loops it predicts unification beyond the Planck scale. In principle the inclusion of (one- or two-loop) threshold corrections could lead to unification below or at the Planck scale. However, these are dependent upon the compactification and may be beyond the current technical understanding of F-theory, so we do not consider this possibility further.
\item We could add additional  states at the supersymmetry-breaking scale such that unification is restored. One direct way to achieve this is to add ``bachelor'' fields that complement the $\mathbf{8_0} + \mathbf{3_0}  + \mathbf{1_0}$ set of adjoint multiplets such that together they form a complete representation of a unified gauge group. The two simplest choices to add the adjoint  are $SU(5)$ and $(SU(3))^3$: %
\begin{enumerate}[(a)]
\item The $SU(5)$ case consists of adding states in the representation $(\mathbf{3}, \mathbf{2})_{-5/6} +(\ov{\mathbf{3}}, \mathbf{2})_{5/6} $ at the SUSY-breaking scale. We plot the running of the gauge couplings for this scenario in figure \ref{FIG:SU5UNI}: although at one loop unification at a perturbative coupling appears (just) possible, unfortunately at two loops we find that the couplings diverge at an intermediate scale just below $10^{11}$ GeV indicating
the breakdown of perturbation theory. 
\item The $(SU(3))^3$ case consists of adding states in the representation $ (\mathbf{1}, \mathbf{2})_{1/2}  + (\mathbf{1}, \mathbf{2})_{-1/2} + 2 \times (\mathbf{1}, \mathbf{1})_{\pm 1}  $ at the SUSY-breaking scale in addition to some singlets which do not affect the running of gauge couplings (at one loop). We plot the running of the gauge couplings for this scenario in figure \ref{FIG:SU33UNI}, and in this case we find precise unification at a scale of $1 \div 2 \times 10^{17}$ GeV.  
\end{enumerate}
\end{enumerate}

Of these options, if we would like to preserve the usual definition of unification without introducing new fixed energy scales then it is clear that the simplest choice is to simply add fields in the representations
\begin{align}
(\mathbf{1}, \mathbf{2})_{1/2}  + (\mathbf{1}, \mathbf{2})_{-1/2} + 2 \times (\mathbf{1}, \mathbf{1})_{\pm 1} 
\label{EQ:ExtraReps}\end{align}
of the Standard Model groups $(SU(3), SU(2))_Y$. This is the approach that we shall adopt in this paper, and build a distinct model from it. 
We then consider the boundary conditions at the unification
scale so that we can infer predictions at low energies after RGE evolution. In this way we can obtain 
constraints on our spectrum which could be useful for collider studies, by embedding the scenario into a spectrum generator. 
These are the goals of this paper. 

In section \ref{SEC:MUDGSSM} we consider the construction of a \emph{low-energy} theory with Dirac gaugino masses and the extra field content in equation (\ref{EQ:ExtraReps}) that can, in principle, be embedded into a GUT (we make some comments about possible embeddings in appendix \ref{APP:EMBEDDINGS}). The consequences of these new states for naturalness (and the subsequent implications for flavour physics) are discussed in section \ref{SEC:NATURAL}. We propose that the new states are charged under lepton number (we do not, as in \cite{Frugiuele:2012kp,Frugiuele:2012pe,Bertuzzo:2012su}, identify lepton number with R-symmetry -- in fact, our model explicitly breaks R-symmetry in the Higgs sector) which has consequences for charged lepton flavour violation that we describe in section \ref{SEC:LFV} and in more detail in appendix \ref{APP:CLFV}. 

Having determined the low-energy constraints, we propose constrained boundary conditions at the GUT scale, thus defining a new 
constrained minimal Dirac gaugino supersymmetric Standard Model in section \ref{SEC:CMDGSSM}.  We implement this model in the 
spectrum generator generator \SARAH \cite{Staub:2008uz,Staub:2010jh,Staub:2009bi,Staub:2012pb,Dreiner:2012dh,Staub:2013tta}, 
details of which we provide in appendix \ref{APP:SARAH}, produce and modify the subsequent \SPheno 
\cite{Porod:2003um,Porod:2011nf} code, and use it to perform some exploratory
scans of the parameter space. Our main results are given in the form of ``generic predictions'' in section \ref{SEC:GenericPredictions}. We also provide simplified renormalisation group equations for the model in appendix \ref{APP:RGEs}.


\newpage

\section{Minimal unified Dirac gaugino model}
\label{SEC:MUDGSSM}

\subsection{Superpotential}

The results of the previous section make it clear that the preferred scenario is to add the field content at low energies consistent with a $SU(3)^3$ GUT group  to the minimal Dirac gaugino supersymmetric Standard Model. In this way, we predict perturbative gauge coupling unification at two loops without introducing any additional scales, and unification should thus be considered to be as natural as in the MSSM. 
If we wish to construct a fully-fledged GUT model from this, we should then consider how the fields are embedded in GUT representations; this may in principle allow us to determine GUT relations for some of the other couplings. However, what is required for unification is not the whole $\mathbf{8} \oplus \mathbf{8} \oplus\mathbf{8}$ adjoint multiplet of $(SU(3))^3$ but actually only states in the representations
$$(\mathbf{1}, \mathbf{2})_{1/2}  + (\mathbf{1}, \mathbf{2})_{-1/2} + 2 \times (\mathbf{1}, \mathbf{1})_{\pm 1} $$
of the Standard Model groups $(SU(3), SU(2))_Y$. The full $(SU(3))^3$ adjoint multiplet would contain in addition four singlets, but the above fields could equally fit into incomplete representations of $SU(5)$ or even different representations of $SU(3)^3$ (such as $(\mathbf{3}, \mathbf{\ov{3}}) \oplus (\mathbf{\ov{3}}, \mathbf{3})$). 

The exact field content below the GUT scale also depends in principle upon the way in which the GUT group is broken. For example, whether by the expectation values of two pairs of $(\mathbf{3}, \mathbf{\ov{3}}) \oplus (\mathbf{\ov{3}}, \mathbf{3})$ fields in a four-dimensional $SU(3)^3$ GUT; or, more interestingly, we could consider a string-theory construction where the breaking is performed by higher-dimensional fluxes (so that, from the four-dimensional point of view, there is never actually a GUT group but the unification still holds). Due to the wealth of possibilities here we postpone such top-down model-building to future work, although we make a few pertinent comments about this in appendix \ref{APP:EMBEDDINGS}: instead, it is our intention to construct the simplest and most phenomenologically appealing scenario from the bottom up. To this end, we desire a minimal field content, and thus we shall suppose that there are no additional singlets (beyond the one which gives a Dirac mass to the bino) at the scale of the superpartner masses $M_S$ (by which we mean approximately the TeV scale).

By the above reasoning, in the following we shall consider our unified model to have, in addition to the fields of the MSSM and an adjoint for each gauge group, only the additional states given in equation (\ref{EQ:ExtraReps}). In fact we do not even need to specify whether the unified gauge group is $SU(5)$ or $(SU(3))^3$. We can then write down the superpotential of the theory, which should have the most general gauge-invariant form.
However, this contains many additional couplings which would violate our desire for minimality, and we are thus faced with a choice:
\begin{enumerate}
\item We could impose an R-symmetry on the model. In this way, the $(\mathbf{1}, \mathbf{2})_{1/2}  \oplus (\mathbf{1}, \mathbf{2})_{-1/2} $ fields would become the $R_{u,d}$ fields of the MRSSM. 
\item We could charge the additional fields under \emph{lepton number}. We would thus have a vector-like pair of left-handed leptons and two vector-like pairs of right-handed leptons. 
\end{enumerate}

Although $(1)$ may be theoretically appealing, phenomenologically it has the issue that the additional fermionic fields that couple to the Higgs tend to \emph{decrease} its mass rather than increase it\footnote{We thank P. Slavich for mentioning this result to us based on unpublished work.}, so such a model would seem to typically require rather heavy stops (particularly since there will be little stop mixing due to the absence of $A$-terms) or rather large adjoint-Higgs couplings and adjoint scalar masses that generate a one-loop boost as in \cite{Bertuzzo:2014bwa}. 
In contrast, scenario $(2)$ would allow additional couplings to the Higgs, enabling a boost to the \emph{tree-level} mass, and, if the extra leptons are not very light compared to the other supersymmetric particles, could be considered to be a UV completion of the phenomenological models studied elsewhere in \cite{Belanger:2009wf,Benakli:2011kz,Goodsell:2012fm,Benakli:2012cy}. Hence this is the approach we shall take here. 

Out of a desire to have a unified notation, we shall label the new states by 
\begin{equation}
\begin{array}{|c|c|} \hline 
\mathrm{Field} & (SU(3), SU(2))_{Y} \\ \hline
R_u & (\mathbf{1}, \mathbf{2})_{-1/2} \\
R_d & (\mathbf{1}, \mathbf{2})_{1/2}\\ \hline
\hE_{1,2} & (\mathbf{1}, \mathbf{1})_1 \\
\hEt_{1,2} & (\mathbf{1}, \mathbf{1})_{-1} \\ \hline
\end{array}
\label{eq:newstates}
\end{equation}
We label the usual MSSM fields $Q, L, E, U, D, H_u, H_d$: we write the usual Yukawa couplings as
\begin{align}
W_{Yukawa} =& Y_u^{ij} U_i Q_j H_u - Y_d^{ij} D_i Q_j H_d - Y_e^{ij} E_i L_j H_d .
\end{align}
Then the superpotential for scenario $(1)$ is  
\begin{align}
W_{R-symmetric} =& W_{Yukawa} \nn\\ 
& +(\mu_d + \lambda_{d,S}S)  H_d R_d + 2\lambda_{d,T} H_d T R_d \nn\\
&+ (\mu_u + \lambda_{u,S}S)  R_u H_u + 2\lambda_{u,T} R_u T H_u \nn\\
&+ (\mu_{\hE\, ij} + \lambda_{S\hat{E}ij}S)\hE_i \hEt_j + Y_{\hEt i} R_d H_u \hEt_i .
\end{align}
We shall not discuss this further. 

The superpotential of our theory, that of scenario $(2)$, is:
\begin{align}
W =& W_{Yukawa} + W_{adjoint} \nn\\ 
& +(\mu + \lambda_{S} S ) H_d H_u + 2\lambda_T H_d T H_u \nn\\
&+ (\mu_R + \lambda_{SR} S )R_u R_d + 2\lambda_{TR} R_u T R_d + (\mu_{\hE\, ij} + \lambda_{S\hat{E}\, ij}S)\hE_i \hEt_j\nn\\
&+ \lambda_{SLRi} S L_i R_d + 2\lambda_{TLRi} L_i T R_d + \lambda_{SEij}S E_i \hEt_j\nn\\
& - Y_{\hE i} R_u H_d \hE_i - Y_{\hEt i} R_d H_u \hEt_i  \nn\\
& - Y_{LFV}^{ij} L_i \cdot H_d \hE_j - Y_{EFV}^{j} R_u H_d E_j
\label{EQ:Superpotential}\end{align}
where 
\begin{align}
W_{adjoint} \equiv\,& L S  + \frac {M_S}{2}S^2 + \frac{\kappa}{3}
S^3 + M_T \textrm{tr}(TT) + M_O \textrm{tr}(OO) \nn\\
&+\lambda_{ST} S\textrm{tr}(TT) +\lambda_{SO} S\textrm{tr}(OO)
 + \frac{\kappa_O}{3} \textrm{tr}(OOO) \nn\\
\underset{\mathrm{R-symmetry}}{\longrightarrow}& 0.
\end{align}
where $T$ and $O$ are in the adjoint representation of $SU(2)$ and $SU(3)$, respectively.
$S$ is a gauge singlet allowing the Dirac mass term of the bino. 
We have written the above in a basis where there are no mass terms $L_i R_d, E_i \hEt_j$ until $S, T$ develop expectation 
values; after electroweak symmetry breaking the couplings $ \lambda_{SLR},  \lambda_{TLR} , \lambda_{SE}$ will give small non-diagonal vector-like mass terms. In addition to the off-diagonal mass terms generated by $ Y_{\hE}, Y_{\hEt}, Y_{LFV}, Y_{EFV}$ these will potentially cause charged lepton flavour violation via rare lepton decays, which we shall describe in section \ref{SEC:LFV} and appendix \ref{APP:CLFV}; the new terms should be considered 
to be merely off-diagonal Yukawa couplings which ought therefore to be small, 
and thus not relevant for the mass spectrum of the model (although relevant for the decays of the new leptons). 
However, note 
that none of the new fields $R_{u,d}, \hE, \hEt$ will obtain a vacuum expectation value and the electroweak symmetry breaking 
proceeds as in \cite{Benakli:2011kz,Benakli:2012cy}.

Hence, in summary, the new Yukawa couplings that are introduced (compared to the models in \cite{Belanger:2009wf,Benakli:2011kz,Benakli:2012cy}) are unimportant for the mass spectrum of the model. Only $\lambda_{SR}, \lambda_{TR}, \lambda_{S\hat{E}\, ij} $ are possibly substantial, and these only affect the masses of the new fields, which are massive and essentially spectators. However,  the reader should bear in mind that the new couplings should still not be exactly zero: their presence is expected, and indeed required to allow the new fields to decay. We have at this point defined the supersymmetric data of a unified model which should ultimately accommodate Dirac gaugino masses.

\subsection{Soft terms}

We can now write down the soft terms allowed in the theory; this subsection serves to establish notation. 
Suppressing gauge indices but retaining generation indices, and denoting complex conjugation of a field by a raised index, the usual soft terms are
\begin{align}
\label{potential4}
- \Delta\mathcal{L}^{\rm scalar\ soft}_{\rm MSSM} =& [ T_u^{ij} U_i Q_j H_u - T_d^{ij} D_i Q_j H_d - T_e^{ij} E_i L_j H_d  + h.c. ]\nn\\
&+ m_{H_u}^2 |H_u|^2 +
m_{H_d}^2 |H_d|^2
 + [B_{\mu} H_u\cdot H_d + h.c. ]\nn\\
&+ Q^i (m_Q^2)_i^j Q_j + U_i (m_u^2)^i_j U^j + D_i (m_d^2)^i_j D^j + L^i (m_l^2)_i^j L_j  + E_i (m_e^2)^i_j E^j  
\end{align}
and there are  soft terms involving the adjoint scalars
\begin{eqnarray}
- \Delta\mathcal{L}^{\rm scalar\ soft}_{\rm adjoints} &= &  (t_S S + h.c.) \nn\\
&&+ m_S^2  |S|^2 + \frac{1}{2} B_S
(S^2 + h.c.)  + 2 m_T^2 \textrm{tr}(T^\dagger T) + (B_T \textrm{tr}(T T)+ h.c.)
\nonumber \\  &&+ 
[A_S \lambda_S SH_u\cdot H_d +  2 A_T \lambda_T H_d \cdot T H_u +
\frac{1}{3} \kappa  A_{\kappa} S^3 + h.c.] \nonumber \\ &&+ 2 m_O^2 \textrm{tr}(O^\dagger O) 
+ (B_O \textrm{tr}(OO)+ h.c.)
\label{Lsoft-DGAdjoint}
\end{eqnarray}
Similarly there are the A-terms for $W_{adjoint}$. We also add the supersoft terms which include the Dirac gaugino masses, coming from holomorphic operators:
\begin{align}
W_{\rm{supersoft}} =& \int d^2 \theta \sqrt{2} \theta^\alpha \bigg[ m_{D1} \mathbf{S} W_{Y\,\alpha} + 2m_{D2} \tr (\mathbf{T} W_{2\,\alpha}) + 2m_{D3} \tr (\mathbf{O} W_{3\,\alpha})    \bigg].
\end{align}
The above are all identical to the model in \cite{Benakli:2012cy}. 

Finally we have the soft terms involving the new vector-like leptons:
\begin{align}
- \Delta\mathcal{L}^{\rm scalar\ soft}_{\rm vector-like} =& m_{R_u}^2 |R_u|^2 + m_{R_d}^2 |R_d|^2 + [ B_R R_d R_u + h.c.] \nn\\
& + \hE_i (m_{\hE}^2)^i_j \hE^j +\hEt^i (m_{\hEt}^2)_i^j \hEt_j + [ B_{\hE}^{ij} \hE_i \hEt_j + h.c.]  
\end{align}

\subsection{Electroweak symmetry breaking}

Since the new states $R_{u,d}, \hE_i, \hEt_j$ carry lepton number, they do not develop vacuum expectation values. Hence electroweak-symmetry breaking occurs exactly as in \cite{Belanger:2009wf,Benakli:2011kz,Benakli:2012cy}, to which we refer the reader, but recall some notation  relevant for the following: we redefine the singlet and triplet scalars in terms of real components $S \equiv \frac{1}{\sqrt{2}} ( v_S + s_R + i s_I), T^0 \equiv \frac{1}{\sqrt{2}} ( v_T + T_R + i T_I) $ and have an ``effective $\mu$-term'' $\tilde{\mu} \equiv \mu + \frac{1}{\sqrt{2}} ( v_S \lambda_S + v_T \lambda_T) $. There are then \emph{four} non-trivial Higgs-potential minimisation conditions, which we must use to fix four low-energy real parameters. The scalars $s_R, T_R$ mix with the Higgs fields, with a $4\times 4$ mass matrix, while $s_I, T_I$ mix with the pseudoscalars to give three pseudoscalar Higgs fields and one would-be goldstone boson.

\section{Minimal ``Natural'' SUSY}
\label{SEC:NATURAL}

Having defined the low-energy parameters of our model, we now turn to considering the effects of running from higher scales. Before turning to specific ultra-violet boundary conditions, we will make some very general comments about what the most \emph{natural} spectrum of masses for our model can be.

Since the model of \ref{SEC:MUDGSSM} can be considered to be a completion into a GUT theory of the work of \cite{Belanger:2009wf,Benakli:2011kz,Benakli:2012cy}, the discussion there about the size of the couplings $\lambda_S, \lambda_T$ and the soft masses $m_S, m_T$ also hold for this model, which we shall not repeat here. 
Furthermore, some of the analysis regarding fine-tuning of \cite{Arvanitaki:2013yja} will also qualitatively apply to our case, although they focussed mostly on the gauge-mediation-inspired case where the Dirac gaugino masses were much larger than the scalar masses, and furthermore since the specific particle content of our model is different we expect the quantitative conclusions to differ. However, since their conclusion was that a certain degree of fine-tuning will be necessary, we shall instead ask in this section: 
\begin{quote}\it
In light of recent bounds on squark masses from the LHC, can we decouple the first two generations from the light spectrum while retaining some vestiges of naturalness?
\end{quote}
This is also relevant for a discussion of flavour, since it is tempting to place the first two generations of squarks to be as heavy as possible in order to avoid flavour constraints (see the recent discussion in \cite{Dudas:2013gga}).  To provide an answer, we shall adapt the approach of \cite{ArkaniHamed:1997ab} (see also \cite{Binetruy:1994bn}) to our model. In doing so, we can draw some rather general conclusions which should apply both for the specific model that we are considering, but also to the completion of the MRSSM into a GUT model (scenario (1) of section \ref{SEC:MUDGSSM}).

We divide the squarks and sleptons into ``light'' and ``heavy'' flavours which we divide into $SU(5)$ multiplets; a complete generation will thus be one $10$-plet and one $5$-plet, but we allow generally $N_5$ heavy $5$-plets and $N_{10}$ heavy $10$-plets. While we allow the ``light'' masses to differ, we shall take all of the heavy states to have a common soft mass $m_{1,2}$. In the interest of naturalness, the Higgs masses should be considered to be ``light''. Then we can examine the RGEs of the model and neglect all ``light'' masses. We also neglect the Yukawa couplings for the first and second generations. With these assumptions, the RGEs are simple enough to be solved analytically; we assume that the heavy states (assuming that the third generation states are light) do not run (we comment on this further below) and find that the two-loop RGEs for the light squarks/sleptons are 
\begin{align}
\frac{d}{dt} m_{\tilde{f}}^2 =& \frac{8}{16\pi^2} \bigg[ m_{1,2}^2 \bigg(\frac{1}{2}( N_5 + 3 N_{10}) \sum_i \alpha_i^2 C_i^f + (N_{5} - N_{10}) \frac{3}{5} Y_f \alpha_1^2 (\frac{4}{3} \alpha_3 -  \frac{3}{4} \alpha_2  -  \frac{1}{12} \alpha_1) \bigg)\nn\\
& + 2 m_T^2 \alpha_2^2 C_2^f  +  3 m_O^2\alpha_3^2 C_3^f \bigg] 
\end{align}
with $t = \log \mu/M_{GUT}$, which differ from the MSSM case by (a) the absence of the one-loop Majorana gaugino mass contribution, and (b) the presence of the adjoint scalars. The RGEs for the adjoint scalars become
\begin{align}
\frac{d}{dt} m_T^2 =& \frac{16\alpha_2^2}{16\pi^2} \bigg[ \frac{1}{2}( N_5 + 3 N_{10}) m_{1,2}^2 + 2 m_T^2  \bigg] \nn\\
\frac{d}{dt} m_O^2 =& \frac{24 \alpha_3^2}{16\pi^2} \bigg[ \frac{1}{2}( N_5 + 3 N_{10}) m_{1,2}^2 + 3 m_O^2  \bigg]. 
\end{align}
These both place limits on how heavy the first two generations can be if we demand that there is not large tuning of the initial masses; in the case of the adjoint scalars, if they are initially much lighter than the first two generations then they may be driven tachyonic, and if they are of comparable mass then they will help to drive the lighter generations tachyonic. 
We next require the  gauge couplings, which we solve at one-loop order  assuming gauge coupling unification:
\begin{align}
\alpha_{1,2} (t) =& \frac{\alpha_1 (0)}{ 1 - \frac{b_{1,2} t }{2\pi} \alpha_{1,2}(0)} \nn\\ 
\alpha_{3} (t) \simeq& \alpha_{3} (0)
\end{align}
where $b_1 = 16\times \frac{3}{5}, b_2 = 4$ and hence, writing $C_g^f $ for the quadratic casimir of fermion $f$ under group $g$ we have
\begin{align} 
m_O^2 (\mu) =& - \frac{1}{6} ( N_5 + 3 N_{10}) m_{1,2}^2 + \bigg( m_O^2 (0) +  \frac{1}{6} ( N_5 + 3 N_{10}) m_{1,2}^2 \bigg) \left( \frac{\mu}{M_{GUT}} \right)^{\frac{72 \alpha_3^2 (0)}{16\pi^2}} \nn\\
m_O^2 (\mathrm{10\ TeV}) \simeq&  - 0.030 ( N_5 + 3 N_{10}) m_{1,2}^2 +  0.81  m_O^2 (0) \underset{\small \mathrm{two\ heavy\ generations}}{\longrightarrow} -0.24 m_{1,2}^2 +  0.81  m_O^2 (0) \nn\\
m_T^2 (\mu) =& - \frac{1}{4} ( N_5 + 3 N_{10}) m_{1,2}^2 + \bigg( m_T^2 (0) +  \frac{1}{4} ( N_5 + 3 N_{10}) m_{1,2}^2 \bigg) \exp \bigg[ \frac{8\alpha_{G} }{16\pi^2} \frac{2^2 \times 4\pi}{2b_2} (\alpha_2(\mu) - \alpha_{G}) \bigg]\nn\\
m_T^2 (\mathrm{10\ TeV}) \simeq& -0.0008 ( N_5 + 3 N_{10}) m_{1,2}^2 +  0.997  m_T^2 (0) \underset{\small \mathrm{two\ heavy\ generations}}{\longrightarrow} 0.006 m_{1,2}^2 +  0.997  m_T^2 (0) \nn\\ 
m_{\tilde{f}}^2 (\mu) =& m_{\tilde{f}}^2 (0) \nn\\
&- \frac{8 m_{1,2}^2}{16\pi^2} \bigg[ \frac{1}{2}( N_5 + 3 N_{10}) \bigg( \alpha_3^2 C_3^f \log \frac{M_{GUT}}{\mu} +  \sum_{i=1}^2 \frac{4\pi}{2b_i} ( \alpha_{GUT} - \alpha_i (\mu)) C_i^f  \bigg) \nn\\
&- (N_5 - N_{10}) \frac{3}{5} 4 \pi Y_f \bigg( \frac{4}{3} \frac{\alpha_{GUT}}{b_1 - b_3} \frac{1}{2} \log \frac{\alpha_1(\mu)}{\alpha_3 (\mu)} - \frac{3}{4} \frac{\alpha_{GUT}}{b_2 - b_3} \frac{1}{2} \log \frac{\alpha_1(\mu)}{\alpha_2 (\mu)} + \frac{1}{12} \frac{1}{2b_1} (\alpha_{GUT} - \alpha_1 (\mu)\bigg)\bigg] \nn\\
& + C_3^f \frac{1}{3} \bigg[ (m_O^2(\mu) - m_O^2(0)) +  \frac{1}{2}( N_5 + 3 N_{10}) m_{1,2}^2  \frac{24}{16\pi^2} \alpha_3^2 \log \frac{M_{GUT}}{\mu}\bigg] \nn\\
& +C_2^f \frac{1}{2}\bigg[ (m_T^2(\mu) - m_T^2(0)) +  \frac{1}{2}( N_5 + 3 N_{10})  m_{1,2}^2\frac{16}{4\pi} \frac{1}{2b_2} ( \alpha_{GUT} - \alpha_2 (\mu)) \bigg].
\end{align}
This gives
\begin{align}
m_{\tilde{f}}^2 (\mathrm{10\ TeV}) =& m_{\tilde{f}}^2 (M_{GUT}) \nn\\
& - 0.06 C_f^3 m_O^2 (M_{GUT}) - 0.003 C_f^2 m_T^2 (0) \nn\\
& -  N_{10} m_{1,2}^2 (M_{GUT}) \times 10^{-2} \times ( 0.5 C_f^1 + 0.12 C_f^2 + 3.0 C_f^3 - 0.3 Y_f) \nn\\
& - N_5 m_{1,2}^2 (M_{GUT}) \times 10^{-2} \times  ( 0.17 C_f^1 + 0.04 C_f^2 + 1.0 C_f^3 + 0.3 Y_f).
\end{align}
The dependence on $m_{1,2}^2$ is increased for strongly coupled particles by a factor of about $3$ compared to the Majorana case; the largest contribution actually comes from feeding the running of $m_O$ into the sfermion masses, but there is also a significant contribution from the $ m_{1,2}^2 \alpha_3^2 C_f^3$ term. In addition, there is no Majorana contribution that can lift the masses - and so we genuinely have an upper bound on the splitting between $m_{1,2}^2$ and the third generation masses, at least at the GUT scale. We also see that in fact there is some running even for the first two generations, at the order of $10\%$ variation in the mass-squareds (so this is not enough to badly affect the approximation of no running for $m_{1,2}$). 

Clearly the most significant contributions come from $m_O$ and from the strong sector; we can write
\begin{align}
m_{\tilde{Q}}^2 (\mathrm{10\ TeV}) \simeq& m_{\tilde{Q}}^2 (M_{GUT})  - 0.08 m_O^2 (M_{GUT}) -0.04 m_{1,2}^2 N_{10} - 0.014 m_{1,2}^2 N_5 \nn\\
m_{\tilde{U}}^2 (\mathrm{10\ TeV}) \simeq& m_{\tilde{U}}^2 (M_{GUT})  - 0.08 m_O^2 (M_{GUT}) -0.04 m_{1,2}^2 N_{10} - 0.012 m_{1,2}^2 N_5 \nn\\
m_{\tilde{D}}^2 (\mathrm{10\ TeV}) \simeq& m_{\tilde{D}}^2 (M_{GUT})  - 0.08 m_O^2 (M_{GUT}) -0.04 m_{1,2}^2 N_{10} - 0.015 m_{1,2}^2 N_5
\end{align}
and when there are two heavy generations we can put that $\Delta m_{\tilde{Q},\tilde{U},\tilde{D}}^2 \equiv m_{\tilde{Q},\tilde{U},\tilde{D}}^2 (\mathrm{10\ TeV}) -  m_{\tilde{Q},\tilde{U},\tilde{D}}^2 (M_{GUT})  \simeq -0.1 m_{1,2}^2 -0.08 m_O^2 $. Hence if the octet scalars are light then the maximum splitting we can have between the first two generations and the third without tuning is 
$$m_{\tilde{f}_3}^2(\mathrm{10\ TeV}) \gtrsim 0.1 \,m_{1,2}^2,$$
corresponding to a hierarchy of only a factor of about $3$, in contrast to the MSSM where a hierarchy of about $10$ is naturally allowed \cite{ArkaniHamed:1997ab}. This has potentially significant consequences for flavour constraints from meson oscillations; we refer the reader to \cite{kpw, Fok:2012me, Dudas:2013gga} for discussion of this in the context of Dirac gaugino models to which this should apply. However, it more significantly suggests that models without a hierarchy between the generations at the GUT scale are preferred: we shall take this as a motivation for our constrained construction in section \ref{SEC:CMDGSSM}.

\section{Charged lepton constraints}
\label{SEC:LFV}

Our model contains new vector-like leptons and it is important to consider what the constraints on these should be. Indeed, the direct search constraints are surprisingly weak and come from LEP;  see e.g. \cite{Achard:2001qw,Carpenter:2010bs}. They are comparable to other electroweak fermion searches and are of the order of $100$ GeV. However, there are much stronger indirect constraints that we shall discuss below. 

Supersymmetric models typically predict some flavour-violation at the one-loop level via interactions mediated via scalar superpartners of Standard Model fields and gauginos. The strongest flavour constraints usually come from $\Delta F = 2$ meson oscillations  (see \cite{kpw, Dudas:2013gga} for discussion of this in the context of Dirac gaugino models). In addition, there are the $\Delta F = 1$ amplitudes which induce processes such as $\mu \rightarrow e \gamma$ and electric dipole moments. These have also been discussed in the context of the MRSSM in \cite{kpw, Fok:2012me}. These provide constraints on the flavour structure of the soft terms (in particular, any CP violating phases are strongly constrained) and  also apply to our model defined in section \ref{SEC:MUDGSSM}. However, in this model there are additional contributions to charged lepton flavour violation (cLFV) induced by the presence of the additional vector-like leptons. Here we shall discuss the constraints imposed by these. 

The extra particle content of our model relevant for cLFV is similar to the models considered in \cite{Kannike:2011ng,Dermisek:2013gta,Ishiwata:2013gma}, except we have an additional right-handed electron pair -- and also our model is supersymmetric. However, the novel contributions are very similar to those models and we shall adapt the approach of \cite{Ishiwata:2013gma} to our case. We provide a detailed discussion in appendix \ref{APP:CLFV} and summarise the results here.

\subsection{$\mu \rightarrow 3e$}

One of the interesting consequences of new vector-like leptons is the possibility of processes such as $\mu \rightarrow 3e$ at \emph{tree level}. This occurs because the diagonalisation of the lepton mass matrices no longer leads to exactly diagonal neutral currents - i.e. we can induce to a small degree $\mu \rightarrow e Z^* \rightarrow 3e$. 

As described in appendix \ref{APP:mu3e}, if we put all masses to a TeV we obtain
\begin{align}
BR (\mu \rightarrow 3e) =&  2\times 10^{-4} c_\beta^2 \bigg[ |Y^{2k}_{LFV} \ov{Y}^{1k}_{LFV}|^2 + |Y^{2}_{EFV} \ov{Y}^{1}_{EFV}|^2 \bigg]\left( \frac{\mathrm{TeV}}{\mu_{E,R}} \right)^{4} .
\end{align}
Comparing to the experimental bound
\begin{align}
BR (\mu \rightarrow 3e) < 1.0 \times 10^{-12}
\end{align}
we find 
\begin{align}
|Y^{2k}_{LFV} \ov{Y}^{1k}_{LFV}|^2  \lesssim& 10^{-8}\left( \frac{\mathrm{TeV}}{\mu_{E,R}} \right)^{4} \nn\\
Y^{ik}_{LFV} \sim Y^{j}_{EFV} \lesssim& 10^{-2} \left( \frac{\mathrm{TeV}}{\mu_{E,R}} \right),
\end{align}
which can be relaxed a little for large $\tan \beta$.

\subsection{$\mu \rightarrow e \gamma$}

The current experimental bound on the branching ratio for the process $\mu \rightarrow e \gamma $ is 
\begin{align}
BR (\mu \rightarrow e \gamma) < 2.4 \times 10^{-12}.
\end{align}
This occurs via an amplitude having the structure \cite{Lavoura:2003xp}
\begin{align} 
\mathcal{A}_{ij} =& e \epsilon_\mu^* \ov{e}_i \bigg[ i \sigma^{\mu\nu} q_\nu ( \sigma_L P_L + \sigma_R P_R) + \delta_L \Delta_L^\mu + \delta_R \Delta_R^\mu \bigg] e_j, 
\end{align}
where $e$ is the electromagnetic coupling and $e_i$ are the fermions; $P_{L,R}$ are chiral projectors, $q_\nu$ is the momentum of the outgoing photon and $\Delta_{L,R}^\mu $ vanish on shell. The constraint on the quantities $\sigma_{L,R}$ is 
\begin{align}
\sigma_{L,R} <& 5.8 \times 10^{-13} \mathrm{GeV}^{-1}.
\end{align}
These processes are induced by many diagrams. However, the interesting new contributions come from loops containing a higgs (neutral or charged) and heavy lepton, which are enhanced compared to the usual supersymmetric diagrams due to the possibility of a chirality flip (due to the vector-like mass of the new fermions). In the limit of small mixing between the MSSM Higgs fields and the singlet/triplet scalars, we can understand the contributions as either coming from the couplings giving off-diagonal mass terms $Y_{EFV}, Y_{LFV}, Y_{\hE}, Y_{\hEt} $ involving the MSSM Higgses, or from the direct cLFV couplings involving the heavy singlet and triplet scalars $\lambda_{SLE}, \lambda_{SR}, \lambda_{TLR} $. In the first case, in the limit of large vector-like lepton mass $\mu_E \sim \mu_R \equiv \mu_{E,R} \sim$ TeV and all the couplings of a similar order of magnitude  we obtain
\begin{align}
Y_{EFV} Y_{LFV} Y_{\hE}  \lesssim& 1.8 \times 10^{-7} \left(\frac{\mu_{E,R}}{\mathrm{TeV}} \right)^2
\end{align}
or
\begin{align}
Y_{EFV} \sim Y_{LFV} \sim Y_{\hE} \sim Y_{\hEt} \lesssim& 10^{-2} \left(\frac{\mu_{E,R}}{\mathrm{TeV}}\right)^{2/3},
\end{align} 
which is very mild when we consider that these should be equivalent to off-diagonal Yukawa couplings. 

In the second case, under the same assumptions, we find
\begin{align}
\lambda_{SLE} \lambda_{SE} Y_{\hEt}  \lesssim& 10^{-7} \left(\frac{\mu_{E,R}}{\mathrm{TeV}}\right)^2 \left( \frac{v}{v_S} \right) \nn\\
\lambda_{TLR} Y_{EFV} Y_{\hE} \lesssim& 10^{-5} \left(\frac{\mu_{E,R}}{\mathrm{TeV}}\right)^2 \left( \frac{\mathrm{GeV}}{v_T} \right)
\end{align}
or 
\begin{align}
\lambda_{SLE} \sim \lambda_{SE} \lesssim& 10^{-2}  \left(\frac{\mu_{E,R}}{\mathrm{TeV}}\right)^{2/3} \left( \frac{v}{v_S} \right) \nn\\
\lambda_{TLR}  \lesssim& \mathcal{O}(1)  \left( \frac{\mu_{E,R}}{\mathrm{TeV}} \right)^{2/3} \left( \frac{\mathrm{GeV}}{v_{T}} \right).
\end{align} 
The coupling $\lambda_{TLR}$ is not constrained by this diagram in the limit of no mixing with the lighter Higgs states or equivalently $v_T = 0$.

\subsection{Electron electric dipole moment}

The recently improved upper bound on
the electron dipole moment  \cite{Baron:2013eja} of 
\begin{align}
|d_e| <& 8.9 \times 10^{-29} e\ \mathrm{cm } = 4.5 \times 10^{-15}e\ \mathrm{GeV}^{-1}
\end{align}
places a restriction on our model similar in nature to that from $\mu \rightarrow e \gamma$:
\begin{align}
|\mathrm{Im}(\sigma_{L,R})| <& 2.3 \times 10^{-15} \mathrm{GeV}^{-1},
\end{align}
which for arbitrary phases in the new couplings corresponds to 
\begin{align}
Y_{EFV} \sim Y_{LFV} \sim Y_{\hE} \sim Y_{\hEt} \lesssim& 10^{-3} \left( \frac{\mu_{E,R}}{\mathrm{TeV}} \right)^{2/3} 
\end{align}
and
\begin{align}
\lambda_{SLR} \sim \lambda_{SE} \lesssim& 10^{-3} \left( \frac{\mu_{E,R}}{\mathrm{TeV}} \right)^{2/3} \left( \frac{v}{v_{S}} \right) \nn\\
\lambda_{TLR}  \lesssim& \mathcal{O}(0.1)  \left( \frac{\mu_{E,R}}{\mathrm{TeV}} \right)^{2/3} \left( \frac{\mathrm{GeV}}{v_{T}} \right),
\end{align}
i.e. this is now the strongest constraint on the model, although note that for purely real couplings the constraint disappears.

Thus, the new couplings (with the possible exception of $\lambda_{TLR}$) are constrained to a level that they cannot play a significant role in the RGE evolution of the other parameters -- as we would expect, being as they are off-diagonal Yukawa couplings -- while they may still be large enough to allow prompt decay of the new states.

\section{The CMDGSSM}
\label{SEC:CMDGSSM}

Now that we have defined the fields and couplings of the model and determined the constraints upon them 
we are in a position to introduce a constrained model at the GUT scale for phenomenological anaylses. 
The aim of this section 
is to present a model analagous to the constrained MSSM or minimal SUGRA boundary conditions but with 
Dirac gaugino masses. The motivation for this is that we can give predictions for the many low-energy 
parameters that we know can come from a well-defined and natural high-energy completion, and this 
then can facilitate meaningful phenomenological studies. 

We define the following parameters at the GUT scale: 
\begin{itemize}
\item $m_0$, familiar to the CMSSM or mSUGRA scenario, is a common supersymmetry-breaking scalar mass for the new vector-like leptons $R_{u,d}, \hE_i, \hEt_i$ and all scalar superpartners of Standard Model fields with the exception of the Higgs.
\item We take for expediency the Higgs to have non-universal mass-squareds (NUHM), choosing instead to fix $\mu$ and $B\mu$ via the tadpole conditions, which is the only source of $R$-symmetry violation in the model. This could in principle be dropped to make a more constrained model. Note that we set  all $A$-terms and Majorana gaugino masses to zero since they violate $R$-symmetry; this condition is perfectly consistent and preserved by the RGE running. 
\item $m_{D0}$ is a common Dirac mass for all gauginos.
\item $\lambda_S, \lambda_T$ are the important supersymmetric couplings to the Higgs. In principle they could be set equal to their $N=2$ values at the GUT scale where $\lambda_T = g_{GUT}/\sqrt{2}, \lambda_S = g_{GUT} \sqrt{3/10}$ (for the unified gauge coupling $g_{GUT}$) although $\lambda_S$ would depend on the actual GUT embedding of the singlet; we choose not to take these values. 
\item We set a common adjoint scalar mass $m_\Sigma$ for the triplet and octet, but allow a mass $m_S$ to differ for the singlet. In principle these could be restricted to be equal in scenarios where $\lambda_S, \lambda_T$ are small (or indeed in a unified version of the MRSSM) but there is no top-down reason to suppose that the singlet mass should be equal to the other adjoint scalar masses so we retain the more general case. For simplicity we take the $B$-type adjoint masses to be zero. 
\end{itemize}
In addition, there are several additional parameters that have little impact on the phenomenology of the Higgs or coloured particles. We must define
\begin{itemize}
\item $\mu_E, \mu_R$ are the supersymmetric mass parameters for the new vector-like leptons; and their couplings $\lambda_{(S,T) E}, \lambda_{(S, T) R} $, in addition to the lepton-flavour violating Yukawa couplings $Y_{LFV}, Y_{EFV}$. Due to the constraints on $Y_{\hE},Y_{\hEt},Y_{LFV}, Y_{EFV} $ to be small (which, since they are essentially off-diagonal Yukawa couplings, we would expect) we can choose to neglect these couplings in anaylses of the spectrum of the model, and set $\mu_E, \mu_R$ to a reference value (we shall take $1$ and $1.5$ TeV respectively in the scans below). However, for the purposes of the phenomenology of the model, we should understand that $Y_{\hE},Y_{\hEt},Y_{LFV}, Y_{EFV}$ are non-zero as they allow the additional leptons to decay. Since they should be $\lesssim 10^{-3}$ their values are not substantially affected by the running from the GUT scale so in collider studies appropriate choices can be made a posteriori to allow prompt decays. 
\end{itemize}

\subsection{First forays}

\begin{figure}[!h]
\includegraphics[width=0.75\textwidth]{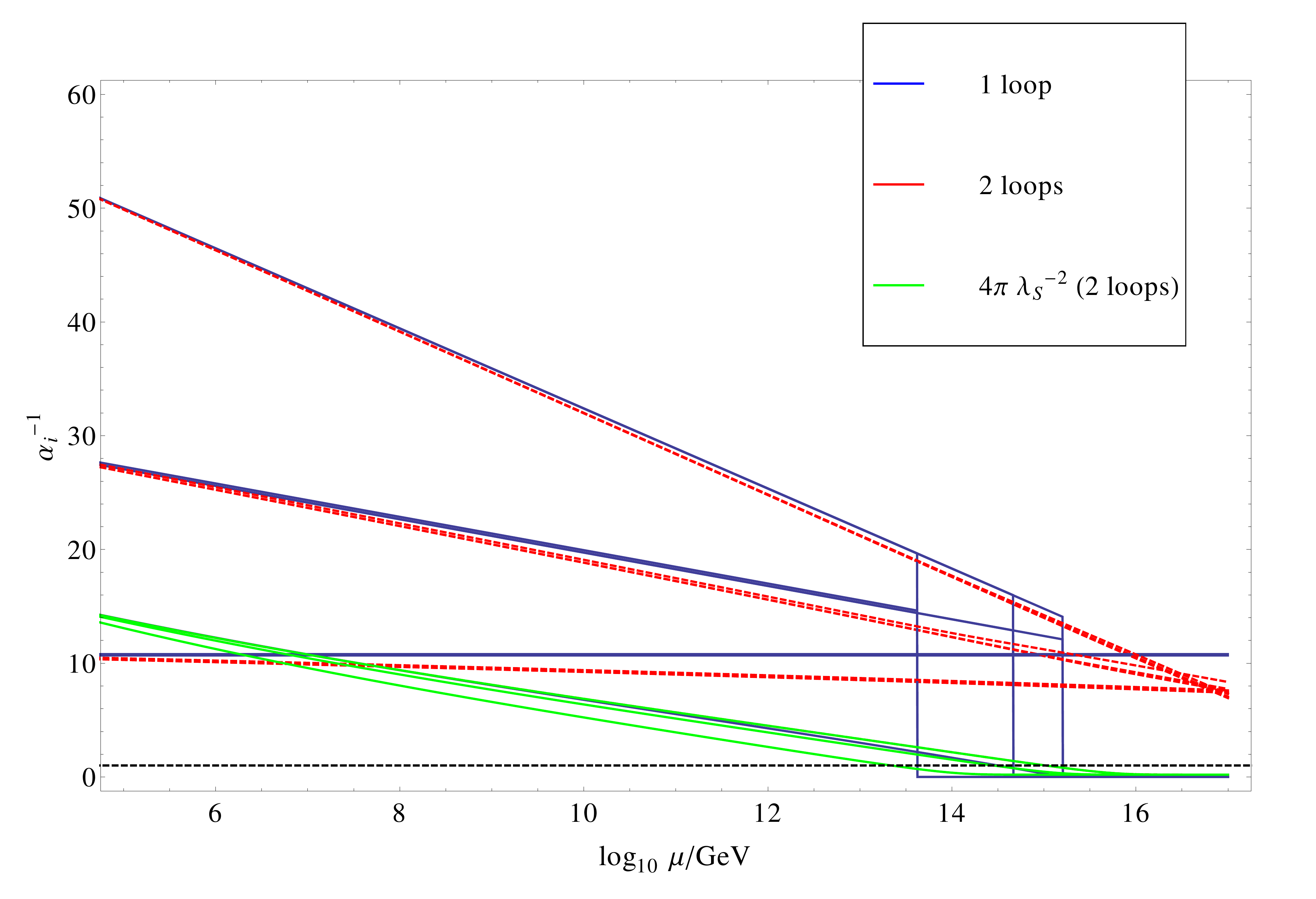} 
\caption{Unification varying $\lambda_S$: this plot as in figure \ref{FIG:SU33UNI}, shows the running of the gauge couplings with energy for the field content described in section \ref{SEC:MUDGSSM}, but with $\lambda_S = 0.85$ at the SUSY scale and showing the running of $4\pi/\lambda_S^2$ in green. The couplings become nonperturbative around $10^{14}$ GeV; note that in the one-loop case $\lambda_S$ becomes divergent whereas corresponding to abrupt changes in the gauge couplings, whereas in the two-loop case $\lambda_S$ merely becomes large, giving the misleading impression that the curves for the gauge couplings are smooth. }
\label{FIG:SU33UNIlambda}\end{figure}

We now describe the results of a first probe of the parameter space of the CMDGSSM, in the corner where $\lambda_S$ is large at low energies to provide a significant tree-level boost to the Higgs mass. We scan over random values within limited ranges (with a flat distribution) of the following parameters:
\begin{itemize}
\item $\tan \beta \in [1.5,3]$
\item $m_0 \in [1000,6000]$  GeV
\item $m_{D0} \in [500,1700]$  GeV
\item $m_{S} (M_S) \in [100,1400]$  GeV
\item $m_{\Sigma} (M_{GUT}) \in [1200,4200]$  GeV
\item $\lambda_S (M_S) \in [0.65,0.78]$
\item $\mu (M_{GUT})\in [150,1000]$ GeV
\item $\sqrt{B_\mu} (M_{GUT}) \in [200,1200]$  GeV
\end{itemize}
while fixing $\mu_E, \mu_R$ to be $1$ and $1.5$ TeV respectively. All other new Yukawa couplings ($Y_{\hE},Y_{\hEt},Y_{LFV}, Y_{EFV},\lambda_{SR}, \lambda_{SE}, \lambda_{SLR}, \lambda_{TLR} $) are taken to be zero, with the exception of $\lambda_T$, which for computational reasons is set to $10^{-7}$ (effectively zero).  The above ranges of parameters were chosen to find points such that the major contribution to the Higgs mass comes from the tree-level contribution via $\lambda_S$, necessitating small values of $\tan \beta$. We then have an effective ``$\lambda$SUSY in disguise'' scenario. Similar to the usual $\lambda$SUSY case \cite{Barbieri:2006bg,Hall:2011aa} there is an upper bound on $\lambda_S$ consistent with perturbative unification of about $\lambda_S (M_S) \simeq 0.7$ (this is illustrated in figure \ref{FIG:SU33UNIlambda}). This requirement leads to us only finding models for which $\tan \beta \gtrsim 2$. In the scans we keep only points where perturbative unification occurs, but we choose to specify $\lambda_S, m_S $ at the SUSY scale (rather than the GUT scale) due to the sensitivity of the parameter to the running: small changes in $\lambda_S \sim 0.7$ at $M_S$ lead to large changes at $M_{GUT}$ and so it is easier to locate the desired values by specifying them at the low scale.

To perform the scans we implemented the model in the spectrum-generator generator \SARAH and produced \SPheno code. We then modified the \SPheno code so that the low-energy solution of the tadpole equations would be for the parameters $m_{H_u}^2, m_{H_d}^2, v_S, v_T$. In particular, the equations for $v_S, v_T$ are not in general linear -- this required numerical solution, which we implemented via Broyden's method; this possibility has now been included in the latest version of \SARAH. Further details of the implementation are given in appendix \ref{APP:SARAH}.  

In the scans presented below we only retain points for which the one-loop mass for the Standard Model-like Higgs mass is between $123$ and $127$ GeV (which is probably  overly restrictive given that two-loop corrections are not yet available \cite{Goodsell:2014inprep}), and satisfies all Higgs constraints which we check using \HiggsBounds \cite{Bechtle:2008jh,Bechtle:2009ic,Bechtle:2011sb,Bechtle:2013wla}. 
Furthermore, the above ranges of masses were chosen so as to be safe for LHC searches. We have been very conservative in that, rather than performing a full collider check for the points of our model point by point, by ensuring a lower value of $m_{D0}$ of $500$ GeV we find the gluino to be heavier than $1750$ GeV (see below), easily above the current bounds; and furthermore, since it has a Dirac mass, this suppresses the sensitivity of searches to the squarks. Finally, the remaining limits are on electroweak-charged particles, for which the best lower bound still comes from LEP of $105$~GeV (since we find the sleptons to be heavy in this model), and we show this in the plots where appropriate.

\begin{center}
\begin{figure}[!h]
\includegraphics[width=\textwidth]{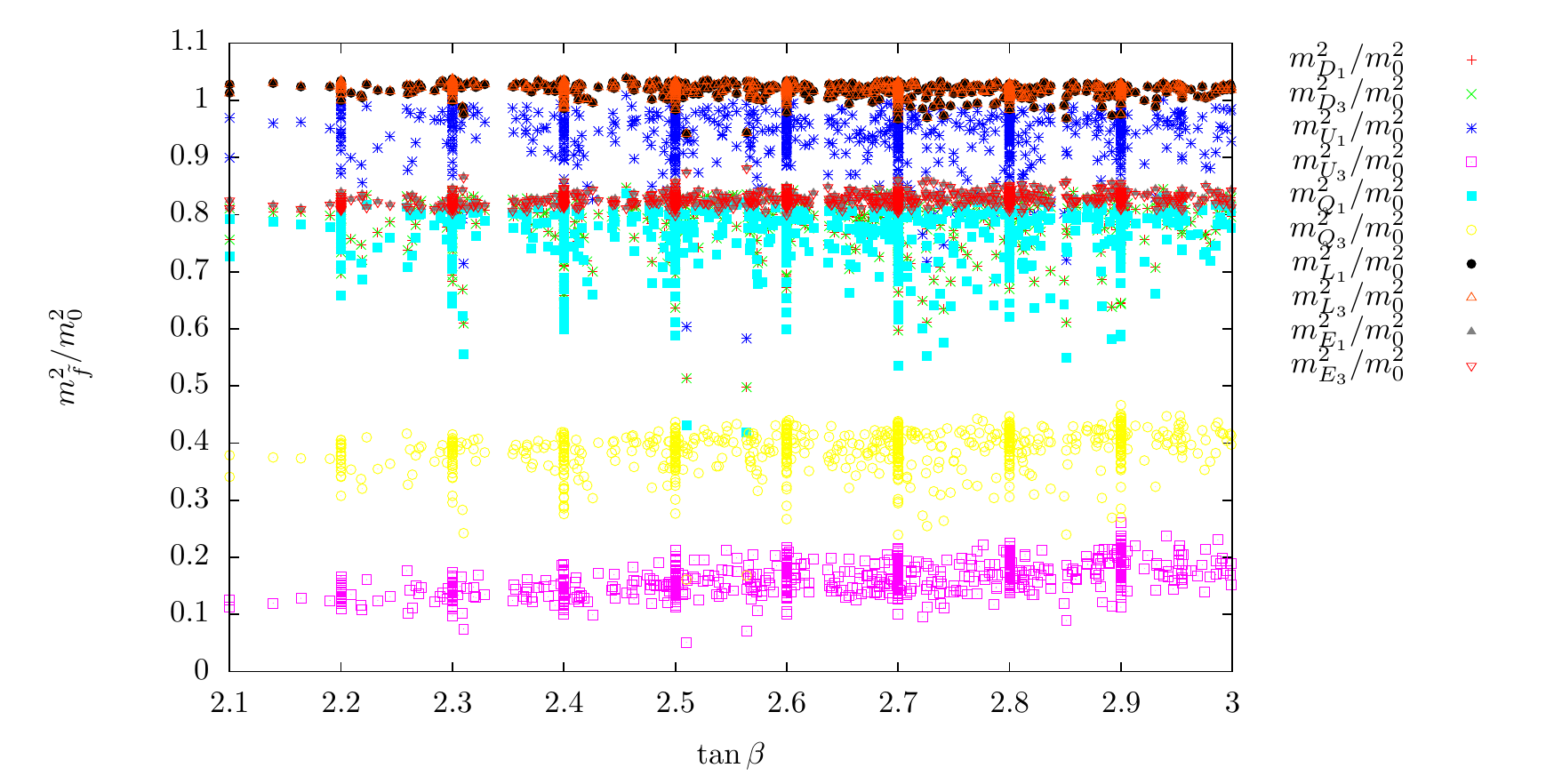}
\caption{Ratio of sfermion mass-squareds to $m_0^2$ against $\tan \beta$.}
\label{FIG:p1}\end{figure}
\end{center}

\begin{center}
\begin{figure}[!h]
\includegraphics[width=\textwidth]{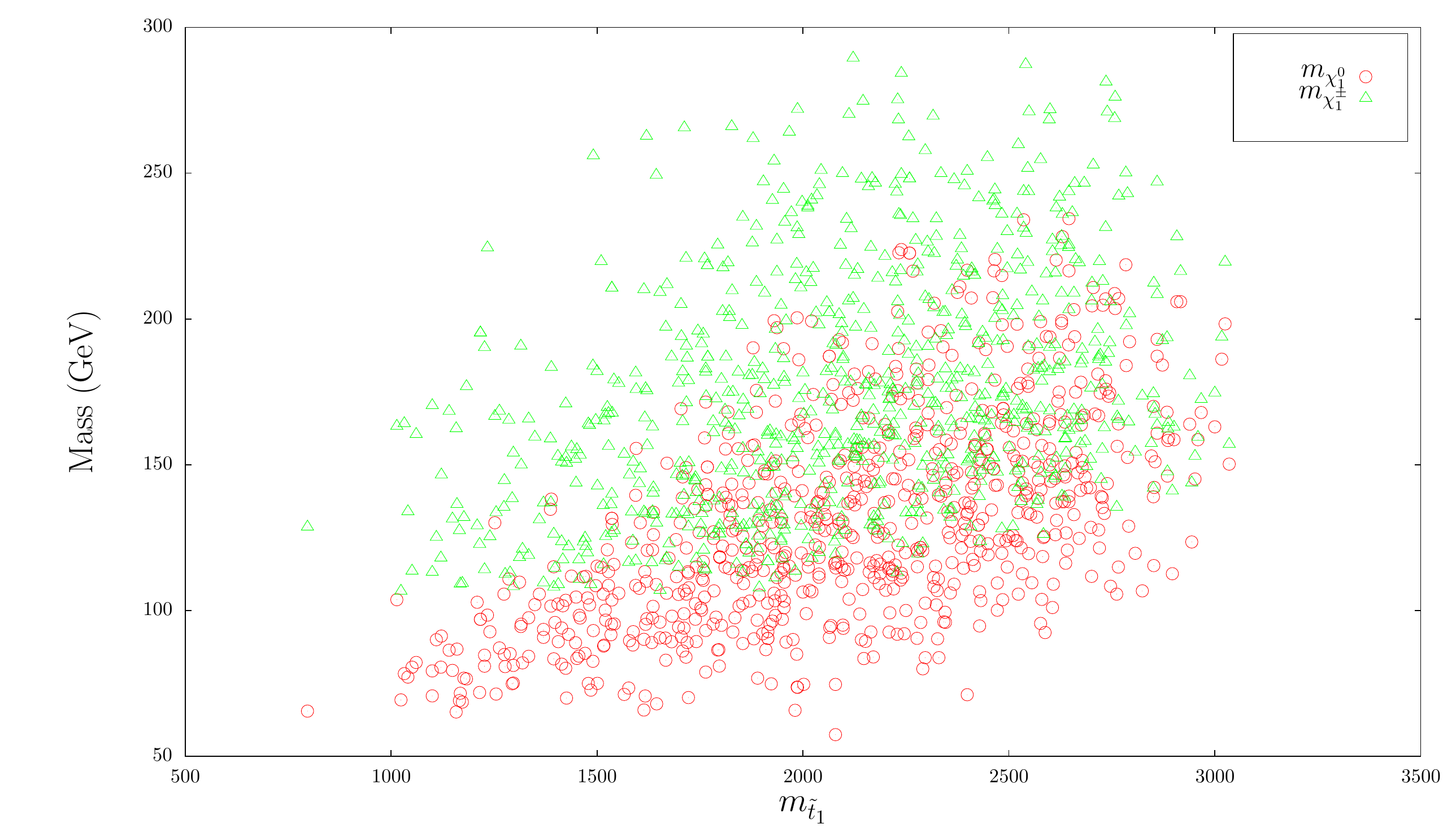}
\caption{Distribution of lightest neutralino and chargino masses against lightest stop mass.}
\label{FIG:p2}\end{figure}
\end{center}

\begin{center}
\begin{figure}[!h]
\includegraphics[width=\textwidth]{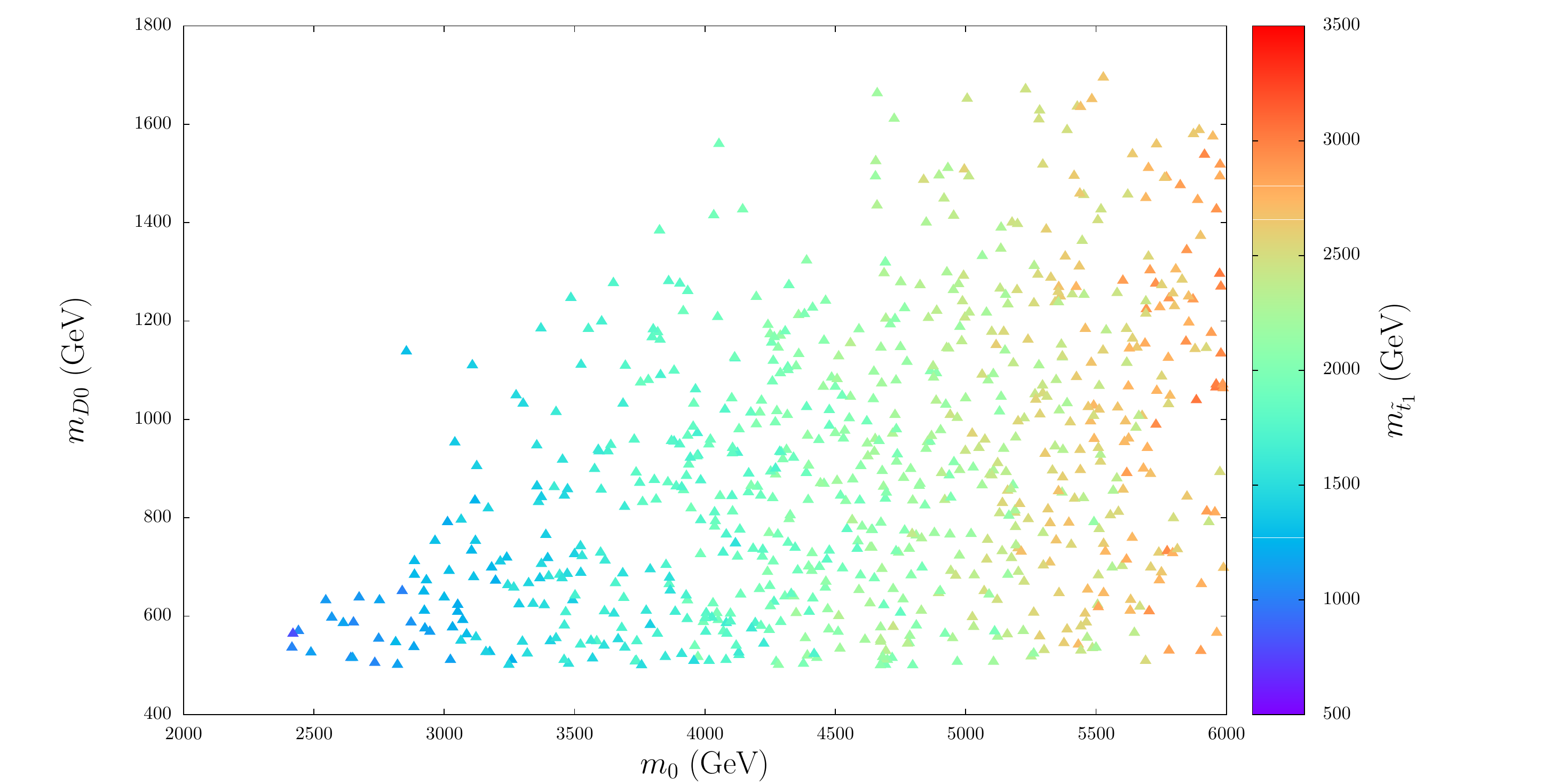}
\caption{Valid model points in the $m_0-m_{D0}$ plane, colour showing the lightest stop mass.}
\label{FIG:p3}\end{figure}
\end{center}

\begin{center}
\begin{figure}[!h]
\includegraphics[width=\textwidth]{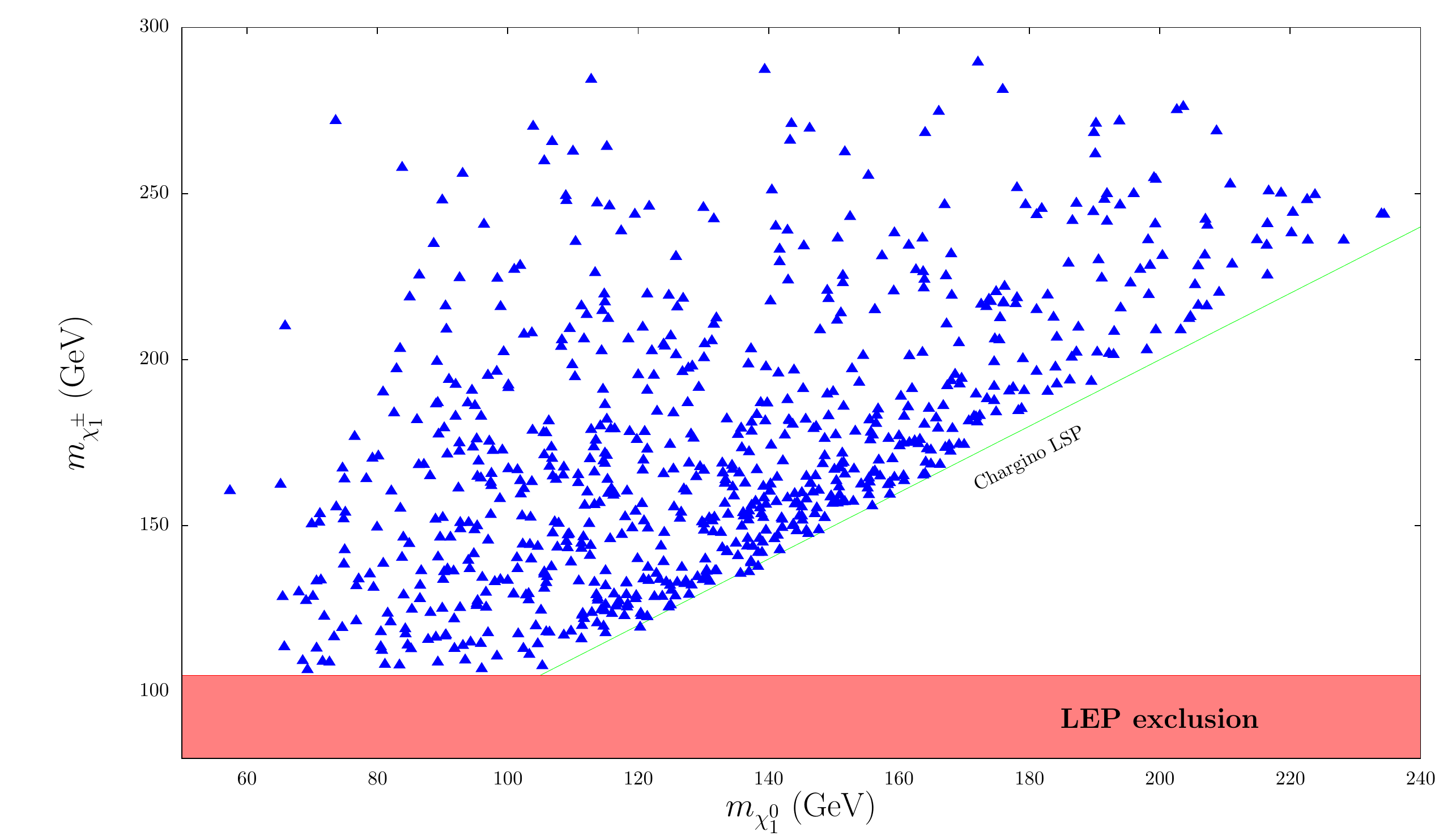}
\caption{Lightest neutralino versus lightest chargino mass. Points lying near the line of equality are primarily Higgsino-like; otherwise we find a bino-like LSP.}
\label{FIG:p4a}\end{figure}
\end{center}

\begin{center}
\begin{figure}[!h]
\includegraphics[width=0.9\textwidth]{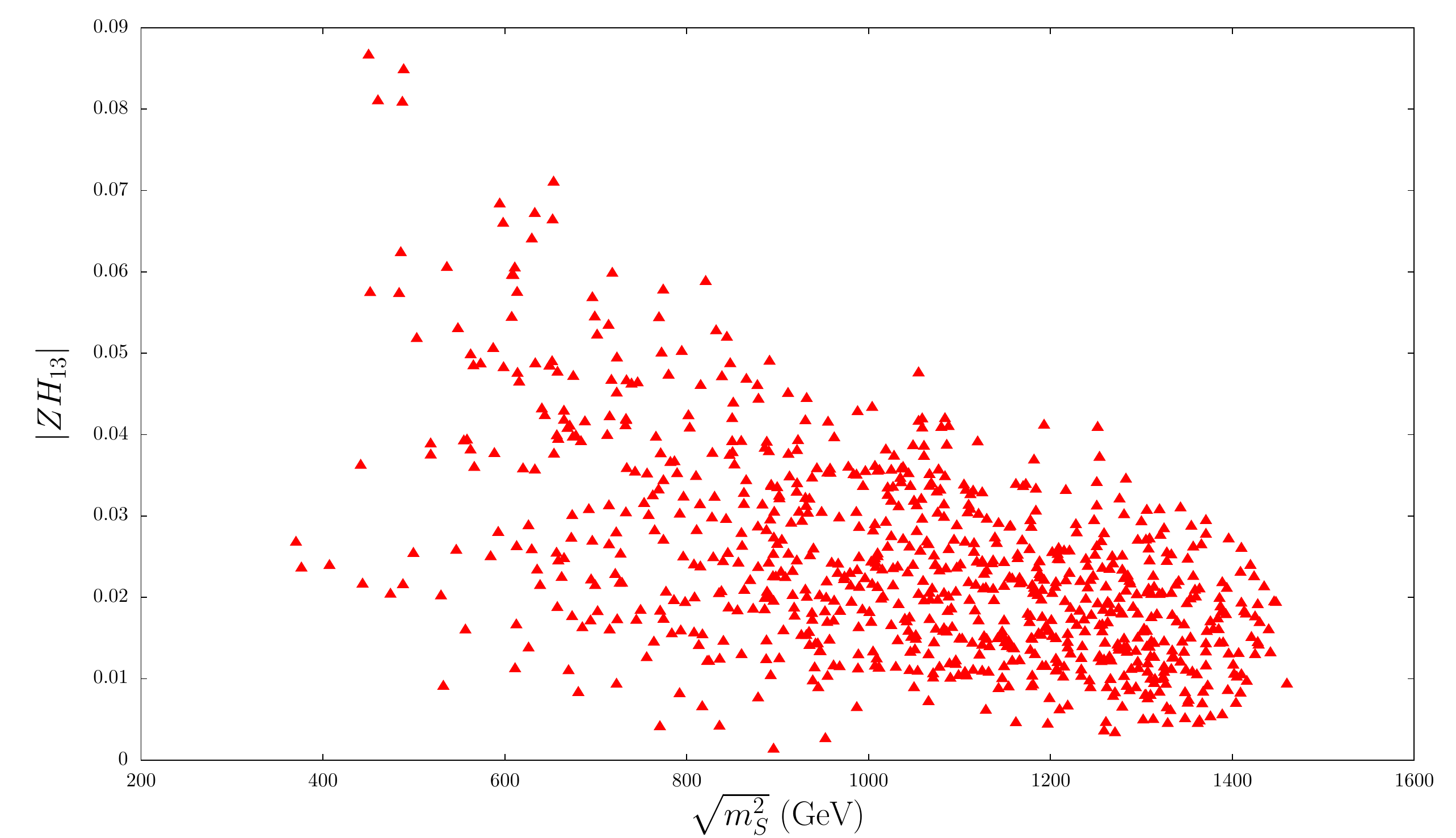}\vspace{-0.5cm}
\caption{Soft singlet mass $m_S$ against mixing between lightest Higgs and the singlet (given as the absolute value of the mixing matrix element $ZH_{13}$). There is only small mixing for all model points found, but this of course decreases as the singlet becomes heavier.}
\label{FIG:p5}\end{figure}
\end{center}

\begin{center}
\begin{figure}[!h]
\includegraphics[width=\textwidth]{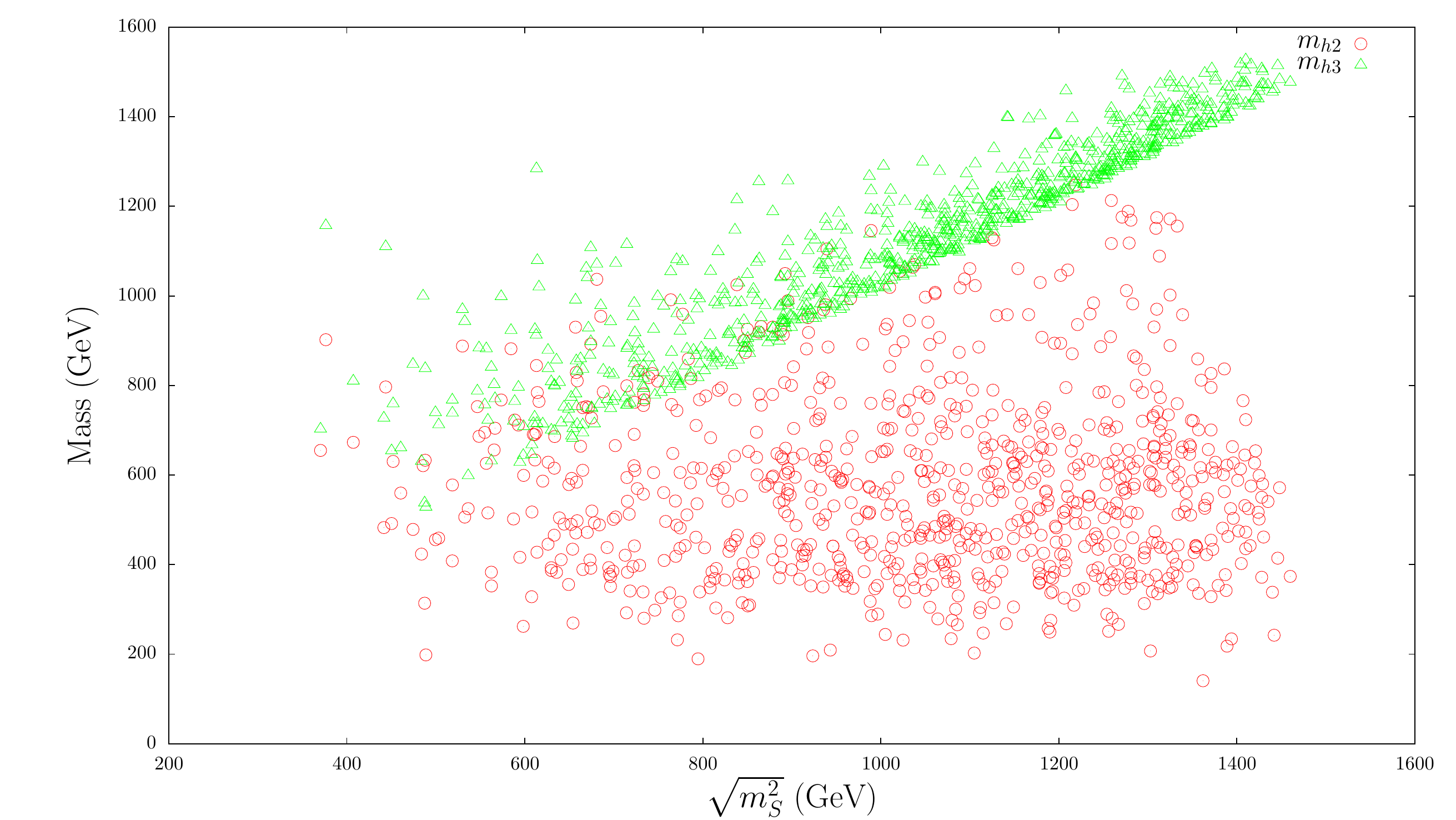}\vspace{-0.5cm}
\caption{Soft singlet mass $m_S$ the mass of the second and third Higgses. The third Higgs is to a very good approximation singlet-like for heavier $m_S$.}
\label{FIG:p6}\end{figure}
\end{center}

We present the results of the scans in a series of figures:
\begin{enumerate}
\item In figure \ref{FIG:p1} we show the ratios of all the sfermion mass-squareds to the initial value $m_0^2$ 
against $\tan \beta$, for which there is only weak dependence. The results can be compared to those of section 
\ref{SEC:NATURAL}. The points show only small deviations from the averages, although there are significant 
outliers which come from finely-tuned points with large Higgs mass-squareds: in these cases the large 
contributions of the Higgs soft masses enter into the RGEs for sfermions charged under $U(1)_Y$ and $SU(2)$ at one loop.

\item In figure \ref{FIG:p2} we show the distribution of the lightest neutralino and chargino masses against the lightest stop mass. We do find stops as light as $800$ GeV, although their typical value is above 1 TeV. 
\item In figure \ref{FIG:p3} we show our equivalent of the classic $m_0-m_{D0}$ plane, with here  colour showing the lightest stop mass. Strikingly we find that the upper left portion of the plane is unpopulated, because in that region the bino mass becomes large enough to destabilise the Higgs potential. This arises because of our desire for naturalness: we have placed an upper limit on $m_S$ of $1400$ GeV. The off-diagonal mixing term in the Higgs mass-squared matrix between the light Higgs and the singlet is, at tree-level, approximately $ (M_{h}^2)_{13} \simeq - g_Y v m_{DY} c_{2\beta}$ and so the upper limit on $m_S$ means that when $ m_{DY}^2  \sim m_S^2$ the tree-level potential is no longer stable. This corresponds to the well-known D-flatness of the supersoft limit of the potential \cite{fnw,Belanger:2009wf}.
\item In figure \ref{FIG:p4a} we show the lightest neutralino versus lightest chargino mass. The portion of the parameter space that we have chosen should give many electroweakinos that can be searched for in future runs of the LHC; these correspond to the relatively small values of $\mu$ that we have chosen, for reasons of preserving some remnant of naturalness (and expediency in the model search). 
\item In figures \ref{FIG:p5} and \ref{FIG:p6} we demonstrate that we are searching in a portion of the parameter space where there is little mixing between the singlet $S$ and the Higgs: in figure \ref{FIG:p5} we directly show the (anti-)correlation between the soft singlet mass $m_S$ and mixing between lightest Higgs and the singlet; in figure \ref{FIG:p6} we show the soft singlet mass $m_S$ against the mass of the second and third Higgses: for larger $m_S$, the third Higgs is almost entirely singlet. 
\end{enumerate}

\newpage

\newpage

\newpage
\subsection{Benchmark points}
\label{SEC:benchmarks}

In table \ref{TAB:BENCHMARKS} we show the input parameters and spectrum of three benchmark points, which we have chosen to have relatively light stop masses while being otherwise representative of the sample set. 
\begin{table}[!h]
{\scriptsize
\begin{minipage}[b]{0.43\textwidth}\begin{center}
$
\begin{array}{|c|c|c|c|} \hline \hline
 \mathrm{Model} &  1\_299 &  8\_79 &  10\_77 \\ \hline
  \tan\beta &  2.300 &  2.264 &  2.574 \\
  \lambda_S(M_{S}) &  0.690 &  0.669 &  0.724 \\
  m_{D0} &  835 &  864 &  1110 \\
  m_0 &  3118 &  3356 &  3108 \\
  m_{S0} &  952 &  1183 &  1093 \\
  m_{\Sigma} &  3851 &  3213 &  3572 \\
  \mu &  164 &  154 &  120 \\
  \sqrt{B_\mu} &  365 &  588 &  389 \\ \hline\hline
  g_3(M_{GUT}) &  1.160 &  1.155 &  1.143 \\
  M_{GUT} &  1.49 \times 10^{17} &  1.53 \times 10^{17} &  1.26 \times 10^{17} \\
  v_S &  -1.173 &  -2.729 &  -5.617 \\
  v_T &  0.620 &  0.874 &  0.905 \\
  \Delta \rho &  2.2 \times 10^{-4} &  2.0 \times 10^{-4} &  2.6 \times 10^{-4} \\
  m_h &  123 &  124 &  125 \\
  m_{h_2} &  596 &  960 &  663 \\
  m_{h_3} &  1122 &  1301 &  1257 \\
  m_{h_4} &  3954 &  3359 &  3921 \\
\hline
\end{array}
$
\end{center} 
\end{minipage}
\begin{minipage}[b]{0.26\textwidth}\begin{center}
$
\begin{array}{|c|c|c|c|} \hline \hline
 \mathrm{Model} &  1\_299 &  8\_79 &  10\_77 \\ \hline
 m_{Ah} &  598 &  961 &  666 \\
  m_{Ah_2} &  1058 &  1232 &  1178 \\
  m_{Ah_3} &  3654 &  2967 &  3358 \\
  m_{H^\pm_1} &  591 &  958 &  658 \\
  m_{H^\pm_2} &  3655 &  2967 &  3359 \\
  m_{H^\pm_3} &  3954 &  3360 &  3921 \\
  m_{\tilde{t}_1} &  1252 &  1387 &  1395 \\
  m_{\tilde{t}_2} &  1918 &  2180 &  2048 \\
  m_{\tilde{b}_1} &  1913 &  2175 &  2044 \\
  m_{\tilde{\tau}_1} &  2870 &  3067 &  2852 \\
  m_{\tilde{e}_{12}} &  4692 &  5079 &  5493 \\
  m_{\tilde{N}_{1}} &  3122 &  3384 &  3116 \\
  m_{\tilde{N}_{5}} &  4691 &  5078 &  5491 \\\hline
\end{array}$
\end{center} 
\end{minipage}
\begin{minipage}[b]{0.26\textwidth}\begin{center}
$
\begin{array}{|c|c|c|c|} \hline \hline
 \mathrm{Model} &  1\_299 &  8\_79 &  10\_77 \\ \hline
  m_{\chi^0_1} &  130 &  135 &  115 \\
  m_{\chi^0_2} &  176 &  168 &  137 \\
  m_{\chi^0_3} &  208 &  228 &  242 \\
  m_{\chi^0_4} &  254 &  260 &  263 \\
  m_{\chi^0_5} &  800 &  829 &  1061 \\
  m_{\chi^0_6} &  801 &  829 &  1061 \\
  m_{\chi^\pm_1} &  167 &  159 &  126 \\
  m_{\chi^\pm_2} &  798 &  827 &  1059 \\
  m_{\chi^\pm_3} &  803 &  831 &  1062 \\
  m_{\tilde{g}} &  3110 &  3215 &  3932 \\
  m_{O_1} &  6367 &  6174 &  7687 \\
  m_{O_2} &  4653 &  4178 &  5331 \\\hline
\end{array}
$\end{center} 
\end{minipage}
}
\caption{Parameters and selected masses from the spectrum of benchmark points. All masses are in GeV and are accurate to one-loop order. The model numbers refer to filenames for full model files available upon request. The mass labels are standard under the Les Houches accord \cite{Skands:2003cj,Allanach:2008qq}, except for $m_{O_{1,2}}$ which refer to the two components of the scalar octet.}
\label{TAB:BENCHMARKS}\end{table}

\pagebreak
\subsection{Generic predictions}
\label{SEC:GenericPredictions}

From the scans that we have performed it is possible to draw several interesting conclusions about the spectrum of the model:
\begin{itemize}
\item Unification takes place at $(1.8 \pm 0.4) \times 10^{17}$ GeV.
\item We have a compressed pattern of soft masses (with deviations of a few percent upon varying the input parameters): 
\begin{align}
&m_{U 33}^2 : m_{Q 33}^2 : m_{Q 11}^2 : m_{D ii }^2 : m_{E ii}^2 : m_{U 11}^2 : m_{L ii}^2  \nn\\
=&0.16 : 0.39 : 0.77 : 0.79 : 0.83 : 0.93 : 1.02\nn
\end{align}
where the ratios are normalised with respect to the common mass at the GUT scale $m_0$. 
\item Sleptons are heavy and quasi-degenerate with the first two generations of squarks. This is because the Dirac gaugino masses do not enter the squark RGEs. 
\item The gaugino masses are in the ratio 
$$m_{DY}/m_{D0} : m_{D2}/m_{D0} : m_{D3}/m_{D0} =  0.22: 0.9 : 3.5,$$
i.e. the Wino barely runs from $m_{D0}$ (as can be seen from the one-loop RGE, which is zero for small $\lambda_T$). 
\item The lightest stop masses are $2.1 \pm 0.4, 3.1 \pm 0.6 $ TeV, i.e. we still typically require some contributions from the stops to the Higgs mass to obtain the experimental value -- the tree-level contribution from $\lambda_S$ is typically not quite enough.  We find that the major contribution to the shift in the pole mass of coloured squarks comes from integrating out the gluino and octet via the usual supersoft term \cite{fnw}:
\begin{equation}\delta m_{\tilde{q}}^2 \simeq \frac{4 \alpha_3 m_{D3}^2}{3\pi} \log \frac{m_O^2 + 4 m_D^2}{m_D^2} \simeq 0.6 m_{D0}^2. \label{EQ:SUPERSOFT}\end{equation}
This can be viewed as providing a lower bound on models with lighter stop masses, such as, in particular,  the benchmark points in table \ref{TAB:BENCHMARKS}. 

However, it is important to distinguish this contribution from the usual discussion of naturalness: this contribution does not enter into the Higgs mass calculation (at one loop), but rather the one-loop pole masses. Hence the stop masses which enter the Higgs mass calculation in \SARAH are actually lighter: for the benchmark models in table \ref{TAB:BENCHMARKS} we have $m_{\tilde{U}_{3,3}} (\ov{\mathrm{DR}}^\prime) \sim $ TeV. This makes the models much more natural than would otherwise appear.\footnote{On the other hand, over the whole sample set the $\ov{\mathrm{DR}}^\prime$ stop masses which \SARAH uses to calculate the Higgs mass are $1.9 \pm 0.4, 2.9 \pm 0.6$ TeV, i.e. typical models have heavier stops, and the benchmark points are special cases.} Thus we would expect have models with rather (or substantially) \emph{lighter} stop masses (and thus much more natural) once the two-loop $\alpha_s \alpha_t$ contributions are included; this is work in progress \cite{Goodsell:2014inprep}.
\end{itemize}

\section{Conclusions}

We have identified a set of fields which, having masses at the SUSY-breaking scale, allow unification of gauge couplings and Dirac masses for the gauginos. The lack of unification has in the past been an objection to this class of models which now can be considered resolved. We discussed to what extent these models can have a ``natural SUSY'' spectrum, important for flavour physics and LHC bounds, and found that in fact models without large hierarchies between the soft masses of different generations are favoured. Furthermore, we proposed two different symmetries to simplify the couplings of the model at low energies, and of these, identified one scenario that can be viewed as a unified completion of the models previously considered by (some of) the authors \cite{Belanger:2009wf,Benakli:2011kz,Goodsell:2012fm,Benakli:2012cy}. 

Within the context of our favoured low-energy scenario we discussed the constraints from lepton flavour violation and then proposed a constrained set of boundary conditions at the unification scale, containing a minimal number of eight key parameters: $m_0, m_{D0}, \tan \beta, m_S ,m_\Sigma, \lambda_S, \mu,  B_\mu$. We then implemented this scenario in \SARAH and produced a customised spectrum generator based on \SPheno code (as described in appendix \ref{APP:SARAH}) to perform a first exploration of a portion of the model parameter space, and found some interesting predictions for the pattern of masses.   

It is logical to compare the spectrum in our scenario to other constrained models. Of these, the CMSSM/mSUGRA is very popular and useful as a ``spherical cow'' model of SUSY for collider studies. Indeed, it is known that only a very small number of mass hierarchies are possible in the CMSSM compared to the general MSSM \cite{Konar:2010bi,Dreiner:2012wm}, which has biased experimental searches in the past. For example, one finds for low $\tan\beta$ for all sfermion mass parameters at $Q=1$~TeV except the stops
\begin{eqnarray}
\frac{m^2_Q}{m^2_0} \underset{\rm CMSSM}{\simeq} 1 +  4.3 \frac{M^2_{1/2}}{m^2_0} &\,\,,\,\,& 
         \frac{m^2_D}{m^2_0} \underset{\rm CMSSM}{\simeq}\frac{m^2_U}{m^2_0}\underset{\rm CMSSM}{\simeq}  1 +  4  \frac{M^2_{1/2}}{m^2_0}\\
\frac{m^2_L}{m^2_0}\underset{\rm CMSSM}{\simeq} 1 +  0.44 \frac{M^2_{1/2}}{m^2_0} &\,\,,\,\,& 
     \frac{m^2_E}{m^2_0} \underset{\rm CMSSM}{\simeq} 1 +  0.13 \frac{M^2_{1/2}}{m^2_0},
\end{eqnarray}
where $m_0, M_{1/2}$ are the scalar and gaugino mass parameters of the CMSSM. However, for our models we find
\begin{align}
&m_{U 33}^2 : m_{Q 33}^2 : m_{Q 11}^2 : m_{D ii }^2 : m_{E ii}^2 : m_{U 11}^2 : m_{L ii}^2  \nn\\
=&0.16 : 0.39 : 0.77 : 0.79 : 0.83 : 0.93 : 1.02\nn .
\end{align}
Since the RGEs of the scalars are independent of the Dirac gaugino masses \cite{Jack:1999ud,Jack:1999fa,Goodsell:2012fm}, to compare with the CMSSM it is instead instructive to consider the physical masses and add the contribution of equation (\ref{EQ:SUPERSOFT}); this gives
\begin{align}
\frac{m^2_{\tilde{u}_L,\tilde{d}_L}}{m^2_0} \underset{\rm CMDGSSM}{\simeq}  \frac{m^2_{\tilde{d}_R}}{m^2_0} \underset{\rm CMDGSSM}{\simeq}& 0.8 +  0.6 \frac{m^2_{D0}}{m^2_0},\quad \frac{m^2_{\tilde{u}_R}}{m^2_0} \underset{\rm CMDGSSM}{\simeq}  0.9 + 0.6 \frac{m^2_{D0}}{m^2_0},   \\
\frac{m^2_{\tilde{l}_L}}{m^2_0} \underset{\rm CMDGSSM}{\simeq} 1, \qquad  & \frac{m^2_{\tilde{l}_R}}{m^2_0} \underset{\rm CMDGSSM}{\simeq} 0.8.
\end{align}
Note that the above does conceal the dependence on the octet scalar mass $m_O$. However, this is rather weak, as we determined in section \ref{SEC:NATURAL}, entering only at two loops (i.e. we have a correction of $\sim 0.08 m_O^2/m_{0}^2$ to the masses of coloured squarks) -- and in the scans we typically find $m_\Sigma \lesssim m_0$.  

The above then show that the spectrum of the first two generations of squarks alone may not be enough to distinguish the model from the CMSSM; it would resemble a model with small $M_{1/2}$ since the masses correlate with $M_{1/2}/m_0$ in a similar way, although we could easily distinguish them if the gaugino masses are known. However, we should be able to distinguish the spectrum of fermions by themselves. The prediction for the Majorana masses $M_i$ at 1-loop assuming a unified Majorana mass $M_{1/2}$ 
at the GUT scale gives for the CMSSM:
$$M_1/M_{1/2} : M_2/M_{1/2} : M_3/M_{1/2} = 0.44 : 0.84 : 2.34.$$
In comparison, our model yields 
$$m_{DY}/m_{D0} : m_{D2}/m_{D0} : m_{D3}/m_{D0} =  0.22: 0.9 : 3.5.$$

One main difference between our model and the CMSSM would be the mass of the \emph{third} generation of squarks, which would be considerably lighter. In section \ref{SEC:GenericPredictions} we gave some generic predictions for these masses in our model and some comments about how this affects naturalness. This highlighted that one very important future development is the calculation of two-loop corrections to the Higgs mass \cite{Goodsell:2014inprep}, which we argued will lead to us predicting lighter stops and thus more natural models. 

Importantly, in our model we have additional electroweak-charged states for which we can to some extent constrain the spectrum and which may be detectable in the next run of the LHC. This paper therefore hopefully opens the way for more detailed collider studies; in particular, since our parameter choices may have been overly conservative such studies would allow us to identify the most natural models consistent with data. In addition, we have not computed any dark matter observables. If we assume a standard thermal history of the universe, then it would be very interesting to find how the dark matter relic density and lack of direct detection constrains the parameters of our model, extending the work of \cite{Belanger:2009wf}.

As an interesting aside, the field content of our model can be connected with the recent ``Fake Split Supersymmetry'' proposal of \cite{Dudas:2013gga,Benakli:2013msa}. If we take the supersymmetry-breaking scale to be high, such that all scalar superpartners are heavy, and add Majorana masses for the gauginos at the same scale (but not the adjoint fermions $\chi_\Sigma$), then below the supersymmetry-breaking scale we can have a model that resembles Split SUSY (with different couplings). Then above the SUSY-breaking scale, we can have the field content of our model. The change in the one-loop beta-function coefficients\footnote{By which we mean the $b_i $ in the RGEs $\frac{d}{dt} \alpha_i^{-1} = -\frac{b_i}{2\pi} t $.} is the same for each gauge group: apart from the sfermions of the MSSM (which come in complete $SU(5)$ multiplets, except for the one heavy Higgs which contributes negligibly as in standard Split SUSY), we add one adjoint scalar, one adjoint fermion (this time the usual gaugino of the MSSM) and the fields of equation (\ref{EQ:ExtraReps}) which shift the beta function coefficients by three. Thus we preserve unification \emph{for any SUSY scale}. Moreover, the FSSM model of   \cite{Benakli:2013msa}  called for two additional Fake Higgs multiplets, which we can recognise as our $R_{u,d}$. If, instead of our choice in section \ref{SEC:MUDGSSM} of R-symmetry or lepton number, we assign charges of the new fields under a new (broken) global symmetry $U(1)_F$ of
$$
\begin{array}{|c|c|} \hline 
\mathrm{Field} & U(1)_F \\ \hline
S, T, O & 1 \\
H_{u,d} & 0 \\ 
R_{u,d} & 1 \\
\hE_i, \hEt_i & 0 \\ \hline
\end{array}
$$
we find the required UV completion of the FSSM up to the GUT scale. It would be interesting to explore this connection further.

\section*{Acknowledgments}

MDG would like to thank Emilian Dudas for helpful discussions.
He would like to thank Universit\"at W\"urzburg and the Galileo Galilei Institute for Theoretical Physics for hospitality, and the INFN for partial support at different stages of this work. MDG was supported by the European ERC Advanced Grant 226371 MassTeV and the Marie-Curie contract no. PIEF-GA-2012-330060 BMM@3F.
KB acknowledges support from the Institut Lagrange de Paris and the ERC grant Higgs@LHC.
FS is supported by the BMBF PT DESY Verbundprojekt 05H2013-THEORIE 'Vergleich von LHC-Daten mit supersymmetrischen Modellen', and would like to thank the LPTHE in Paris for hospitality. WP is supported by the DFG, project No. PO-1337/3-1. 

\newpage
\appendix

\section{GUT embeddings}
\label{APP:EMBEDDINGS}

In this appendix we briefly summarise the embedding of the adjoint superfields and extra superfields in the representations of equation (\ref{EQ:ExtraReps}) into $SU(5)$ and $(SU(3))^3$ gauge groups. 

\subsection{$SU(5)$}

The adjoint superfields $\mathbf{8_0} + \mathbf{3_0}  + \mathbf{1_0} $ can fit into the adjoint $\mathbf{24}$ of $SU(5)$, which we shall denote $\mathbf{\Sigma_{24}}$, although with this gauge group we must exclude the bachelor states $(\mathbf{3},\mathbf{2})_{\mathbf{5/6}} \oplus (\mathbf{\ov{3}},\mathbf{2})_{\mathbf{-5/6}} $. Then it is clear that $(\mathbf{1}, \mathbf{2})_{1/2}  \oplus (\mathbf{1}, \mathbf{2})_{-1/2} \supset \mathbf{5}_R\oplus \mathbf{\ov{5}}_R$,  $(\mathbf{1}, \mathbf{1})_{\pm 1} \supset \mathbf{10}_{\hE}, \mathbf{\ov{10}}_{\hEt}$; we must simply exclude the extra triplet and other unwanted states from the spectrum by the UV embedding (such as, in IIB/F-theory, by an appropriate choice of fluxes) in much the same way as Higgs triplets are excluded; it would be interesting to attempt to embed this in a UV-complete model. We can then write down our superpotential (\ref{EQ:Superpotential}) from these GUT representations schematically as:
\begin{align}
W =& W_{Yukawa} \nn\\ 
& +\mu\mathbf{5}_H \ov{\mathbf{5}}_H + \lambda \mathbf{5}_H \mathbf{\Sigma_{24}} \ov{\mathbf{5}}_H + \mu_E \mathbf{10}_{\hE}\ov{\mathbf{10}}_{\hEt} + \lambda_{\hE} \mathbf{10}_{\hE} \mathbf{\Sigma_{24}} \ov{\mathbf{10}}_{\hEt} \nn\\
&+ \mu_R \mathbf{5}_R\ov{\mathbf{5}}_R  +\lambda_R  \mathbf{5}_R \mathbf{\Sigma_{24}} \ov{\mathbf{5}}_R  + \lambda_{LR}  \mathbf{5}_L \mathbf{\Sigma_{24}} \ov{\mathbf{5}}_R + \lambda_{SE}\mathbf{10}_{E} \mathbf{\Sigma_{24}} \ov{\mathbf{10}}_{\hEt} \nn\\
& +  Y_{\hE} \ov{\mathbf{5}}_H  \ov{\mathbf{5}}_R  \mathbf{10}_{\hE}+Y_{\hEt} \mathbf{5}_H \mathbf{5}_R \mathbf{\ov{10}}_{\hEt}  + Y_{LFV} \ov{\mathbf{5}}_L  \ov{\mathbf{5}}_H  \mathbf{10}_{E} + Y_{EFV} \ov{\mathbf{5}}_H  \ov{\mathbf{5}}_R  \mathbf{10}_{E}.
\end{align} 
If, after the breaking of the GUT group, the singlet which remains to give a mass to the Bino is only the one surviving from the $\mathbf{\Sigma_{24}}$, then this would predict $\lambda_S = -\sqrt{3/5} \lambda_T = -\frac{1}{2} \sqrt{3/5} \lambda$ at the GUT scale. Moreover, natural choices of values for $\lambda$ would be the $N=2$ value $\sqrt{2} g_{GUT}$, giving $\lambda_T = g_{GUT}/\sqrt{2}, \lambda_S = -\sqrt{3/10}\  g_{GUT}$, or zero (if, for example, the adjoints were located apart from the matter fields in a higher-dimensional theory). In section \ref{SEC:CMDGSSM} we rather take this latter choice: the relationship with the singlet is easily broken if there are additional singlets that mix at the GUT scale, leaving only one light. We suppose that one of these singlets couples strongly with the Higgs, giving us a substantial $\lambda_S$, and the other adjoint fields couple very weakly, in one stroke also explaining the different value of $m_S$ to the other adjoint masses. 

\subsection{$(SU(3))^3$}

\begin{center}
\begin{figure}
\includegraphics[width=0.6\textwidth]{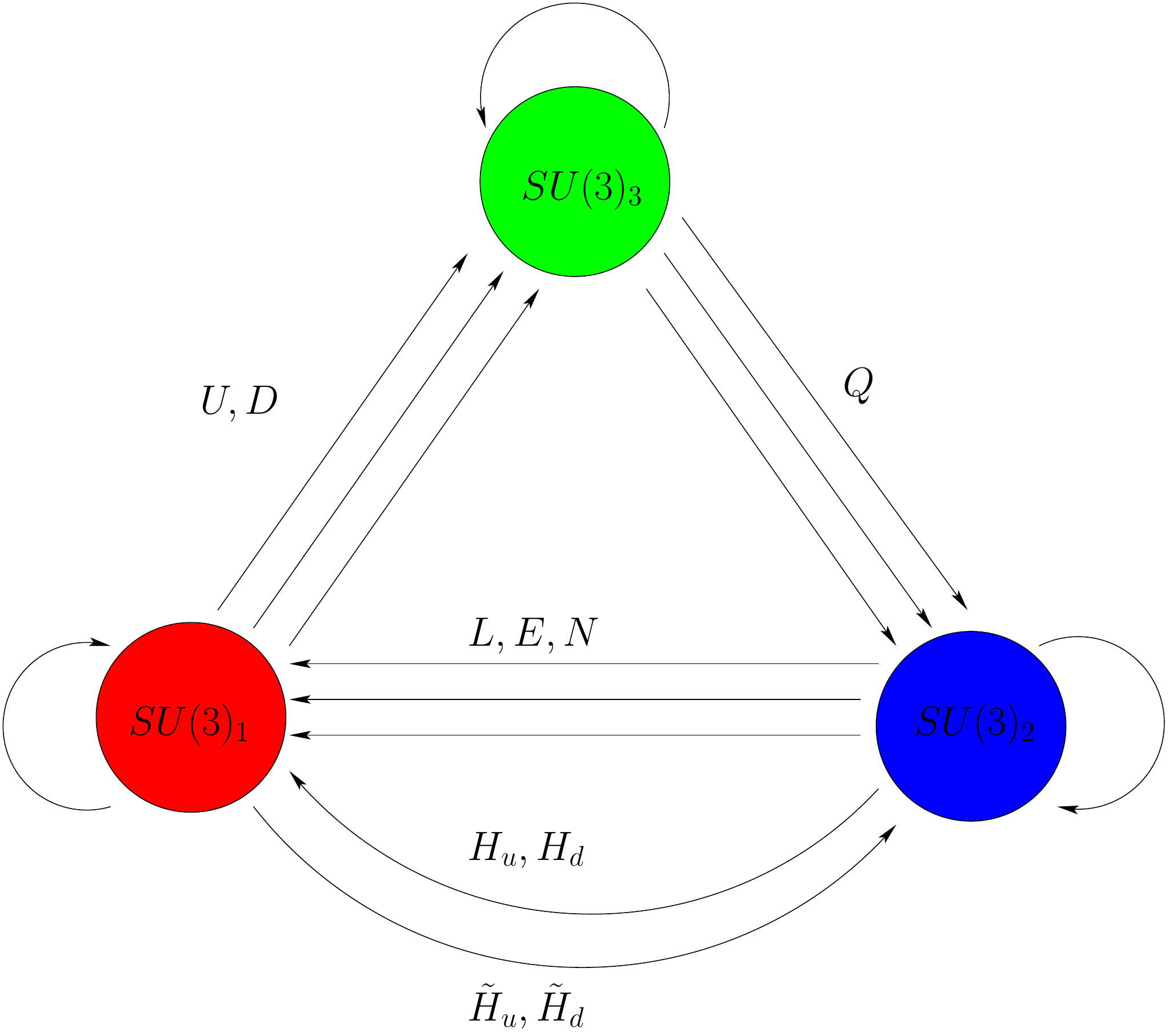} 
\caption{Naive $(SU(3))^3$ quiver.}\label{FIG:SU33quiver}\end{figure}
\end{center}

In the unified group $(SU(3))^3$ the Higgs fields $H_u, H_d$ are conventionally in the \emph{same} representation $(\mathbf{3}, \mathbf{\bar{3}},\mathbf{1})$ \cite{Shafi:1978gg, Dvali:1994vj,Dvali:1994wj}: we can write $H_u, H_d \subset \mathcal{H}$. Let us write the conjugate representation $(\mathbf{\bar{3}}, \mathbf{3}, \mathbf{1})$ as $\mathcal{\tilde{H}}$. Then under the breaking of $(SU(3))^3 \rightarrow SU(3) \times SU(2) \times U(1)_Y $  we have
\begin{align}
\C{H} \rightarrow& (\mathbf{1}, \mathbf{2})_{1/2} +2\times (\mathbf{1}, \mathbf{2})_{-1/2} + (\mathbf{1}, \mathbf{1})_{1} + 2\times (\mathbf{1}, \mathbf{1})_{0} \nn\\
\C{\tilde{H}} \rightarrow& (\mathbf{1}, \mathbf{2})_{-1/2} +2\times (\mathbf{1}, \mathbf{2})_{1/2} + (\mathbf{1}, \mathbf{1})_{-1} + 2\times (\mathbf{1}, \mathbf{1})_{0}
\end{align}
In fact, the leptons $L, E$ fit into identical multiplets to $\mathcal{H}$ (up to charges under lepton number). On the other hand, the adjoints $\mathbf{8}^{(1)} \oplus \mathbf{8}^{(2)} \oplus \mathbf{8}^{(3)}$ (where we have labelled the group factors) decompose as
\begin{align}
\mathcal{O}^{(3)}  \equiv \mathbf{8}^{(3)} \rightarrow& (\mathbf{8}, \mathbf{1})_{0} \nn\\
\mathcal{O}^{(2)} \equiv\mathbf{8}^{(2)} \rightarrow& (\mathbf{1}, \mathbf{3})_{0} + (\mathbf{1}, \mathbf{2})_{\pm 1/2} + (\mathbf{1}, \mathbf{1})_{\mathbf{0}}\nn\\
\mathcal{O}^{(1)} \equiv\mathbf{8}^{(1)} \rightarrow& 2 \times (\mathbf{1}, \mathbf{1})_{\mathbf{\pm 1}} + 4 \times (\mathbf{1}, \mathbf{1})_{\mathbf{0}}.
\end{align}
We find that there is hence more than one way to embed the extra states in equation (\ref{EQ:ExtraReps}) into $(SU(3))^3 $: they could come from the adjoint, as originally envisaged, or from bifundamental states. These have implications for the couplings:

\begin{itemize}
\item If we keep all of the states in the full adjoint and keep only $H_u, H_d$ from $\mathcal{H}$, then we find that we cannot write the couplings $\lambda_S, \lambda_T$ in the GUT theory until the group is broken, meaning that these couplings should be severely suppressed. In fact, the only unsuppressed couplings involving the adjoints with non-singlets would come from $W \supset \mathrm{tr} [(O^2)^3] \rightarrow S R_d \cdot R_u, R_d \cdot T R_u$ and $\mathrm{tr} [(O^1)^3] \rightarrow S\hE \hEt$. Moreover we could not charge the extra states under lepton number, so this embedding would be more appropriate for the MRSSM with $\lambda_{Su}, \lambda_{Sd}, \lambda_{Tu}, \lambda_{Td} $ all small (suppressed by at least $M_{GUT}/\Lambda$, where $\Lambda$ is the fundamental scale of the theory) which would imply rather heavy stops. The quiver diagram for this naive model is shown in figure \ref{FIG:SU33quiver}.
\item If we instead build our extra states from the bifundamentals, so that $R_u, R_d \supset \mathcal{\tilde{H}} $, then we can happily build our model, given in equation (\ref{EQ:Superpotential}), by endowing the extra states with lepton number. However, all couplings to the adjoints (such as $\lambda_S, \lambda_T$)  would still be suppressed as above. This would give us a phenomenologically interesting model but not from the point of view of a minimum of important parameters, but again would require stops to enhance the Higgs mass. Alternatively we could write an MRSSM model with unsuppressed couplings $\lambda_{Su}, \lambda_{Sd}, \lambda_{Tu}, \lambda_{Td} $, appropriate, for example for \cite{Bertuzzo:2014bwa}.
\end{itemize}
In summary, neither of these options is suitable for the embedding of the boundary conditions in section \ref{SEC:CMDGSSM} into $(SU(3))^3$; instead we favour an embedding into $SU(5)$ -- if a full GUT structure is at all required. 

\section{LFV constraints}
\label{APP:CLFV}

Here we compute the constraints on our model arising from the new contributions to charged lepton flavour violation, partially following the approach of \cite{Ishiwata:2013gma}. 
We start with a superpotential of equation (\ref{EQ:Superpotential}) which contains additional vector-like fermions carrying lepton number. Recall that $R_u = (R_u^0, R_u^-), R_d = (R_d^+, R_d^0)$, after electroweak symmetry breaking we have the mass terms for charged states 
\begin{align}
W \supset& -(\mu_R + \frac{1}{\sqrt{2}} \lambda_{SR} v_S ) R_u^- R_d^+ + (\mu_{\hE\, ij} +  \frac{1}{\sqrt{2}}\lambda_{S\hE ij} v_S) \hE_i^+ \hEt_j^-\nn\\
& -\frac{1}{\sqrt{2}} \lambda_{SLRi} v_S L_i^- R_d^+ + \frac{1}{\sqrt{2}} \lambda_{SE ij} v_S E_i \hEt_j^- -  \frac{1}{\sqrt{2}} \lambda_{TLRi} v_T L_i^- R_d^+   \\
& - \frac{1}{\sqrt{2}} v c_\beta Y_{\hE i} R_u^- \hE_i^+ + \frac{1}{\sqrt{2}} v s_\beta Y_{\hEt i} R_d^+ \hEt_i^- - \frac{1}{\sqrt{2}} v c_\beta Y_{LFV}^{ij} e_i^L \hE_j^+ - \frac{1}{\sqrt{2}} v c_\beta Y_{EFV}^{j} r_u^- e_j^R. \nn
\end{align}
The mass matrix for the fermions becomes
\begin{align}
\mathcal{L}_{leptons} =& - (\begin{array}{ccc} r_u^- & \hat{\tilde{e}}_i  & e^L_i \end{array}) \left( \begin{array}{ccc} \mu_R & \frac{v c_\beta}{\sqrt{2}}  Y_{\hE_i} & -\frac{v c_\beta}{\sqrt{2}} Y_{EFV} \\ \frac{v s_\beta}{\sqrt{2}}  Y_{\hEt_i} & \mu_E &  \frac{1}{\sqrt{2}} \lambda_{SE ij} v_S\\ -\frac{1}{\sqrt{2}} \lambda_{SLRi} v_S  - \frac{1}{\sqrt{2}} \lambda_{TLR i} v_T& -\frac{v c_\beta}{\sqrt{2}} Y_{LFV} &-\frac{v c_\beta}{\sqrt{2}} Y_{E} \end{array} \right) \left( \begin{array} {c} r_d^+ \\ \hat{e}_i \\ e_{R i} \end{array} \right) 
\end{align}

This is then diagonalised in two stages: first we must perform the usual rotations to diagonalise the Standard Model Yukawa couplings, where $e_R = E_R^{SM} e_R^\prime, e_L = E_L^{SM} e_L^\prime $, while also going to a basis where $\mu_E$ and $\mu_R$ are diagonal, giving
\begin{align}
\mathcal{M}^\prime =& \left( \begin{array}{ccc} \mu_R & \frac{v c_\beta}{\sqrt{2}}  Y_{\hE_i} & -\frac{v c_\beta}{\sqrt{2}} Y_{EFV} E_R^{SM} \\ \frac{v s_\beta}{\sqrt{2}}  Y_{\hEt_i} & \mu_E &  \frac{1}{\sqrt{2}} \lambda_{SE} v_S E_R^{SM}\\  -\frac{1}{\sqrt{2}}  (E_L^{SM})^\dagger \lambda_{SLRi} v_S  - \frac{1}{\sqrt{2}}  (E_L^{SM})^\dagger \lambda_{TLR i} v_T & -\frac{v c_\beta}{\sqrt{2}} (E_L^{SM})^\dagger Y_{LFV} & d_e \end{array} \right)
\end{align}
and then a further rotation such that $\mathcal{M}_{\mathrm{diag}} = E_L^\dagger \mathcal{M} E_R $ so that $E_L^\dagger \mathcal{M}\mathcal{M}^\dagger E_L = E_R^\dagger \mathcal{M}^\dagger \mathcal{M} E_R  = \mathcal{M}_{\mathrm{diag}} \mathcal{M}_{\mathrm{diag}}^\dagger $ so that to leading order $E_L$ diagonalises (dropping the Standard Model factors and ignoring the Standard Model lepton masses)
\begin{align}
 \left( \begin{array}{ccc} \mu_R \mu_R^\dagger & \frac{v c_\beta}{\sqrt{2}}  Y_{\hE_i} \mu_E^\dagger + \frac{v s_\beta}{\sqrt{2}} \mu_R Y_{\hEt_i}^\dagger & -\frac{\mu_R }{\sqrt{2}} ( \lambda_{SLRi}^\dagger v_S  + \lambda_{TLR i}^\dagger v_T) \\  \frac{v c_\beta}{\sqrt{2}} \mu_E Y_{\hE_i}^\dagger +  \frac{v s_\beta}{\sqrt{2}}  Y_{\hEt_i} \mu_R^\dagger & \mu_E \mu_E^\dagger &  -\frac{v c_\beta}{\sqrt{2}} \mu_E Y_{LFV}^\dagger  \\ -\frac{1}{\sqrt{2}} ( \lambda_{SLRi} v_S  + \lambda_{TLR i} v_T) \mu_R^\dagger & -\frac{v c_\beta}{\sqrt{2}}Y_{LFV} \mu_E^\dagger & 0 \end{array} \right)
\end{align}
so to leading order 
\begin{align}
E_L =& \threematrix[1, p_L,r_L][-p_L^\dagger,1,-q_L^\dagger][-r_L^\dagger,q_L,1] \nn\\
p_L^{ij} =& -\frac{v}{\sqrt{2}} \frac{1}{(\mu_R^{ii})^2 - (\mu_E^{jj})^2}\bigg[ c_\beta Y_{\hE_i} \mu_E^\dagger + s_\beta \mu_R Y_{\hEt_i}^\dagger \bigg]^{ij} \nn\\
=& -\delta^{ij}\frac{v}{\sqrt{2}} \frac{1}{(\mu_R^{ii})^2 - (\mu_E^{ii})^2}\bigg[ c_\beta Y_{\hE_i} \ov{\mu}_E^{ii} + s_\beta \mu_R^{ii} Y_{\hEt_i}^\dagger \bigg] \nn\\
-(q_L^{\dagger})^{ij} =& \frac{1}{(\mu_E^{ii})^2} \frac{v c_\beta}{\sqrt{2}} \bigg[\mu_E Y_{LFV}^\dagger  \bigg]^{ij} \nn\\
=&  \frac{1}{\ov{\mu}_E^{ii}} \frac{v c_\beta}{\sqrt{2}} \bigg[ Y_{LFV}^\dagger  \bigg]^{ij} \nn\\
r_L^{ij} =& \frac{1}{\mu_R^{ii}} \frac{1}{\sqrt{2}} \bigg[  \lambda_{SLRi}^\dagger v_S  + \lambda_{TLR i}^\dagger v_T\bigg]^{ij}
\end{align}
whereas $E_R$ diagonalises to leading order
\begin{align}
 \left( \begin{array}{ccc}  \mu_R^\dagger \mu_R  & \frac{v s_\beta}{\sqrt{2}}  Y_{\hEt_i}^\dagger \mu_E + \frac{v c_\beta}{\sqrt{2}} \mu_R^\dagger Y_{\hE_i}   &  -\frac{v c_\beta}{\sqrt{2}} \mu_R^\dagger Y_{EFV}  \\  \frac{v c_\beta}{\sqrt{2}}Y_{\hE_i}^\dagger \mu_R  +  \frac{v s_\beta}{\sqrt{2}} \mu_E^\dagger Y_{\hEt_i}  & \mu_E^\dagger\mu_E  &  \frac{\mu_E^\dagger}{\sqrt{2}} \lambda_{SE ij} v_S  \\  -\frac{v c_\beta}{\sqrt{2}}  Y_{EFV}^\dagger \mu_R & \frac{1}{\sqrt{2}} \lambda_{SE ij}^\dagger v_S\mu_E & 0 \end{array} \right)
\end{align}
giving to leading order
\begin{align}
E_R =& \threematrix[1, -p_R,q_R][p_R^\dagger,1,r_R][-q_R^\dagger,-r_R^\dagger,1] \nn\\
-p_R^{ij} =& -\frac{v}{\sqrt{2}} \frac{1}{(\mu_R^{ii})^2 - (\mu_E^{jj})^2} \bigg[s_\beta   Y_{\hEt_i}^\dagger \mu_E +  c_\beta \mu_R^\dagger Y_{\hE_i} \bigg]^{ij} \nn\\ 
=&  - \delta^{ij} \frac{v}{\sqrt{2}} \frac{1}{(\mu_R^{ii})^2 - (\mu_E^{ii})^2} \bigg[s_\beta   Y_{\hEt_i}^\dagger \mu_E^{ii} +  c_\beta (\ov{\mu}_R^{ii}) Y_{\hE_i} \bigg]\nn\\ 
q_R^{ij} =&  -\frac{v c_\beta}{\sqrt{2}} \frac{1}{(\mu_R^{ii})^2} \bigg[\mu_R^\dagger Y_{EFV} \bigg]^{ij} \nn\\
=&  -\frac{v c_\beta}{\sqrt{2}} \frac{1}{\mu_R^{ii}} Y_{EFV}^{j} \nn\\
r_R^{ij} =& - \frac{v_S}{\sqrt{2} \mu_E^{ii}} \lambda_{SEij}.
\end{align}

\subsection{Couplings}

To calculate the flavour constraints we must determine the couplings which lead to LFV. The new ones come from the Yukawa couplings where Higgs and charged Higgs exchanges contribute, and via gauge couplings when the sfermions contribute. There is also a contribution from Z exchange. To better compute these, we need a basis of Dirac fermions: label 
\begin{equation}
e_{4,5}^L \equiv \hat{\tilde{e}}_i, \hspace{1cm}
e_{6}^L \equiv r_u^-, \hspace{1cm}
e_{4,5}^R \equiv \hat{e}_i^+, \hspace{1cm}
e_{6}^R \equiv r_d^+ \,\,,
\end{equation}
and then we write
\begin{align}
\ov{e}_i P_L e_j =& e_i^R e_j^L \nn\\
\ov{e}_i P_R e_j =& \ov{e}_i^L \ov{e}_j^R \nn\\
(r_u^- \hat{e}^+_i) =& \ov{e}_{i+3} P_L e_6 .
\end{align}

Since in this paper we have very little mixing between the lightest Higgs and the other neutral eigenstates, and moreover the charged/heavy Higgs have substantially higher masses, we can to a first approximation just take the lightest Higgs' couplings. There we have
\begin{align}
W \supset&   \frac{1}{\sqrt{2}} (v+h) c_\beta Y_{\hE i} R_u^- \hE_i^+ - \frac{1}{\sqrt{2}} (v+h) s_\beta Y_{\hEt i} R_d^+ \hEt_i^- \nonumber \\
 & + \frac{1}{\sqrt{2}} (v+h) c_\beta Y_{LFV}^{ij} e_i^L \hE_j^+ + \frac{1}{\sqrt{2}} (v+h) c_\beta Y_{EFV}^{j} r_u^- e_j^R .
\end{align}
Expanding we have for the first line
\begin{align}
\L\supset& - \frac{1}{\sqrt{2}} (v+h) c_\beta \bigg[ Y_{\hE i} \ov{e}_{i+3} P_L e_6 -  t_\beta Y_{\hEt i} \ov{e}_6 P_L e_{i+3} + Y_{\hE i}^\dagger \ov{e}_6 P_R e_{i+3} - t_\beta Y_{\hEt i}^\dagger \ov{e}_{i+3} P_R e_6 \bigg]
\end{align}
and for the second
\begin{align}
\L\supset&- \frac{1}{\sqrt{2}} h c_\beta \bigg[ Y_{LFV}^{ij} \ov{e}_{j+3} P_L e_{i}  + \ov{Y}_{EFV}^{i} \ov{e}_6 P_R e_{i} +  Y_{EFV}^{i} \ov{e}_{i} P_L e_{6}  + \ov{Y}_{LFV}^{ij} \ov{e}_i P_R e_{j+3} \bigg].
\end{align}
We write the usual Yukawas in this basis
\begin{align}
\L \supset & - \frac{h c_\beta}{\sqrt{2}} d_{i}\ov{e}_i P_L e_{i}  - \frac{h c_\beta}{\sqrt{2}} d_{i}\ov{e}_i P_R e_{i}
\end{align}

Let us write this in matrix form
\begin{align}
\L \supset& -\frac{h c_\beta}{\sqrt{2}} \bigg[ L_{ij} \ov{e}_i P_L e_j + R_{ij} \ov{e}_i P_R e_j \bigg]
\end{align}
where now
\begin{equation}
L =  \threematrix[0,-t_\beta Y_{\hEt},0][Y_{\hE},0,Y_{LFV}^T][Y_{EFV},0,d] \, \hspace{1cm} 
R =  \threematrix[0, Y_{\hE}^\dagger,Y_{EFV}^\dagger][-t_\beta Y_{\hEt}^\dagger,0,0][Y_{LFV}^*,0,d].
\end{equation}
We must then transform these according to $L \rightarrow E_L^\dagger L E_L, R \rightarrow E_R^\dagger R E_R  $ and consider the off-diagonal couplings:
\begin{align}
\L \supset& -\frac{h c_\beta}{\sqrt{2}} \bigg[ L^\prime_{i+3,j} \ov{e}_{i+3} P_L e_j + R_{i+3,j}^\prime \ov{e}_{i+3} P_R e_j + h.c. \bigg].
\end{align}
So
\begin{align}
L^\prime_{6j} = t_\beta Y_{\hEt} q_L^\dagger - p_L Y_{LFV}^T, \qquad &L^\prime_{i+3 j} \rightarrow Y_{LFV}^T + Y_{\hE} r_L +  ... \nn\\
R^\prime_{6j} = Y_{EFV}^\dagger + Y_{\hE}^\dagger r_R + ... ..., \qquad &R^\prime_{i+3 j} = -p_R^\dagger Y_{EFV}^\dagger - t_\beta Y_{\hEt} q_R .
\end{align}

\subsection{Constraints}

\subsubsection{$\mu \rightarrow e \gamma$}

The relevant amplitude/effective operator, using $\sigma^{\mu\nu} \equiv \frac{i}{2} [\gamma^\mu,\gamma^\nu]$, has the structure \cite{Lavoura:2003xp}:
\begin{align} 
\mathcal{A}_{ij} =& e \epsilon_\mu^* \ov{e}_i \bigg[ i \sigma^{\mu\nu} q_\nu ( \sigma_L P_L + \sigma_R P_R) + \delta_L \Delta_L^\mu + \delta_R \Delta_R^\mu \bigg] e_j,
\end{align}
where $e$ is the electromagnetic coupling and $e_i$ are the fermions; $q_\nu$ is the momentum of the outgoing photon and $\Delta_{L,R}^\mu $ vanish on shell. 

Now we require the results of the amplitudes. While there are two types of amplitudes, namely involving $L^* L$ or $R^* R$ couplings and the $L^* R$ or $R^* L$ amplitudes, the latter are enhanced by a mass insertion of the heavy fermion; the former are proportional to the light fermion mass while the latter are proportional to the mass of the heavy fermion. Hence the Higgs-mediated contribution gives generically
\begin{align}
\sigma_L =&  \frac{2i}{16 \pi^2 m_h^2} \sum_k L^*_{k2} R_{k1} m_k \bigg[ \frac{(t_k-3)(t_k-1) + 2 \log t_k}{4 (t_k-1)^3} \bigg] \nn\\
\underset{m_f \gg m_h}{\rightarrow}&  \frac{i}{32 \pi^2 }  \sum_k L^*_{k2} R_{k1} \frac{1}{m_k} \nn\\
\sigma_R \underset{m_f \gg m_h}{\rightarrow}&  \frac{i}{32\pi^2 }  \sum_k R^*_{k2} L_{k1} \frac{1}{m_k},
\end{align}
where $t_k = m_k^2/m_h^2, x_k = m_h^2/m_k^2$.  
Then this gives 
\begin{align}
\Gamma (\mu \rightarrow e \gamma) =& \frac{(m_\mu^2 - m_e^2)^3( |\sigma_L|^2 + |\sigma_R|^2) }{16\pi m_\mu^3} e^2 \nn\\
\simeq& \frac{\alpha}{4} m_\mu^3  ( |\sigma_L|^2 + |\sigma_R|^2) .
\end{align}
Since the limit is
\begin{align}
\mathrm{Br} (\mu \rightarrow e \gamma) <& 2.4 \times 10^{-12} \nn\\
\mathrm{Br} (\mu \rightarrow 3e) <& 1.0 \times 10^{-12}\nn\\
\Gamma (\mu \rightarrow X) \simeq &  \Gamma (\mu \rightarrow e \nu_\mu \bar{\nu}_e) \nn\\
=& 3 \times 10^{-19} \mathrm{GeV}
\end{align}
we have
\begin{align}
\sigma_{L,R} <& 5.8 \times 10^{-13} \mathrm{GeV}^{-1}.
\end{align}
Hence if the heavy fermions all have similar masses $\mu_E \sim \mu_R \equiv \mu_{E,R} \sim$ TeV, we need
\begin{align}
R^* L \lesssim 1.8 \times 10^{-7} \left( \frac{\mu_{E,R}}{\mathrm{TeV}} \right) \sim& Y_{EFV} Y_{LFV} Y_{\hE} \frac{v}{\mu_{E,R}} \nn\\
\sim& Y_{EFV} Y_{\hE} (\lambda_{SLR} \frac{v_S}{\mu_{E,R}} + \lambda_{TLR} \frac{v_T}{\mu_{E,R}}) \nn\\
\sim& Y_{LFV} Y_{\hE} \lambda_{SE} \frac{v_S}{\mu_{E,R}}
\end{align}
and hence if all the couplings are of a similar order of magnitude then
\begin{align}
Y_{EFV} \sim Y_{LFV} \sim Y_{\hE} \sim Y_{\hEt} \lesssim& 10^{-2} \left( \frac{\mu_{E,R}}{\mathrm{TeV}} \right)^{2/3} \nn\\
\lambda_{SLR} \sim \lambda_{SE} \lesssim& 10^{-2} \left( \frac{\mu_{E,R}}{\mathrm{TeV}} \right)^{2/3} \left( \frac{v}{v_{S,T}} \right) \nn\\
\lambda_{TLR}  \lesssim& \mathcal{O}(1)  \left( \frac{\mu_{E,R}}{\mathrm{TeV}} \right)^{2/3} \left( \frac{\mathrm{GeV}}{v_{T}} \right),
\end{align} 
which is very mild when we consider that these should be equivalent to off-diagonal Yukawa couplings. In particular, since we know that $v_{T} < \mathcal{O}(\mathrm{GeV})$ from electroweak precision data, we find that $\lambda_{TLR}$ is essentially unconstrained by this process.

\subsubsection{Electron electric dipole moment}

The recently improved bound  of the electron dipole moment  \cite{Baron:2013eja} of 
\begin{align}
|d_e| <& 8.9 \times 10^{-29} e\ \mathrm{cm } = 4.5 \times 10^{-15}e\ \mathrm{GeV}^{-1}
\end{align}
places a restriction on our model similar in nature to that from $\mu \rightarrow e \gamma$:
\begin{align}
|\mathrm{Im}(\sigma_{L,R})| <& 2.3 \times 10^{-15} \mathrm{GeV}^{-1},
\end{align}
which then corresponds to a bound on the imaginary part of the product of three couplings. For arbitrary complex phases we have 
\begin{align}
Y_{EFV} \sim Y_{LFV} \sim Y_{\hE} \sim Y_{\hEt} \lesssim& 10^{-3}  \left( \frac{\mu_{E,R}}{\mathrm{TeV}} \right)^{2/3},
\end{align}
and
\begin{align}
\lambda_{SLR} \sim \lambda_{SE} \sim \lambda_{TLR}  \lesssim& 10^{-3}  \left( \frac{\mu_{E,R}}{\mathrm{TeV}} \right)^{2/3} \left( \frac{v}{v_{S}} \right) \nn\\
\lambda_{TLR}  \lesssim& \mathcal{O}(0.1)  \left( \frac{\mu_{E,R}}{\mathrm{TeV}} \right)^{2/3} \left( \frac{\mathrm{GeV}}{v_{T}} \right),
\end{align}
i.e. this is now the strongest constraint on the model, although again $\lambda_{TLR}$ is barely constrained -- but note that for purely real couplings the constraint disappears.

\subsubsection{$\mu \rightarrow 3e$}
\label{APP:mu3e}

For this process, we need to consider the couplings to the Z and the photon. These come from
\begin{align}
\L \supset& \sum_i Q_i e A_\mu  \ov{\psi}_i  \gamma^\mu \psi_i + \frac{e}{c_W s_W} Z_\mu  \ov{\psi}_i  \gamma^\mu [ (T_3 - s_W^2 )P_L + (T_3 - s_W^2 )P_R ]\psi_i .
\end{align}
Now, however, not all of the left-handed or right-handed leptons have the same $T_3$, since we now have $e_{4,5}^L$ with $T_3 = 0$. We then write
\begin{align}
j_Z^\mu  \supset&  \ov{e}_i  \gamma^\mu [ (\frac{1}{2} - s_W^2 )P_L - s_W^2 P_R ] e_i + (\frac{1}{2} - s_W^2 ) \ov{e}_6  \gamma^\mu e_6 + \sum_{i=4,5} - s_W^2 \ov{e}_{i}  \gamma^\mu  e_i \nn\\
\supset& \sum_i^6  \ov{e}_{i}  \gamma^\mu (\frac{1}{2} P_L - s_W^2)  e_i  + \frac{1}{2} \ov{q}^i_R q^j_R \ov{e}_{i}  \gamma^\mu  P_R  e_j -  \frac{1}{2} q_L^{ik} \ov{q}_L^{jk} \ov{e}_{i}  \gamma^\mu P_L  e_j .
\end{align}
These then give two seperate sources of flavour violation. 

If we write
\begin{align}
\L \supset& -\frac{e}{2c_W s_W} \bigg[ c_L \ov{e}_{2}  \gamma^\mu P_L  e_1 +  c_R \ov{e}_{2}  \gamma^\mu P_R  e_1 +  c_L \ov{e}_{1}  \gamma^\mu P_L  e_2 +  c_R \ov{e}_{1}  \gamma^\mu P_R  e_2 \bigg] ,
\end{align}
we find
\begin{align}
\Gamma (\mu \rightarrow 3e) =& \frac{m_\mu^5 \alpha^2}{1536 \pi M_Z^2 c_W^4 s_W^4} \bigg[ 2 |c_L|^2 (1- 4 s_W^2 + 6 s_W^2) + |c_R|^2 (1- 4 s_W^2 + 12 s_W^2)\bigg].
\end{align}
Defining
\begin{align}
\kappa_L \equiv& -1/2 + s_W^2 \nn\\
\kappa_R \equiv& s_W^2 
\end{align}
this becomes
\begin{align}
\Gamma (\mu \rightarrow 3e) =& \frac{m_\mu^5 \alpha^2}{1536 \pi M_Z^4 c_W^4 s_W^4} \bigg[ 4 |c_L|^2(2\kappa_L^2 +  \kappa_R^2)  + 4 |c_R|^2 (\kappa_L^2 + 2 \kappa_R^2)\bigg].
\end{align}
Applying to our case, we have 
\begin{align}
c_L =&  q_L^{2k} \ov{q}_L^{1k} \nn\\
=& \frac{1}{2} v^2 c_\beta^2 \frac{Y^{2k}_{LFV} \ov{Y}^{1k}_{LFV}}{(\mu^k_E)^2} \nn\\
c_R =& - \ov{q}_R^{2} q_R^{1} \nn\\
=& - \frac{1}{2} v^2 c_\beta^2  \frac{Y^{2}_{EFV} \ov{Y}^{1}_{EFV}}{\mu_R^2},
\end{align}
and thus, using $M_Z^2 = \frac{v^2}{4} \frac{e^2}{c_W^2 s_W^2} = \frac{\pi \alpha }{c_W^2 s_W^2}$
\begin{align}
\Gamma (\mu \rightarrow 3e) =& \frac{m_\mu^5 c_\beta^2}{1536 \pi^3 } \bigg[ |\frac{Y^{2k}_{LFV} \ov{Y}^{1k}_{LFV}}{(\mu^k_E)^2}|^2(2\kappa_L^2 +  \kappa_R^2)  +  |\frac{Y^{2}_{EFV} \ov{Y}^{1}_{EFV}}{\mu_R^2}|^2 (\kappa_L^2 + 2 \kappa_R^2)\bigg].
\end{align}
This provides a similar constraint on the couplings to before. Putting all masses to a TeV, we find
\begin{align}
BR (\mu \rightarrow 3e) =&  2\times 10^{-4} c_\beta^2 \bigg[ |Y^{2k}_{LFV} \ov{Y}^{1k}_{LFV}|^2 + |Y^{2}_{EFV} \ov{Y}^{1}_{EFV}|^2 \bigg] \left( \frac{\mathrm{TeV}}{\mu_{E,R}} \right)^{4}
\end{align}
and so
\begin{align}
|Y^{2k}_{LFV} \ov{Y}^{1k}_{LFV}|^2  \lesssim& 10^{-8} \left( \frac{\mathrm{TeV}}{\mu_{E,R}} \right)^{4}\nn\\
Y^{ik}_{LFV} \sim Y^{j}_{EFV} \lesssim& 10^{-2} \left( \frac{\mathrm{TeV}}{\mu_{E,R}} \right),
\end{align}
which can even be relaxed a little for large $\tan \beta$.

\section{Implementation in SARAH}
\label{APP:SARAH}
\lstset{basicstyle=\scriptsize, frame=shadowbox}
To perform the phenomenological studies described in \ref{SEC:CMDGSSM} we implemented the model in the spectrum generator \SARAH \cite{Staub:2008uz,Staub:2010jh,Staub:2009bi,Staub:2012pb,Dreiner:2012dh,Staub:2013tta} from which we produced \SPheno \cite{Porod:2003um,Porod:2011nf} code. The new states of eq.~\ref{eq:newstates} together with the MSSM particle contents and the adjoints superfields are defined in the 
\SARAH mode file by
\begin{lstlisting}
Fields[[1]] = {{uL,  dL},  3, q,   1/6, 2, 3};  
...

Fields[[8]] = {s, 1, S, 0, 1, 1};
Fields[[9]] = {{{T0/Sqrt[2],Tp},{Tm, -T0/Sqrt[2]}}, 1, T, 0, 3, 1};
Fields[[10]] = {Oc, 1, oc, 0, 1, 8}; 

Fields[[11]] = {{Ru0,Rum}, 1, ru, -1/2, 2, 1};
Fields[[12]] = {{Rdp,Rd0}, 1, rd, 1/2, 2, 1};
Fields[[13]] = {conj[ER1], 2, er1, 1, 1, 1};
Fields[[14]] = {conj[ER2], 2, er2, -1, 1, 1};
\end{lstlisting}
and the superpotential of eq.~(\ref{EQ:Superpotential}) is given as
\begin{lstlisting}
SuperPotential = { {{1, Yu},{u,q,Hu}}, {{-1,Yd},{d,q,Hd}},
                   {{-1,Ye},{e,l,Hd}}, {{1,\[Mu]},{Hu,Hd}},
                   {{1,\[Lambda]},{S,Hd,Hu}}, {{1,LT},{Hd,T,Hu}},
                   {{1,L1},{S}}, {{1/2,MS},{S,S}}, {{1/3, \[Kappa]},{S,S,S}}, 
                   {{1/2,MT},{T,T}}, {{1/2,LS},{S,T,T}}, {{1/2,MO},{oc,oc}},
{{1, LambdaSR},{S,ru,rd}},{{1, MuR},{ru,rd}}, 
{{1, LambdaTR},{ru,T,rd}},
{{1, LambdaSE},{S,er1,er2}},{{1, MuE},{er1,er2}}, 
{{-1, YLFV},{l,Hu,er1}},
{{-1, YEFV},{e,ru,Hd}},
{{-1, YE1},{ru,Hd,er1}},{{-1, YE2},{rd,Hu,er2}}
 }; 
\end{lstlisting}
Finally, one has to enable the Dirac gaugino mass terms which are suppressed by default:
\begin{lstlisting}
AddDiracGauginos = True; 
\end{lstlisting}
This information is already sufficient for \SARAH to calculate the entire two-loop RGEs. 
Thus, with small modifications it was possible to check the behavior of the other models
mentioned in the introduction with respect to the running at one- and two-loop. 

For a study of the mass spectrum one has to define in addition the decomposition of complex
scalars after EWSB and the rotation to the mass eigenstates. The corresponding lines in 
the model file read 
\begin{lstlisting}
DEFINITION[EWSB][VEVs]= 
  {{SHd0, {vd, 1/Sqrt[2]}, {sigmad, \[ImaginaryI]/Sqrt[2]},{phid,1/Sqrt[2]}},
   {SHu0, {vu, 1/Sqrt[2]}, {sigmau, \[ImaginaryI]/Sqrt[2]},{phiu,1/Sqrt[2]}},
   {ST0, {vT, 1/Sqrt[2]}, {sigmaT, \[ImaginaryI]/Sqrt[2]},{phiT,1/Sqrt[2]}},
   {Ss, {vS, 1/Sqrt[2]}, {sigmaS, \[ImaginaryI]/Sqrt[2]},{phiS,1/Sqrt[2]}},
   {SOc, {0,0}, {sigmaO, \[ImaginaryI]/Sqrt[2]},{phiO,1/Sqrt[2]}}
 };

DEFINITION[EWSB][MatterSector]= 
{    ...
     {{SRu0, SRd0}, {RSn, ZR}},
     {{SRum, conj[SRdp],SER1,conj[SER2]}, {RSc, ZRc}},
     {{FRu0, FRd0}, {RN, Zf}},
     {{{FRum,conj[FER2]}, {FRdp,conj[FER1]}}, {{RC1, Zf1},{RC2,Zf2}}},
     {{phid, phiu, phiS, phiT}, {hh, ZH}},
     {{sigmad, sigmau, sigmaS, sigmaT}, {Ah, ZA}},
     {{SHdm,conj[SHup],STm,conj[STp]},{Hpm,ZP}},
     {{fB, fW0, FHd0, FHu0, Fs, FT0}, {L0, ZN}}, 
     {{{fWm, FHdm,FTm}, {fWp, FHup,FTp}}, {{Lm,UM}, {Lp,UP}}},
...
}; 
\end{lstlisting}
Based on this input \SARAH analytically derives  the minimum conditions of the vacuum, 
the mass matrices and the full one-loop corrections of all states. For a numerical study 
one can export this information to \SPheno using the {\tt MakeSPheno[]} command of \SARAH. 

{\tt MakeSPheno} needs as additional input the free parameters in the model, 
the choosen boundary conditions as well as the 
parameters which are fixed by the minimisation of the vacuum. 
The canonical choices of solutions for the MSSM tadpole equations are $m_{H_u}^2, m_{H_d}^2$ or $\mu, B_\mu$ because they
are linear and in the latter case do not enter into the RGEs for the soft masses. However, in our model the equivalent choices would be $m_S^2, m_T^2$ or $B_S, B_T$. We find that both of these are inconvenient: 
\begin{itemize}
\item In the first case, because we are searching for models with large $\lambda_S$, the value of $m_S$ has a very large impact upon the RGEs so the solution we find will vary dramatically from one iteration to the next -- so much so that it becomes difficult to find solutions at all. 
\item In the second case,  there will often be no stable solution, because large values of $B_S, B_T$ lead to tachyonic adjoint scalars. 
\end{itemize}
We chose instead to solve the tadpole equations for $m_{H_u}^2, m_{H_d}^2, v_S$ and $v_T$. Since the equations are 
cubic in the VEVs this demands a numerical method which is now available from version 4 
of \SARAH. The corresponding lines read
\begin{lstlisting}
MINPAR={ {1,m0},   
         {3,TanBeta},
         {7,MDirac},  
         {8,MBilinear}};

EXTPAR = {
       { 61, LambdaInput}, 
       {161, LambdaTInput},
       { 71, LambdaSRInput}, 
       { 171, LambdaTRInput},
       { 81, LambdaSTInput}};

ParametersToSolveTadpoles = {mHu2,mHd2,vT,vS}; 

BoundaryHighScale={
...
{\[Lambda], LambdaInput},
{LT, LambdaTInput},
{LS, LambdaSTInput},
...
{mq2, DIAGONAL m0^2},
{ml2, DIAGONAL m0^2},
{md2, DIAGONAL m0^2},
{mu2, DIAGONAL m0^2},
...
{MDWBT,MDirac},
{MDBS,MDirac},
{MDGoc,MDirac}
};
\end{lstlisting}
Adjusting these lines allows the numerical setup to be easily changed and to study different variants 
of the model. We will make full input files for \SARAH public in the future 
by including them in the \SARAH distribution. 

{\tt MakeSPheno} generates Fortran code which is copied to a new subdirectory of \SPheno 
version 3.1 or above. This new module provides a precise spectrum generator for our new 
model with interesting features:
\begin{itemize}
 \item Running of all RGEs at two-loop including flavour effects.
 \item Including of all thresholds in the given model at $M_Z$ to obtain the 
  initial values of the gauge and  Yukawa coupling for the RGE running.
 \item Calculation of the mass spectrum at one-loop in 't Hooft gauge including the 
 effect of the external momenta.
 \item Calculation of precision observables like $\Delta \rho$, $b\to s \gamma$ 
 and $B^0 \to l \bar{l}$.
\end{itemize}
Compared to the precision which public spectrum generators provide for the MSSM the main
drawback is the missing two-loop corrections. However, this will also be improved in the 
future \cite{Goodsell:2014inprep}. In addition, we have modified the produced source code to 
increase the numerical stability. The changes are:
\begin{itemize}
 \item Increasing the maximum number of iterations before \SPheno stops because of tachyons; it often occurs that early iterations include tachyons while the code can eventually settle on a stable solution. Indeed, changing the values of $v_S, v_T$ changes the values of couplings, and so this is to be expected. However, to avoid needless looping over hopeless parameter points which will never converge, we set a limit on the number of consecutive tachyonic iterations.  
 \item Input of the parameters $\lambda_S, m_S$ at the SUSY scale rather than the GUT scale. As mentioned in the text, this is due to the large sensitivity to these paramters on running. However, we ensure that \emph{all} couplings are perturbative at the GUT scale. 
\item Hybrid version of solution of tadpole equations. Since we set many couplings to zero in the scans (particularly the R-symmetry-violating superpotential couplings) then attempting to solve all four tadpole equations by brute force with Broyden's method fails due to a degenerate Jacobian: the equations for $v_S, v_T$ do not depend on $m_{H_{u,d}}^2$. Hence instead we solve first for $v_S, v_T$ and then solve the resulting linear equations for $m_{H_{u,d}}^2$ analytically (this meant in fact originally setting the solutions to the tadpoles in \SARAH to be {\tt \{mHu2,mHd2,mS2,mT2\}} and adding the Broyden routine by hand). Since in the limit we are considering the equations for $v_S, v_T$ are almost linear this guarantees an accurate and fast solution. 
\item Addition of single line interface so that the code can be called by passing command-line parameters rather than a Les Houches Input file; the major advantage of our scenario is that we have only a few parameters, and this method of calling the program removes the need for reading and writing input files, potentially accelerating the scan but also simplifying the checking of parameter points.  
\end{itemize}
Because of these modifications the Fortran code and model files can be obtained upon request from the authors.\footnote{We will endeavour to make a version available for general release in the future. }

\section{RGEs}
\label{APP:RGEs}

In this appendix we present the RGEs for the model after making the simplification that the Yukawa couplings are only non-zero for the third generation (so $y_t, y_b$ and $y_\tau$) and that the first two generations of soft masses are degenerate, but differing from the third, denoting the soft squark and slepton masses $m_{Q[i]}, m_{U[i]}, m_{D[i]}, m_{L[i]}, m_{E[i]} $ where $i$ is the generation; for brevity we also set the couplings $Y_{\hE i},  Y_{\hEt i}, Y_{LFV}^{ij} $ to zero. 

The purpose of these is to understand the RG flows; for example, it is well known that it is not possible to completely decouple the first two generations of squarks or sleptons due to their contribution at two-loops (via the trace terms). In this model we also find a similar effect for the octet scalar mass - which does not contribute to other RGEs at one loop. 

\input{simplerges_noLFV.tex}

\bibliography{Dirac}
\bibliographystyle{h-physrev5}
\end{document}

%% file: simplerges_noLFV.tex
\subsection{Gauge Couplings}
{\allowdisplaybreaks  \begin{align} 
\beta_{g_1}^{(1)} = &  
\frac{48}{5} g_{1}^{3} \\ 
\beta_{g_1}^{(2)} = &  
\frac{1}{25} g_{1}^{3} \Big(352 g_{1}^{2} +5 \Big(-14 y_{b}^{2}  -18 y_{\tau}^{2}  -26 y_{t}^{2}  + 36 g_{2}^{2}  -6 \lambda_{S}^{2}  -6 \lambda_{SR}^{2}  \nonumber \\ 
 &+ 88 g_{3}^{2}  -9 \lambda_{T}^{2}  -9 \lambda_{TR}^{2} \Big)-60 \mbox{Tr}\Big({\lambda_{SE} \lambda_{SE}^{T} }\Big) \Big)\\ 
\beta_{g_2}^{(1)} = &  
4 g_{2}^{3} \\ 
\beta_{g_2}^{(2)} = &  
\frac{1}{5} g_{2}^{3} \Big(12 g_{1}^{2}  + 5 \Big(24 g_{3}^{2}  -2 \lambda_{S}^{2}  -2 \lambda_{SR}^{2}  -2 y_{\tau}^{2}  + 56 g_{2}^{2}  -6 y_{b}^{2}  -6 y_{t}^{2}  -7 \lambda_{T}^{2}  -7 \lambda_{TR}^{2} \Big)\Big)\\ 
\beta_{g_3}^{(1)} = &  
0\\ 
\beta_{g_3}^{(2)} = &  
\frac{1}{5} g_{3}^{3} \Big(11 g_{1}^{2}  + 5 \Big(-4 \Big(y_{b}^{2} + y_{t}^{2}\Big) + 68 g_{3}^{2}  + 9 g_{2}^{2} \Big)\Big)
\end{align}} 
\subsection{Dirac gaugino masses}
{\allowdisplaybreaks  \begin{align} 
\beta_{m_{DY} }^{(1)} = &  
\frac{2}{5} m_{DY}  \Big(24 g_{1}^{2}  + 5 \Big(\lambda_{SR}^{2} + \lambda_{S}^{2}\Big)\Big) + m_{DY}  \mbox{Tr}\Big({\lambda_{SE} \lambda_{SE}^{T} }\Big) \\ 
\beta_{m_{DY} }^{(2)} = &  
\frac{1}{25} m_{DY}  \Big(352 g_{1}^{4} +5 g_{1}^{2} \Big(-14 y_{b}^{2}  -18 y_{\tau}^{2}  -26 y_{t}^{2}  + 36 g_{2}^{2}  + 88 g_{3}^{2}  -9 \lambda_{T}^{2}  -9 \lambda_{TR}^{2} \Big)\nonumber \\ 
 &-50 \Big(2 \lambda_{SR}^{4}  -3 g_{2}^{2} \Big(\lambda_{SR}^{2} + \lambda_{S}^{2}\Big) + 3 \lambda_{SR}^{2} \lambda_{TR}^{2}  + \lambda_{S}^{2} \Big(2 \lambda_{S}^{2}  + 3 \lambda_{T}^{2}  + 3 y_{b}^{2}  + 3 y_{t}^{2}  + y_{\tau}^{2}\Big)\Big)\nonumber \\ 
 &-50 \mbox{Tr}\Big({\lambda_{SE} \lambda_{SE}^{T} \lambda_{SE} \lambda_{SE}^{T} }\Big) \Big)\\ 
\beta_{m_{D2} }^{(1)} = &  
m_{D2}  \Big(\lambda_{TR}^{2} + \lambda_{T}^{2}\Big)\\ 
\beta_{m_{D2} }^{(2)} = &  
\frac{1}{5} m_{D2}  \Big(3 g_{1}^{2} \Big(4 g_{2}^{2}  + \lambda_{TR}^{2} + \lambda_{T}^{2}\Big)\nonumber \\ 
 &+5 \Big(88 g_{2}^{4} -2 \lambda_{SR}^{2} \lambda_{TR}^{2} -3 \lambda_{TR}^{4} -3 y_{b}^{2} \lambda_{T}^{2} - y_{\tau}^{2} \lambda_{T}^{2} -3 y_{t}^{2} \lambda_{T}^{2} -2 \lambda_{S}^{2} \lambda_{T}^{2} -3 \lambda_{T}^{4} \nonumber \\ 
 &+g_{2}^{2} \Big(-2 \Big(3 y_{b}^{2}  + 3 y_{t}^{2}  + 4 \lambda_{T}^{2}  + 4 \lambda_{TR}^{2}  + \lambda_{SR}^{2} + y_{\tau}^{2} + \lambda_{S}^{2}\Big) + 24 g_{3}^{2} \Big)\Big)\Big)\\ 
\beta_{m_{D3} }^{(1)} = &  
-6 g_{3}^{2} m_{D3}  \\ 
\beta_{m_{D3} }^{(2)} = &  
\frac{1}{5} g_{3}^{2} m_{D3}  \Big(11 g_{1}^{2}  + 5 \Big(104 g_{3}^{2}  -4 y_{b}^{2}  -4 y_{t}^{2}  + 9 g_{2}^{2} \Big)\Big)
\end{align}} 
\subsection{Yukawa couplings}
{\allowdisplaybreaks  \begin{align} 
\beta_{y_t}^{(1)} = &  
-3 g_{2}^{2} y_t  + 6 y_{t}^{3}  -\frac{13}{15} g_{1}^{2} y_t  -\frac{16}{3} g_{3}^{2} y_t  + \frac{3}{2} y_t \lambda_{T}^{2}  + y_{b}^{2} y_t  + y_t \lambda_{S}^{2} \\ 
\beta_{y_t}^{(2)} = &  
+\frac{3913}{450} g_{1}^{4} y_t +\frac{33}{2} g_{2}^{4} y_t +8 g_{2}^{2} g_{3}^{2} y_t +\frac{128}{9} g_{3}^{4} y_t -5 y_{b}^{4} y_t - y_{b}^{2} y_{\tau}^{2} y_t +6 g_{2}^{2} y_{t}^{3} +16 g_{3}^{2} y_{t}^{3} \nonumber \\ 
 &-5 y_{b}^{2} y_{t}^{3} -22 y_{t}^{5} +\frac{1}{45} g_{1}^{2} y_t \Big(136 g_{3}^{2}  + 18 y_{b}^{2}  + 45 g_{2}^{2}  + 54 y_{t}^{2} \Big)-2 \lambda_{SR}^{2} y_t \lambda_{S}^{2} -4 y_{b}^{2} y_t \lambda_{S}^{2} \nonumber \\ 
 &- y_{\tau}^{2} y_t \lambda_{S}^{2} -3 y_{t}^{3} \lambda_{S}^{2} -3 y_t \lambda_{S}^{4} +6 g_{2}^{2} y_t \lambda_{T}^{2} -\frac{3}{2} \lambda_{TR}^{2} y_t \lambda_{T}^{2} -6 y_{b}^{2} y_t \lambda_{T}^{2} -\frac{3}{2} y_{\tau}^{2} y_t \lambda_{T}^{2} -\frac{9}{2} y_{t}^{3} \lambda_{T}^{2} \nonumber \\ 
 &-3 y_t \lambda_{S}^{2} \lambda_{T}^{2} -\frac{15}{4} y_t \lambda_{T}^{4} - y_t \lambda_{S}^{2} \mbox{Tr}\Big({\lambda_{SE} \lambda_{SE}^{T} }\Big) \\ 
\beta_{y_b}^{(1)} = &  
y_b \Big(-3 g_{2}^{2}  + 6 y_{b}^{2}  -\frac{16}{3} g_{3}^{2}  + \frac{3}{2} \lambda_{T}^{2}  -\frac{7}{15} g_{1}^{2}  + y_{\tau}^{2} + y_{t}^{2} + \lambda_{S}^{2}\Big)\\ 
\beta_{y_b}^{(2)} = &  
+\frac{413}{90} g_{1}^{4} y_b +\frac{33}{2} g_{2}^{4} y_b +8 g_{2}^{2} g_{3}^{2} y_b +\frac{128}{9} g_{3}^{4} y_b +6 g_{2}^{2} y_{b}^{3} +16 g_{3}^{2} y_{b}^{3} -22 y_{b}^{5} -3 y_{b}^{3} y_{\tau}^{2} \nonumber \\ 
 &-3 y_b y_{\tau}^{4} -5 y_{b}^{3} y_{t}^{2} -5 y_b y_{t}^{4} +\frac{1}{45} g_{1}^{2} y_b \Big(18 \Big(2 y_{t}^{2}  + 3 y_{\tau}^{2}  + y_{b}^{2}\Big) + 40 g_{3}^{2}  + 45 g_{2}^{2} \Big)-2 \lambda_{SR}^{2} y_b \lambda_{S}^{2} \nonumber \\ 
 &-3 y_{b}^{3} \lambda_{S}^{2} -4 y_b y_{t}^{2} \lambda_{S}^{2} -3 y_b \lambda_{S}^{4} +6 g_{2}^{2} y_b \lambda_{T}^{2} -\frac{3}{2} \lambda_{TR}^{2} y_b \lambda_{T}^{2} -\frac{9}{2} y_{b}^{3} \lambda_{T}^{2} -6 y_b y_{t}^{2} \lambda_{T}^{2} -3 y_b \lambda_{S}^{2} \lambda_{T}^{2} \nonumber \\ 
 &-\frac{15}{4} y_b \lambda_{T}^{4} - y_b \lambda_{S}^{2} \mbox{Tr}\Big({\lambda_{SE} \lambda_{SE}^{T} }\Big) \\ 
\beta_{y_{\tau}}^{(1)} = &  
y_{\tau} \Big(-3 g_{2}^{2}  + 3 y_{b}^{2}  + 4 y_{\tau}^{2}  + \frac{3}{2} \lambda_{T}^{2}  -\frac{9}{5} g_{1}^{2}  + \lambda_{S}^{2}\Big)\\ 
\beta_{y_{\tau}}^{(2)} = &  
-\frac{1}{20} y_{\tau} \Big(-378 g_{1}^{4} -4 g_{1}^{2} \Big(-2 y_{b}^{2}  + 6 y_{\tau}^{2}  + 9 g_{2}^{2} \Big)\nonumber \\ 
 &+5 \Big(-66 g_{2}^{4} -64 g_{3}^{2} y_{b}^{2} +36 y_{b}^{4} +36 y_{b}^{2} y_{\tau}^{2} +40 y_{\tau}^{4} +12 y_{b}^{2} y_{t}^{2} +8 \lambda_{SR}^{2} \lambda_{S}^{2} +12 y_{\tau}^{2} \lambda_{S}^{2} +12 y_{t}^{2} \lambda_{S}^{2} \nonumber \\ 
 &+12 \lambda_{S}^{4} +6 \lambda_{TR}^{2} \lambda_{T}^{2} +18 y_{\tau}^{2} \lambda_{T}^{2} +18 y_{t}^{2} \lambda_{T}^{2} +12 \lambda_{S}^{2} \lambda_{T}^{2} +15 \lambda_{T}^{4} -24 g_{2}^{2} \Big(y_{\tau}^{2} + \lambda_{T}^{2}\Big)\Big)\nonumber \\ 
 &+20 \lambda_{S}^{2} \mbox{Tr}\Big({\lambda_{SE} \lambda_{SE}^{T} }\Big) \Big)\\ 
\beta_{\lambda_{S}}^{(1)} = &  
-\frac{3}{5} g_{1}^{2} \lambda_{S}  + \lambda_{S} \Big(2 \lambda_{SR}^{2}  -3 g_{2}^{2}  + 3 \lambda_{T}^{2}  + 3 y_{b}^{2}  + 3 y_{t}^{2}  + 4 \lambda_{S}^{2}  + y_{\tau}^{2}\Big) + \lambda_{S} \mbox{Tr}\Big({\lambda_{SE} \lambda_{SE}^{T} }\Big) \\ 
\beta_{\lambda_{S}}^{(2)} = &  
-\frac{1}{50} \lambda_{S} \Big(-297 g_{1}^{4} -90 g_{1}^{2} g_{2}^{2} -825 g_{2}^{4} +200 \lambda_{SR}^{4} +20 g_{1}^{2} y_{b}^{2} -800 g_{3}^{2} y_{b}^{2} +450 y_{b}^{4} \nonumber \\ 
 &-60 g_{1}^{2} y_{\tau}^{2} +150 y_{\tau}^{4} -40 g_{1}^{2} y_{t}^{2} -800 g_{3}^{2} y_{t}^{2} +300 y_{b}^{2} y_{t}^{2} +450 y_{t}^{4} -60 g_{1}^{2} \lambda_{S}^{2} \nonumber \\ 
 &-300 g_{2}^{2} \lambda_{S}^{2} +450 y_{b}^{2} \lambda_{S}^{2} +150 y_{\tau}^{2} \lambda_{S}^{2} +450 y_{t}^{2} \lambda_{S}^{2} +500 \lambda_{S}^{4} \nonumber \\ 
 &-20 \lambda_{SR}^{2} \Big(3 g_{1}^{2}  + 5 \Big(-2 \lambda_{S}^{2}  + 3 g_{2}^{2}  -3 \lambda_{TR}^{2} \Big)\Big)-600 g_{2}^{2} \lambda_{T}^{2} +150 \lambda_{TR}^{2} \lambda_{T}^{2} +225 y_{b}^{2} \lambda_{T}^{2} \nonumber \\ 
 &+75 y_{\tau}^{2} \lambda_{T}^{2} +225 y_{t}^{2} \lambda_{T}^{2} +600 \lambda_{S}^{2} \lambda_{T}^{2} +375 \lambda_{T}^{4} -120 g_{1}^{2} \mbox{Tr}\Big({\lambda_{SE} \lambda_{SE}^{T} }\Big) +100 \lambda_{S}^{2} \mbox{Tr}\Big({\lambda_{SE} \lambda_{SE}^{T} }\Big) \nonumber \\ 
 &+100 \mbox{Tr}\Big({\lambda_{SE} \lambda_{SE}^{T} \lambda_{SE} \lambda_{SE}^{T} }\Big) \Big)\\ 
\beta_{\lambda_{T}}^{(1)} = &  
-\frac{3}{5} g_{1}^{2} \lambda_{T}  + \lambda_{T} \Big(2 \lambda_{S}^{2}  + 3 y_{b}^{2}  + 3 y_{t}^{2}  + 4 \lambda_{T}^{2}  -7 g_{2}^{2}  + \lambda_{TR}^{2} + y_{\tau}^{2}\Big)\\ 
\beta_{\lambda_{T}}^{(2)} = &  
-\frac{1}{50} \lambda_{T} \Big(-297 g_{1}^{4} -10 g_{1}^{2} \Big(-2 y_{b}^{2}  + 3 \lambda_{T}^{2}  + 3 \lambda_{TR}^{2}  + 4 y_{t}^{2}  + 6 y_{\tau}^{2}  + 9 g_{2}^{2} \Big)\nonumber \\ 
 &-25 \Big(97 g_{2}^{4} -4 \lambda_{SR}^{2} \lambda_{TR}^{2} -6 \lambda_{TR}^{4} +32 g_{3}^{2} y_{b}^{2} -18 y_{b}^{4} -6 y_{\tau}^{4} +32 g_{3}^{2} y_{t}^{2} -12 y_{b}^{2} y_{t}^{2} -18 y_{t}^{4} \nonumber \\ 
 &-8 \lambda_{SR}^{2} \lambda_{S}^{2} -6 y_{b}^{2} \lambda_{S}^{2} -2 y_{\tau}^{2} \lambda_{S}^{2} -6 y_{t}^{2} \lambda_{S}^{2} -12 \lambda_{S}^{4} -6 \lambda_{TR}^{2} \lambda_{T}^{2} -15 y_{b}^{2} \lambda_{T}^{2} -5 y_{\tau}^{2} \lambda_{T}^{2} \nonumber \\ 
 &-15 y_{t}^{2} \lambda_{T}^{2} -16 \lambda_{S}^{2} \lambda_{T}^{2} -21 \lambda_{T}^{4} -2 g_{2}^{2} \Big(-11 \lambda_{T}^{2}  + \lambda_{TR}^{2}\Big)\Big)\nonumber \\ 
 &+100 \lambda_{S}^{2} \mbox{Tr}\Big({\lambda_{SE} \lambda_{SE}^{T} }\Big) \Big)\\ 
\beta_{\lambda_{SR}}^{(1)} = &  
-\frac{3}{5} g_{1}^{2} \lambda_{SR}  + \lambda_{SR} \Big(2 \lambda_{S}^{2}  -3 g_{2}^{2}  + 3 \lambda_{TR}^{2}  + 4 \lambda_{SR}^{2} \Big) + \lambda_{SR} \mbox{Tr}\Big({\lambda_{SE} \lambda_{SE}^{T} }\Big) \\ 
\beta_{\lambda_{SR}}^{(2)} = &  
-\frac{1}{50} \lambda_{SR} \Big(-297 g_{1}^{4} -90 g_{1}^{2} g_{2}^{2} -825 g_{2}^{4} +500 \lambda_{SR}^{4} +375 \lambda_{TR}^{4} -60 g_{1}^{2} \lambda_{S}^{2} -300 g_{2}^{2} \lambda_{S}^{2} \nonumber \\ 
 &+300 y_{b}^{2} \lambda_{S}^{2} +100 y_{\tau}^{2} \lambda_{S}^{2} +300 y_{t}^{2} \lambda_{S}^{2} +200 \lambda_{S}^{4} +300 \lambda_{S}^{2} \lambda_{T}^{2} -75 \lambda_{TR}^{2} \Big(-2 \lambda_{T}^{2}  + 8 g_{2}^{2} \Big)\nonumber \\ 
 &-20 \lambda_{SR}^{2} \Big(3 g_{1}^{2}  + 5 \Big(-2 \lambda_{S}^{2}  + 3 g_{2}^{2}  -6 \lambda_{TR}^{2} \Big) -5 \mbox{Tr}\Big({\lambda_{SE} \lambda_{SE}^{T} }\Big) \Big)-120 g_{1}^{2} \mbox{Tr}\Big({\lambda_{SE} \lambda_{SE}^{T} }\Big) \nonumber \\ 
 &+100 \mbox{Tr}\Big({\lambda_{SE} \lambda_{SE}^{T} \lambda_{SE} \lambda_{SE}^{T} }\Big) \Big)\\ 
\beta_{\lambda_{TR}}^{(1)} = &  
-\frac{3}{5} g_{1}^{2} \lambda_{TR}  + \lambda_{TR} \Big(2 \lambda_{SR}^{2}  + 4 \lambda_{TR}^{2}  -7 g_{2}^{2}  + \lambda_{T}^{2}\Big)\\ 
\beta_{\lambda_{TR}}^{(2)} = &  
-\frac{1}{50} \lambda_{TR} \Big(-297 g_{1}^{4} -30 g_{1}^{2} \Big(3 g_{2}^{2}  + \lambda_{TR}^{2} + \lambda_{T}^{2}\Big)\nonumber \\ 
 &-25 \Big(97 g_{2}^{4} -12 \lambda_{SR}^{4} -16 \lambda_{SR}^{2} \lambda_{TR}^{2} -21 \lambda_{TR}^{4} -8 \lambda_{SR}^{2} \lambda_{S}^{2} -6 \lambda_{TR}^{2} \lambda_{T}^{2} -6 y_{b}^{2} \lambda_{T}^{2} -2 y_{\tau}^{2} \lambda_{T}^{2} \nonumber \\ 
 &-6 y_{t}^{2} \lambda_{T}^{2} -4 \lambda_{S}^{2} \lambda_{T}^{2} -6 \lambda_{T}^{4} +g_{2}^{2} \Big(22 \lambda_{TR}^{2}  -2 \lambda_{T}^{2} \Big)\Big)\nonumber \\ 
 &+100 \lambda_{SR}^{2} \mbox{Tr}\Big({\lambda_{SE} \lambda_{SE}^{T} }\Big) \Big)\\ 
\beta_{\lambda_{SE}}^{(1)} = &  
2 \lambda_{SE} \lambda_{SE}^{T} \lambda_{SE}   + \lambda_{SE} \Big(2 \Big(\lambda_{SR}^{2} + \lambda_{S}^{2}\Big) -\frac{12}{5} g_{1}^{2}  + \mbox{Tr}\Big({\lambda_{SE} \lambda_{SE}^{T} }\Big)\Big)\\ 
\beta_{\lambda_{SE}}^{(2)} = &  
-4 \lambda_{SR}^{2} \lambda_{SE} \lambda_{SE}^{T} \lambda_{SE}  -4 \lambda_{S}^{2} \lambda_{SE} \lambda_{SE}^{T} \lambda_{SE}  -2 \lambda_{SE} \lambda_{SE}^{T} \lambda_{SE} \Big(\lambda_{SE}^{T} \lambda_{SE}  + \mbox{Tr}\Big({\lambda_{SE} \lambda_{SE}^{T} }\Big)\Big) \nonumber \\ 
 &-\frac{2}{25} \lambda_{SE} \Big(-324 g_{1}^{4} +50 \lambda_{SR}^{4} -15 \lambda_{SR}^{2} \Big(5 g_{2}^{2}  -5 \lambda_{TR}^{2}  + g_{1}^{2}\Big)+50 \lambda_{S}^{4} \nonumber \\ 
 &+5 \lambda_{S}^{2} \Big(-3 g_{1}^{2}  + 5 \Big(-3 g_{2}^{2}  + 3 \lambda_{T}^{2}  + 3 y_{b}^{2}  + 3 y_{t}^{2}  + y_{\tau}^{2}\Big)\Big)-30 g_{1}^{2} \mbox{Tr}\Big({\lambda_{SE} \lambda_{SE}^{T} }\Big) \nonumber \\ 
 &+25 \mbox{Tr}\Big({\lambda_{SE} \lambda_{SE}^{T} \lambda_{SE} \lambda_{SE}^{T} }\Big) \Big)
\end{align}} 
\subsection{Soft-Breaking Scalar Masses}
\begin{align} 
\sigma_{1,1} = & \sqrt{\frac{3}{5}} g_1 \Big(2 m_{\mathrm D [1,2]}^2   + 2 m_{\mathrm E [1,2]}^2   -2 m_{\mathrm L [1,2]}^2   + 2 m_{\mathrm Q [1,2]}^2   -2 m_{\mathrm U [3]}^2   -4 m_{\mathrm U[1,2]}^2   - \mbox{Tr}\Big({m_{\hat{\bar{E}}}^2}\Big) \nonumber \\
& - m_{H_d}^2  - m_{\mathrm L [3]}^2   - m_{R_u}^2  + m_{\mathrm D[3]}^2  + m_{\mathrm E [3]}^2  + m_{H_u}^2 + m_{\mathrm Q [3]}^2  + m_{R_d}^2 + \mbox{Tr}\Big({m_{\hat{E}}^2}\Big)\Big)\\ 
\sigma_{2,1} = & \frac{1}{10} g_{1}^{2} \Big(12 m_{\mathrm E [1,2]}^2   + 16 m_{\mathrm U[1,2]}^2   + 2 m_{\mathrm D[3]}^2   + 2 m_{\mathrm Q [1,2]}^2   + 3 m_{H_d}^2  + 3 m_{H_u}^2  + 3 m_{\mathrm L [3]}^2   \nonumber \\
&+ 3 m_{R_d}^2  + 3 m_{R_u}^2  + 4 m_{\mathrm D [1,2]}^2   + 6 \mbox{Tr}\Big({m_{\hat{\bar{E}}}^2}\Big)  + 6 \mbox{Tr}\Big({m_{\hat{E}}^2}\Big)  + 6 m_{\mathrm E [3]}^2   + 6 m_{\mathrm L [1,2]}^2   + 8 m_{\mathrm U [3]}^2   + m_{\mathrm Q [3]}^2 \Big)\\ 
\sigma_{3,1} = & \frac{1}{20} \frac{1}{\sqrt{15}} g_1 \Big(4 g_{1}^{2} \Big(2 m_{\mathrm D [1,2]}^2   + m_{\mathrm D[3]}^2 \Big)+80 g_{3}^{2} \Big(2 m_{\mathrm D [1,2]}^2   + m_{\mathrm D[3]}^2 \Big)\nonumber \\
&+36 g_{1}^{2} \Big(2 m_{\mathrm E [1,2]}^2   + m_{\mathrm E [3]}^2 \Big)-9 g_{1}^{2} m_{H_d}^2 -45 g_{2}^{2} m_{H_d}^2 +9 g_{1}^{2} m_{H_u}^2 \nonumber \\ 
 &+45 g_{2}^{2} m_{H_u}^2 -9 g_{1}^{2} \Big(2 m_{\mathrm L [1,2]}^2   + m_{\mathrm L [3]}^2 \Big)-45 g_{2}^{2} \Big(2 m_{\mathrm L [1,2]}^2   + m_{\mathrm L [3]}^2 \Big)+g_{1}^{2} \Big(2 m_{\mathrm Q [1,2]}^2   + m_{\mathrm Q [3]}^2 \Big)\nonumber \\
&+45 g_{2}^{2} \Big(2 m_{\mathrm Q [1,2]}^2   + m_{\mathrm Q [3]}^2 \Big)+80 g_{3}^{2} \Big(2 m_{\mathrm Q [1,2]}^2   + m_{\mathrm Q [3]}^2 \Big)\nonumber \\ 
 &+9 g_{1}^{2} m_{R_d}^2 +45 g_{2}^{2} m_{R_d}^2 -30 \lambda_{SR}^{2} \Big(- m_{R_u}^2  + m_{R_d}^2\Big)-45 \lambda_{TR}^{2} \Big(- m_{R_u}^2  + m_{R_d}^2\Big)\nonumber \\ 
 &-9 g_{1}^{2} m_{R_u}^2 -45 g_{2}^{2} m_{R_u}^2 -32 g_{1}^{2} \Big(2 m_{\mathrm U[1,2]}^2   + m_{\mathrm U [3]}^2 \Big)-160 g_{3}^{2} \Big(2 m_{\mathrm U[1,2]}^2   + m_{\mathrm U [3]}^2 \Big)\nonumber \\
&-60 m_{\mathrm D[3]}^2  y_{b}^{2} +90 m_{H_d}^2 y_{b}^{2} -30 m_{\mathrm Q [3]}^2  y_{b}^{2} \nonumber \\ 
 &-60 m_{\mathrm E [3]}^2  y_{\tau}^{2} +30 m_{H_d}^2 y_{\tau}^{2} +30 m_{\mathrm L [3]}^2  y_{\tau}^{2} -90 m_{H_u}^2 y_{t}^{2} -30 m_{\mathrm Q [3]}^2  y_{t}^{2} +120 m_{\mathrm U [3]}^2  y_{t}^{2}\nonumber\\
& +30 m_{H_d}^2 \lambda_{S}^{2} -30 m_{H_u}^2 \lambda_{S}^{2} \nonumber \\ 
 &+45 m_{H_d}^2 \lambda_{T}^{2} -45 m_{H_u}^2 \lambda_{T}^{2} +36 g_{1}^{2} \mbox{Tr}\Big({m_{\hat{E}}^2}\Big) -36 g_{1}^{2} \mbox{Tr}\Big({m_{\hat{\bar{E}}}^2}\Big) +30 \mbox{Tr}\Big({\lambda_{SE} m_{\hat{\bar{E}}}^2 \lambda_{SE}^{T} }\Big) \nonumber \\ 
 &-30 \mbox{Tr}\Big({\lambda_{SE} \lambda_{SE}^{T} m_{\hat{E}}^2 }\Big) \Big)\\ 
\sigma_{2,2} = & \frac{1}{2} \Big(2 m_{\mathrm L [1,2]}^2   + 3 m_{\mathrm Q [3]}^2   + 6 m_{\mathrm Q [1,2]}^2   + 4 m_T^2  + m_{H_d}^2 + m_{H_u}^2 + m_{\mathrm L [3]}^2  + m_{R_d}^2 + m_{R_u}^2\Big)\\ 
\sigma_{2,3} = & 2 m_{\mathrm Q [1,2]}^2   + 3 m_O^2  + \frac{1}{2} m_{\mathrm D[3]}^2   + \frac{1}{2} m_{\mathrm U [3]}^2   + m_{\mathrm D [1,2]}^2  + m_{\mathrm Q [3]}^2  + m_{\mathrm U[1,2]}^2 
\end{align} 
{\allowdisplaybreaks  \begin{align} 
\beta_{m_{\mathrm Q [1,2]}^2 }^{(1)} = &  
\frac{1}{\sqrt{15}} g_1 {\bf 1} \sigma_{1,1} \\ 
\beta_{m_{\mathrm Q [1,2]}^2 }^{(2)} = &  
\frac{2}{15} {\bf 1} \Big(45 g_{2}^{4} \sigma_{2,2}  + 80 g_{3}^{4} \sigma_{2,3}  + g_1 \Big(2 \sqrt{15} \sigma_{3,1}  + g_1 \sigma_{2,1} \Big)\Big)\\ 
\beta_{m_{\mathrm Q [3]}^2 }^{(1)} = &  
2 m_{H_d}^2 y_{b}^{2}  + 2 m_{H_u}^2 y_{t}^{2}  + 2 m_{\mathrm D[3]}^2  y_{b}^{2}  + 2 m_{\mathrm Q [3]}^2  y_{b}^{2}  + 2 m_{\mathrm Q [3]}^2  y_{t}^{2}  + 2 m_{\mathrm U [3]}^2  y_{t}^{2}  + \frac{1}{\sqrt{15}} g_1 {\bf 1} \sigma_{1,1} \\ 
\beta_{m_{\mathrm Q [3]}^2 }^{(2)} = &  
+\frac{4}{5} g_{1}^{2} m_{\mathrm D[3]}^2  y_{b}^{2} +\frac{4}{5} g_{1}^{2} m_{\mathrm Q [3]}^2  y_{b}^{2} -14 m_{\mathrm D[3]}^2  y_{b}^{4} -8 m_{H_d}^2 y_{b}^{4} -14 m_{\mathrm Q [3]}^2  y_{b}^{4} -2 m_{\mathrm D[3]}^2  y_{b}^{2} y_{\tau}^{2}  \nonumber \\ 
 &+\frac{8}{5} g_{1}^{2} m_{H_u}^2 y_{t}^{2} +\frac{8}{5} g_{1}^{2} m_{\mathrm Q [3]}^2  y_{t}^{2} +\frac{8}{5} g_{1}^{2} m_{\mathrm U [3]}^2  y_{t}^{2} -20 m_{H_u}^2 y_{t}^{4} -20 m_{\mathrm Q [3]}^2  y_{t}^{4} -20 m_{\mathrm U [3]}^2  y_{t}^{4}  \nonumber \\ 
 &-2 m_{\mathrm Q [3]}^2  y_{b}^{2} \lambda_{S}^{2} -2 m_{H_d}^2 y_{t}^{2} \lambda_{S}^{2} -4 m_{H_u}^2 y_{t}^{2} \lambda_{S}^{2} -2 m_{\mathrm Q [3]}^2  y_{t}^{2} \lambda_{S}^{2} -2 m_S^2 y_{t}^{2} \lambda_{S}^{2} -2 m_{\mathrm U [3]}^2  y_{t}^{2} \lambda_{S}^{2} \nonumber \\ 
 &-3 m_{\mathrm D[3]}^2  y_{b}^{2} \lambda_{T}^{2} -3 m_{\mathrm Q [3]}^2  y_{b}^{2} \lambda_{T}^{2} -3 m_{H_d}^2 y_{t}^{2} \lambda_{T}^{2} -6 m_{H_u}^2 y_{t}^{2} \lambda_{T}^{2} -3 m_{\mathrm Q [3]}^2  y_{t}^{2} \lambda_{T}^{2} -3 m_T^2 y_{t}^{2} \lambda_{T}^{2}  \nonumber \\ 
 &-\frac{1}{5} y_{b}^{2} \Big(2 \Big(-2 g_{1}^{2} m_{H_d}^2 +15 m_{\mathrm D[3]}^2  y_{b}^{2} +30 m_{H_d}^2 y_{b}^{2} +15 m_{\mathrm Q [3]}^2  y_{b}^{2} +5 m_{\mathrm E [3]}^2  y_{\tau}^{2} +10 m_{H_d}^2 y_{\tau}^{2}  \nonumber \\ 
 &+5 m_{\mathrm L [3]}^2  y_{\tau}^{2} -3 m_{\mathrm U [3]}^2  y_{t}^{2} \lambda_{T}^{2} -2 m_{\mathrm Q [3]}^2  y_{b}^{2} y_{\tau}^{2} -2 m_{\mathrm D[3]}^2  y_{b}^{2} \lambda_{S}^{2} \nonumber \\ 
 &+5 \Big(2 m_{H_d}^2  + m_{H_u}^2 + m_S^2\Big)\lambda_{S}^{2} \Big)+15 \Big(2 m_{H_d}^2  + m_{H_u}^2 + m_T^2\Big)\lambda_{T}^{2} \Big)\nonumber \\ 
 &+\frac{2}{15} {\bf 1} \Big(45 g_{2}^{4} \sigma_{2,2}  + 80 g_{3}^{4} \sigma_{2,3}  + g_1 \Big(2 \sqrt{15} \sigma_{3,1}  + g_1 \sigma_{2,1} \Big)\Big)\\ 
\beta_{m_{\mathrm L [1,2]}^2 }^{(1)} = &  
- \sqrt{\frac{3}{5}} g_1 {\bf 1} \sigma_{1,1} \\ 
\beta_{m_{\mathrm L [1,2]}^2 }^{(2)} = &  
{\bf 1} \Big(6 g_{2}^{4} \sigma_{2,2}  + \frac{2}{5} g_1 \Big(-2 \sqrt{15} \sigma_{3,1}  + 3 g_1 \sigma_{2,1} \Big)\Big)\\ 
\beta_{m_{\mathrm L [3]}^2 }^{(1)} = &  
2 m_{H_d}^2 y_{\tau}^{2}  + 2 m_{\mathrm E [3]}^2  y_{\tau}^{2}  + 2 m_{\mathrm L [3]}^2  y_{\tau}^{2}  - \sqrt{\frac{3}{5}} g_1 {\bf 1} \sigma_{1,1} \\ 
\beta_{m_{\mathrm L [3]}^2 }^{(2)} = &  
+\frac{12}{5} g_{1}^{2} m_{\mathrm E [3]}^2  y_{\tau}^{2} +\frac{12}{5} g_{1}^{2} m_{\mathrm L [3]}^2  y_{\tau}^{2} -6 m_{\mathrm E [3]}^2  y_{b}^{2} y_{\tau}^{2} -6 m_{\mathrm L [3]}^2  y_{b}^{2} y_{\tau}^{2} -10 m_{\mathrm E [3]}^2  y_{\tau}^{4} -8 m_{H_d}^2 y_{\tau}^{4} \nonumber \\ 
 &-2 m_{\mathrm L [3]}^2  y_{\tau}^{2} \lambda_{S}^{2} -3 m_{\mathrm E [3]}^2  y_{\tau}^{2} \lambda_{T}^{2} -3 m_{\mathrm L [3]}^2  y_{\tau}^{2} \lambda_{T}^{2}  -10 m_{\mathrm L [3]}^2  y_{\tau}^{4} -2 m_{\mathrm E [3]}^2  y_{\tau}^{2} \lambda_{S}^{2}\nonumber \\ 
 &-\frac{1}{5} y_{\tau}^{2} \Big(2 \Big(-6 g_{1}^{2} m_{H_d}^2 +15 m_{\mathrm D[3]}^2  y_{b}^{2} +30 m_{H_d}^2 y_{b}^{2} +15 m_{\mathrm Q [3]}^2  y_{b}^{2} +5 m_{\mathrm E [3]}^2  y_{\tau}^{2} +10 m_{H_d}^2 y_{\tau}^{2}\nonumber \\ 
 & +5 m_{\mathrm L [3]}^2  y_{\tau}^{2} +5 \Big(2 m_{H_d}^2  + m_{H_u}^2 + m_S^2\Big)\lambda_{S}^{2} \Big)+15 \Big(2 m_{H_d}^2  + m_{H_u}^2 + m_T^2\Big)\lambda_{T}^{2} \Big)\nonumber \\ 
 &+{\bf 1} \Big(6 g_{2}^{4} \sigma_{2,2}  + \frac{2}{5} g_1 \Big(-2 \sqrt{15} \sigma_{3,1}  + 3 g_1 \sigma_{2,1} \Big)\Big)\\ 
\beta_{m_{H_d}^2}^{(1)} = &  
+6 m_{\mathrm D[3]}^2  y_{b}^{2} +6 m_{H_d}^2 y_{b}^{2} +6 m_{\mathrm Q [3]}^2  y_{b}^{2} +2 m_{\mathrm E [3]}^2  y_{\tau}^{2} +2 m_{H_d}^2 y_{\tau}^{2} +2 m_{\mathrm L [3]}^2  y_{\tau}^{2}  \nonumber \\ 
 &+2 \Big(m_{H_d}^2 + m_{H_u}^2 + m_S^2\Big)\lambda_{S}^{2}+3 \Big(m_{H_d}^2 + m_{H_u}^2 + m_T^2\Big)\lambda_{T}^{2} - \sqrt{\frac{3}{5}} g_1 \sigma_{1,1} \\ 
\beta_{m_{H_d}^2}^{(2)} = &  
-\frac{4}{5} g_{1}^{2} m_{\mathrm D[3]}^2  y_{b}^{2} +32 g_{3}^{2} m_{\mathrm D[3]}^2  y_{b}^{2} -\frac{4}{5} g_{1}^{2} m_{H_d}^2 y_{b}^{2} +32 g_{3}^{2} m_{H_d}^2 y_{b}^{2} -\frac{4}{5} g_{1}^{2} m_{\mathrm Q [3]}^2  y_{b}^{2} \nonumber \\ 
 &-36 m_{\mathrm D[3]}^2  y_{b}^{4} -36 m_{H_d}^2 y_{b}^{4} -36 m_{\mathrm Q [3]}^2  y_{b}^{4} +\frac{12}{5} g_{1}^{2} m_{\mathrm E [3]}^2  y_{\tau}^{2} +\frac{12}{5} g_{1}^{2} m_{H_d}^2 y_{\tau}^{2} +\frac{12}{5} g_{1}^{2} m_{\mathrm L [3]}^2  y_{\tau}^{2} \nonumber \\ 
 &-12 m_{H_d}^2 y_{\tau}^{4} -12 m_{\mathrm L [3]}^2  y_{\tau}^{4} -6 m_{\mathrm D[3]}^2  y_{b}^{2} y_{t}^{2} -6 m_{H_d}^2 y_{b}^{2} y_{t}^{2} -6 m_{H_u}^2 y_{b}^{2} y_{t}^{2} -12 m_{\mathrm Q [3]}^2  y_{b}^{2} y_{t}^{2}  \nonumber \\ 
 &-4 \lambda_{SR}^{2} \Big(2 m_S^2  + m_{H_d}^2 + m_{H_u}^2 + m_{R_d}^2 + m_{R_u}^2\Big)\lambda_{S}^{2} -6 m_{H_d}^2 y_{t}^{2} \lambda_{S}^{2} -12 m_{H_u}^2 y_{t}^{2} \lambda_{S}^{2} \nonumber \\ 
 &-6 m_{\mathrm Q [3]}^2  y_{t}^{2} \lambda_{S}^{2} -6 m_S^2 y_{t}^{2} \lambda_{S}^{2} -6 m_{\mathrm U [3]}^2  y_{t}^{2} \lambda_{S}^{2} -12 m_{H_d}^2 \lambda_{S}^{4} -12 m_{H_u}^2 \lambda_{S}^{4} -12 m_S^2 \lambda_{S}^{4} \nonumber \\ 
 &+3 \Big(4 g_{2}^{2} \Big(m_{H_d}^2 + m_{H_u}^2 + m_T^2\Big)- \lambda_{TR}^{2} \Big(2 m_T^2  + m_{H_d}^2 + m_{H_u}^2 + m_{R_d}^2 + m_{R_u}^2\Big)\nonumber \\ 
 &-3 m_{\mathrm Q [3]}^2  y_{t}^{2} -3 m_T^2 y_{t}^{2} -3 m_{\mathrm U [3]}^2  y_{t}^{2} -4 m_{H_d}^2 \lambda_{S}^{2} -4 m_{H_u}^2 \lambda_{S}^{2} -2 m_S^2 \lambda_{S}^{2} -2 m_T^2 \lambda_{S}^{2} \Big)\lambda_{T}^{2} \nonumber \\ 
 &-15 \Big(m_{H_d}^2 + m_{H_u}^2 + m_T^2\Big)\lambda_{T}^{4} +6 g_{2}^{4} \sigma_{2,2} +\frac{6}{5} g_{1}^{2} \sigma_{2,1} -4 \sqrt{\frac{3}{5}} g_1 \sigma_{3,1} \nonumber \\ 
 &-2 m_{H_u}^2 \lambda_{S}^{2} \mbox{Tr}\Big({\lambda_{SE} \lambda_{SE}^{T} }\Big) -4 m_S^2 \lambda_{S}^{2} \mbox{Tr}\Big({\lambda_{SE} \lambda_{SE}^{T} }\Big) -2 \lambda_{S}^{2} \mbox{Tr}\Big({\lambda_{SE} m_{\hat{\bar{E}}}^2 \lambda_{SE}^{T} }\Big) \nonumber \\ 
 &-2 \lambda_{S}^{2} \mbox{Tr}\Big({\lambda_{SE} \lambda_{SE}^{T} m_{\hat{E}}^2 }\Big)  +32 g_{3}^{2} m_{\mathrm Q [3]}^2  y_{b}^{2}-12 m_{\mathrm E [3]}^2  y_{\tau}^{4} -6 m_{\mathrm U [3]}^2  y_{b}^{2} y_{t}^{2} \nonumber \\ 
 &-3 m_{H_d}^2 y_{t}^{2} -6 m_{H_u}^2 y_{t}^{2}-2 m_{H_d}^2 \lambda_{S}^{2} \mbox{Tr}\Big({\lambda_{SE} \lambda_{SE}^{T} }\Big)   \\
\beta_{m_{H_u}^2}^{(1)} = &  
2 \Big(m_{H_d}^2 + m_{H_u}^2 + m_S^2\Big)\lambda_{S}^{2}  + 3 \Big(m_{H_d}^2 + m_{H_u}^2 + m_T^2\Big)\lambda_{T}^{2}  + 6 m_{H_u}^2 y_{t}^{2} \nonumber \\ &  + 6 m_{\mathrm Q [3]}^2  y_{t}^{2}  + 6 m_{\mathrm U [3]}^2  y_{t}^{2}  + \sqrt{\frac{3}{5}} g_1 \sigma_{1,1} \\ 
\beta_{m_{H_u}^2}^{(2)} = &  
+\frac{8}{5} g_{1}^{2} m_{H_u}^2 y_{t}^{2} +32 g_{3}^{2} m_{H_u}^2 y_{t}^{2} +\frac{8}{5} g_{1}^{2} m_{\mathrm Q [3]}^2  y_{t}^{2} +32 g_{3}^{2} m_{\mathrm Q [3]}^2  y_{t}^{2} +\frac{8}{5} g_{1}^{2} m_{\mathrm U [3]}^2  y_{t}^{2} +32 g_{3}^{2} m_{\mathrm U [3]}^2  y_{t}^{2} \nonumber \\ 
 &-6 m_{\mathrm D[3]}^2  y_{b}^{2} y_{t}^{2} -6 m_{H_d}^2 y_{b}^{2} y_{t}^{2} -6 m_{H_u}^2 y_{b}^{2} y_{t}^{2} -12 m_{\mathrm Q [3]}^2  y_{b}^{2} y_{t}^{2} -6 m_{\mathrm U [3]}^2  y_{b}^{2} y_{t}^{2} -36 m_{H_u}^2 y_{t}^{4} \nonumber \\ 
 &-4 \lambda_{SR}^{2} \Big(2 m_S^2  + m_{H_d}^2 + m_{H_u}^2 + m_{R_d}^2 + m_{R_u}^2\Big)\lambda_{S}^{2} -6 m_{\mathrm D[3]}^2  y_{b}^{2} \lambda_{S}^{2} -12 m_{H_d}^2 y_{b}^{2} \lambda_{S}^{2} \nonumber \\ 
 &-6 m_{H_u}^2 y_{b}^{2} \lambda_{S}^{2} -6 m_{\mathrm Q [3]}^2  y_{b}^{2} \lambda_{S}^{2} -6 m_S^2 y_{b}^{2} \lambda_{S}^{2} -2 m_{\mathrm E [3]}^2  y_{\tau}^{2} \lambda_{S}^{2} -4 m_{H_d}^2 y_{\tau}^{2} \lambda_{S}^{2} \nonumber \\ 
 &-2 m_{H_u}^2 y_{\tau}^{2} \lambda_{S}^{2} -2 m_{\mathrm L [3]}^2  y_{\tau}^{2} \lambda_{S}^{2} -2 m_S^2 y_{\tau}^{2} \lambda_{S}^{2} -12 m_{H_d}^2 \lambda_{S}^{4} -12 m_{H_u}^2 \lambda_{S}^{4} -12 m_S^2 \lambda_{S}^{4} \nonumber \\ 
 &+3 \Big(4 g_{2}^{2} m_{H_d}^2 +4 g_{2}^{2} m_{H_u}^2 +4 g_{2}^{2} m_T^2 - \lambda_{TR}^{2} \Big(2 m_T^2  + m_{H_d}^2 + m_{H_u}^2 + m_{R_d}^2 + m_{R_u}^2\Big)-3 m_{\mathrm D[3]}^2  y_{b}^{2} \nonumber \\ 
 &-6 m_{H_d}^2 y_{b}^{2} -3 m_{H_u}^2 y_{b}^{2} -3 m_{\mathrm Q [3]}^2  y_{b}^{2} -3 m_T^2 y_{b}^{2} - m_{\mathrm E [3]}^2  y_{\tau}^{2} -2 m_{H_d}^2 y_{\tau}^{2} - m_{H_u}^2 y_{\tau}^{2} - m_{\mathrm L [3]}^2  y_{\tau}^{2} \nonumber \\ 
 &- m_T^2 y_{\tau}^{2} -4 m_{H_d}^2 \lambda_{S}^{2} -4 m_{H_u}^2 \lambda_{S}^{2} -2 m_S^2 \lambda_{S}^{2} -2 m_T^2 \lambda_{S}^{2} \Big)\lambda_{T}^{2} \nonumber \\ 
 &-15 \Big(m_{H_d}^2 + m_{H_u}^2 + m_T^2\Big)\lambda_{T}^{4} +6 g_{2}^{4} \sigma_{2,2} +\frac{6}{5} g_{1}^{2} \sigma_{2,1} +4 \sqrt{\frac{3}{5}} g_1 \sigma_{3,1} -2 m_{H_d}^2 \lambda_{S}^{2} \mbox{Tr}\Big({\lambda_{SE} \lambda_{SE}^{T} }\Big) \nonumber \\ 
 &-2 m_{H_u}^2 \lambda_{S}^{2} \mbox{Tr}\Big({\lambda_{SE} \lambda_{SE}^{T} }\Big) -4 m_S^2 \lambda_{S}^{2} \mbox{Tr}\Big({\lambda_{SE} \lambda_{SE}^{T} }\Big) -2 \lambda_{S}^{2} \mbox{Tr}\Big({\lambda_{SE} m_{\hat{\bar{E}}}^2 \lambda_{SE}^{T} }\Big) \nonumber \\ 
 &-2 \lambda_{S}^{2} \mbox{Tr}\Big({\lambda_{SE} \lambda_{SE}^{T} m_{\hat{E}}^2 }\Big)  -36 m_{\mathrm Q [3]}^2  y_{t}^{4} -36 m_{\mathrm U [3]}^2  y_{t}^{4} \\ 
\beta_{m_{\mathrm D [1,2]}^2 }^{(1)} = &  
2 \frac{1}{\sqrt{15}} g_1 {\bf 1} \sigma_{1,1} \\ 
\beta_{m_{\mathrm D [1,2]}^2 }^{(2)} = &  
\frac{8}{15} {\bf 1} \Big(20 g_{3}^{4} \sigma_{2,3}  + g_1 \Big(g_1 \sigma_{2,1}  + \sqrt{15} \sigma_{3,1} \Big)\Big)\\ 
\beta_{m_{\mathrm D[3]}^2 }^{(1)} = &  
2 \frac{1}{\sqrt{15}} g_1 {\bf 1} \sigma_{1,1}  + 4 m_{H_d}^2 y_{b}^{2}  + 4 m_{\mathrm D[3]}^2  y_{b}^{2}  + 4 m_{\mathrm Q [3]}^2  y_{b}^{2} \\ 
\beta_{m_{\mathrm D[3]}^2 }^{(2)} = &  
\frac{2}{15} \Big(90 g_{2}^{2} \Big(m_{\mathrm D[3]}^2  + m_{H_d}^2 + m_{\mathrm Q [3]}^2 \Big)y_{b}^{2} \nonumber \\ 
 &-5 \Big(48 m_{\mathrm Q [3]}^2  y_{b}^{4} +6 m_{\mathrm E [3]}^2  y_{b}^{2} y_{\tau}^{2} +6 m_{\mathrm L [3]}^2  y_{b}^{2} y_{\tau}^{2}  +6 m_{\mathrm Q [3]}^2  y_{b}^{2} y_{\tau}^{2} +6 m_{H_u}^2 y_{b}^{2} y_{t}^{2} \nonumber \\
 & +12 m_{\mathrm Q [3]}^2  y_{b}^{2} y_{t}^{2} +6 m_{\mathrm U [3]}^2  y_{b}^{2} y_{t}^{2} +6 m_{H_u}^2 y_{b}^{2} \lambda_{S}^{2} \nonumber \\ 
 &+6 m_{\mathrm Q [3]}^2  y_{b}^{2} \lambda_{S}^{2} +6 m_S^2 y_{b}^{2} \lambda_{S}^{2} +9 m_{H_u}^2 y_{b}^{2} \lambda_{T}^{2} +9 m_{\mathrm Q [3]}^2  y_{b}^{2} \lambda_{T}^{2} +9 m_T^2 y_{b}^{2} \lambda_{T}^{2} \nonumber \\ 
 &+6 m_{H_d}^2 y_{b}^{2} \Big(2 \lambda_{S}^{2}  + 2 y_{\tau}^{2}  + 3 \lambda_{T}^{2}  + 8 y_{b}^{2}  + y_{t}^{2}\Big)+3 m_{\mathrm D[3]}^2  y_{b}^{2} \Big(16 y_{b}^{2}  + 2 \lambda_{S}^{2}  + 2 y_{t}^{2}  + 2 y_{\tau}^{2}  + 3 \lambda_{T}^{2} \Big)\nonumber \\ 
 &-16 g_{3}^{4} {\bf 1} \sigma_{2,3} \Big)+2 g_{1}^{2} \Big(2 {\bf 1} \sigma_{2,1}  + 3 m_{H_d}^2 y_{b}^{2}  + 3 m_{\mathrm D[3]}^2  y_{b}^{2}  + 3 m_{\mathrm Q [3]}^2  y_{b}^{2} \Big)+4 \sqrt{15} g_1 {\bf 1} \sigma_{3,1} \Big)\\ 
\beta_{m_{\mathrm U[1,2]}^2 }^{(1)} = &  
-4 \frac{1}{\sqrt{15}} g_1 {\bf 1} \sigma_{1,1} \\ 
\beta_{m_{\mathrm U[1,2]}^2 }^{(2)} = &  
\frac{16}{15} {\bf 1} \Big(10 g_{3}^{4} \sigma_{2,3}  + g_1 \Big(2 g_1 \sigma_{2,1}  - \sqrt{15} \sigma_{3,1} \Big)\Big)\\ 
\beta_{m_{\mathrm U [3]}^2 }^{(1)} = &  
-4 \frac{1}{\sqrt{15}} g_1 {\bf 1} \sigma_{1,1}  + 4 m_{H_u}^2 y_{t}^{2}  + 4 m_{\mathrm Q [3]}^2  y_{t}^{2}  + 4 m_{\mathrm U [3]}^2  y_{t}^{2} \\ 
\beta_{m_{\mathrm U [3]}^2 }^{(2)} = &  
-\frac{2}{15} \Big(5 \Big(-18 g_{2}^{2} \Big(m_{H_u}^2 + m_{\mathrm Q [3]}^2  + m_{\mathrm U [3]}^2 \Big)y_{t}^{2} +6 m_{\mathrm D[3]}^2  y_{b}^{2} y_{t}^{2} +6 m_{H_d}^2 y_{b}^{2} y_{t}^{2} +6 m_{H_u}^2 y_{b}^{2} y_{t}^{2} \nonumber \\
& +12 m_{\mathrm Q [3]}^2  y_{b}^{2} y_{t}^{2} +6 m_{\mathrm U [3]}^2  y_{b}^{2} y_{t}^{2} +48 m_{H_u}^2 y_{t}^{4}+6 m_{\mathrm U [3]}^2  y_{t}^{2} \lambda_{S}^{2}  \nonumber \\ 
 &+48 m_{\mathrm Q [3]}^2  y_{t}^{4} +48 m_{\mathrm U [3]}^2  y_{t}^{4} +6 m_{H_d}^2 y_{t}^{2} \lambda_{S}^{2} +12 m_{H_u}^2 y_{t}^{2} \lambda_{S}^{2} +6 m_{\mathrm Q [3]}^2  y_{t}^{2} \lambda_{S}^{2} +6 m_S^2 y_{t}^{2} \lambda_{S}^{2} \nonumber \\ 
 &+9 m_{H_d}^2 y_{t}^{2} \lambda_{T}^{2} +18 m_{H_u}^2 y_{t}^{2} \lambda_{T}^{2} +9 m_{\mathrm Q [3]}^2  y_{t}^{2} \lambda_{T}^{2} +9 m_T^2 y_{t}^{2} \lambda_{T}^{2} +9 m_{\mathrm U [3]}^2  y_{t}^{2} \lambda_{T}^{2} -16 g_{3}^{4} {\bf 1} \sigma_{2,3} \Big)\nonumber \\ 
 &+2 g_{1}^{2} \Big(3 m_{H_u}^2 y_{t}^{2}  + 3 m_{\mathrm Q [3]}^2  y_{t}^{2}  + 3 m_{\mathrm U [3]}^2  y_{t}^{2}  -8 {\bf 1} \sigma_{2,1} \Big)+8 \sqrt{15} g_1 {\bf 1} \sigma_{3,1} \Big)\\ 
\beta_{m_{\mathrm E [1,2]}^2 }^{(1)} = &  
2 \sqrt{\frac{3}{5}} g_1 {\bf 1} \sigma_{1,1} \\ 
\beta_{m_{\mathrm E [1,2]}^2 }^{(2)} = &  
\frac{8}{5} g_1 {\bf 1} \Big(3 g_1 \sigma_{2,1}  + \sqrt{15} \sigma_{3,1} \Big)\\ 
\beta_{m_{\mathrm E [3]}^2 }^{(1)} = &  
2 \sqrt{\frac{3}{5}} g_1 {\bf 1} \sigma_{1,1}  + 4 m_{H_d}^2 y_{\tau}^{2}  + 4 m_{\mathrm E [3]}^2  y_{\tau}^{2}  + 4 m_{\mathrm L [3]}^2  y_{\tau}^{2} \\ 
\beta_{m_{\mathrm E [3]}^2 }^{(2)} = &  
-\frac{2}{5} \Big(5 y_{\tau}^{2} \Big(-6 g_{2}^{2} \Big(m_{\mathrm E [3]}^2  + m_{H_d}^2 + m_{\mathrm L [3]}^2 \Big)+6 m_{\mathrm D[3]}^2  y_{b}^{2} +6 m_{\mathrm E [3]}^2  y_{b}^{2} +12 m_{H_d}^2 y_{b}^{2} \nonumber \\
& +6 m_{\mathrm L [3]}^2  y_{b}^{2} +6 m_{\mathrm Q [3]}^2  y_{b}^{2} +8 m_{\mathrm E [3]}^2  y_{\tau}^{2} +8 m_{H_d}^2 y_{\tau}^{2} +8 m_{\mathrm L [3]}^2  y_{\tau}^{2} +2 m_{\mathrm E [3]}^2  \lambda_{S}^{2} \nonumber \\ 
 &+4 m_{H_d}^2 \lambda_{S}^{2} +2 m_{H_u}^2 \lambda_{S}^{2} +2 m_{\mathrm L [3]}^2  \lambda_{S}^{2} +2 m_S^2 \lambda_{S}^{2} +3 m_{\mathrm E [3]}^2  \lambda_{T}^{2} +6 m_{H_d}^2 \lambda_{T}^{2} \nonumber \\
 & +3 m_{H_u}^2 \lambda_{T}^{2} +3 m_{\mathrm L [3]}^2  \lambda_{T}^{2} +3 m_T^2 \lambda_{T}^{2} \Big)\nonumber \\ 
 &+6 g_{1}^{2} \Big(-2 {\bf 1} \sigma_{2,1}  + m_{H_d}^2 y_{\tau}^{2}  + m_{\mathrm E [3]}^2  y_{\tau}^{2}  + m_{\mathrm L [3]}^2  y_{\tau}^{2} \Big)-4 \sqrt{15} g_1 {\bf 1} \sigma_{3,1} \Big)\\ 
\beta_{m_S^2}^{(1)} = &  
2 \Big(2 \lambda_{SR}^{2} \Big(m_{R_d}^2 + m_{R_u}^2 + m_S^2\Big)+2 \Big(m_{H_d}^2 + m_{H_u}^2 + m_S^2\Big)\lambda_{S}^{2} +m_S^2 \mbox{Tr}\Big({\lambda_{SE} \lambda_{SE}^{T} }\Big) \nonumber \\ 
 &+\mbox{Tr}\Big({\lambda_{SE} m_{\hat{\bar{E}}}^2 \lambda_{SE}^{T} }\Big)+\mbox{Tr}\Big({\lambda_{SE} \lambda_{SE}^{T} m_{\hat{E}}^2 }\Big)\Big)\\ 
\beta_{m_S^2}^{(2)} = &  
-16 \lambda_{SR}^{4} \Big(m_{R_d}^2 + m_{R_u}^2 + m_S^2\Big)\nonumber \\ 
 &+\frac{12}{5} \lambda_{SR}^{2} \Big(5 \Big(g_{2}^{2} \Big(m_{R_d}^2 + m_{R_u}^2 + m_S^2\Big) - \lambda_{TR}^{2} \Big(2 m_{R_d}^2  + 2 m_{R_u}^2  + m_S^2 + m_T^2\Big)\Big) \nonumber \\ 
 & + g_{1}^{2} \Big(m_{R_d}^2 + m_{R_u}^2 + m_S^2\Big)\Big)-\frac{4}{5} \Big(20 \Big(m_{H_d}^2 + m_{H_u}^2 + m_S^2\Big)\lambda_{S}^{4} \nonumber \\ 
 &+\lambda_{S}^{2} \Big(-3 g_{1}^{2} \Big(m_{H_d}^2 + m_{H_u}^2 + m_S^2\Big)\nonumber \\ 
 &+5 \Big(-3 g_{2}^{2} \Big(m_{H_d}^2 + m_{H_u}^2 + m_S^2\Big)+3 m_{\mathrm D[3]}^2  y_{b}^{2} +6 m_{H_d}^2 y_{b}^{2} +3 m_{H_u}^2 y_{b}^{2} \nonumber \\
 & +3 m_{\mathrm Q [3]}^2  y_{b}^{2} +3 m_S^2 y_{b}^{2} +m_{\mathrm E [3]}^2  y_{\tau}^{2} +2 m_{H_d}^2 y_{\tau}^{2} \nonumber \\ 
 &+m_{H_u}^2 y_{\tau}^{2} +m_{\mathrm L [3]}^2  y_{\tau}^{2} +m_S^2 y_{\tau}^{2} +3 m_{H_d}^2 y_{t}^{2} +6 m_{H_u}^2 y_{t}^{2} +3 m_{\mathrm Q [3]}^2  y_{t}^{2} +3 m_S^2 y_{t}^{2} \nonumber \\ 
 & +3 m_{\mathrm U [3]}^2  y_{t}^{2} +6 m_{H_d}^2 \lambda_{T}^{2} +6 m_{H_u}^2 \lambda_{T}^{2} +3 m_S^2 \lambda_{T}^{2} +3 m_T^2 \lambda_{T}^{2} \Big)\Big)\nonumber \\ 
 &-6 g_{1}^{2} m_S^2 \mbox{Tr}\Big({\lambda_{SE} \lambda_{SE}^{T} }\Big) -6 g_{1}^{2} \mbox{Tr}\Big({\lambda_{SE} m_{\hat{\bar{E}}}^2 \lambda_{SE}^{T} }\Big) -6 g_{1}^{2} \mbox{Tr}\Big({\lambda_{SE} \lambda_{SE}^{T} m_{\hat{E}}^2 }\Big) \nonumber \\ 
 &+10 m_S^2 \mbox{Tr}\Big({\lambda_{SE} \lambda_{SE}^{T} \lambda_{SE} \lambda_{SE}^{T} }\Big) +5 \mbox{Tr}\Big({\lambda_{SE} m_{\hat{\bar{E}}}^2 \lambda_{SE}^{T} \lambda_{SE} \lambda_{SE}^{T} }\Big) +5 \mbox{Tr}\Big({\lambda_{SE} \lambda_{SE}^{T} \lambda_{SE} m_{\hat{\bar{E}}}^2 \lambda_{SE}^{T} }\Big) \nonumber \\ 
 & +5 \mbox{Tr}\Big({\lambda_{SE} \lambda_{SE}^{T} \lambda_{SE} \lambda_{SE}^{T} m_{\hat{E}}^2 }\Big)+5 \mbox{Tr}\Big({\lambda_{SE} \lambda_{SE}^{T} m_{\hat{E}}^2 \lambda_{SE} \lambda_{SE}^{T} }\Big) \Big)\\ 
\beta_{m_T^2}^{(1)} = &  
2 \Big(\lambda_{TR}^{2} \Big(m_{R_d}^2 + m_{R_u}^2 + m_T^2\Big) + \Big(m_{H_d}^2 + m_{H_u}^2 + m_T^2\Big)\lambda_{T}^{2} \Big)\\ 
\beta_{m_T^2}^{(2)} = &  
\frac{2}{5} \Big(-30 \lambda_{TR}^{4} \Big(m_{R_d}^2 + m_{R_u}^2 + m_T^2\Big)+\lambda_{TR}^{2} \Big(3 g_{1}^{2} \Big(m_{R_d}^2 + m_{R_u}^2 + m_T^2\Big) \nonumber \\ 
 & -5 \Big(2 \lambda_{SR}^{2} \Big(2 m_{R_d}^2  + 2 m_{R_u}^2  + m_S^2 + m_T^2\Big) + g_{2}^{2} \Big(m_{R_d}^2 + m_{R_u}^2 + m_T^2\Big)\Big)\Big)\nonumber \\ 
 &+3 g_{1}^{2} \Big(m_{H_d}^2 + m_{H_u}^2 + m_T^2\Big)\lambda_{T}^{2} +5 \Big(- g_{2}^{2} \Big(m_{H_d}^2 + m_{H_u}^2 + m_T^2\Big)\lambda_{T}^{2} \nonumber \\ 
 &- \lambda_{T}^{2} \Big(3 m_{\mathrm D[3]}^2  y_{b}^{2} +3 m_{H_u}^2 y_{b}^{2} +3 m_{\mathrm Q [3]}^2  y_{b}^{2} +3 m_T^2 y_{b}^{2} +m_{\mathrm E [3]}^2  y_{\tau}^{2} +m_{H_u}^2 y_{\tau}^{2} \nonumber \\ 
 & +m_{\mathrm L [3]}^2  y_{\tau}^{2} +m_T^2 y_{\tau}^{2} +6 m_{H_u}^2 y_{t}^{2} +3 m_{\mathrm Q [3]}^2  y_{t}^{2} +3 m_T^2 y_{t}^{2} \nonumber \\ 
 &+3 m_{\mathrm U [3]}^2  y_{t}^{2} +4 m_{H_u}^2 \lambda_{S}^{2} +2 m_S^2 \lambda_{S}^{2} +2 m_T^2 \lambda_{S}^{2} +6 m_{H_u}^2 \lambda_{T}^{2} +6 m_T^2 \lambda_{T}^{2} \nonumber \\ 
 &+m_{H_d}^2 \Big(2 y_{\tau}^{2}  + 3 y_{t}^{2}  + 4 \lambda_{S}^{2}  + 6 \lambda_{T}^{2}  + 6 y_{b}^{2} \Big)\Big)+8 g_{2}^{4} \sigma_{2,2} \Big)\Big)\\ 
\beta_{m_O^2}^{(1)} = &  
0\\ 
\beta_{m_O^2}^{(2)} = &  
24 g_{3}^{4} \sigma_{2,3} \\ 
\beta_{m_{R_u}^2}^{(1)} = &  
2 \lambda_{SR}^{2} \Big(m_{R_d}^2 + m_{R_u}^2 + m_S^2\Big) + 3 \lambda_{TR}^{2} \Big(m_{R_d}^2 + m_{R_u}^2 + m_T^2\Big) - \sqrt{\frac{3}{5}} g_1 \sigma_{1,1} \\ 
\beta_{m_{R_u}^2}^{(2)} = &  
-12 \lambda_{SR}^{4} \Big(m_{R_d}^2 + m_{R_u}^2 + m_S^2\Big)-15 \lambda_{TR}^{4} \Big(m_{R_d}^2 + m_{R_u}^2 + m_T^2\Big)\nonumber \\ 
 &+3 \lambda_{TR}^{2} \Big(- \Big(2 m_T^2  + m_{H_d}^2 + m_{H_u}^2 + m_{R_d}^2 + m_{R_u}^2\Big)\lambda_{T}^{2}  + 4 g_{2}^{2} \Big(m_{R_d}^2 + m_{R_u}^2 + m_T^2\Big)\Big)\nonumber \\ 
 &+6 g_{2}^{4} \sigma_{2,2} +\frac{6}{5} g_{1}^{2} \sigma_{2,1} -4 \sqrt{\frac{3}{5}} g_1 \sigma_{3,1} -2 \lambda_{SR}^{2} \Big(3 \lambda_{TR}^{2} \Big(2 m_{R_d}^2 \nonumber \\ 
 & + 2 m_{R_u}^2  + m_S^2 + m_T^2\Big)+2 \Big(2 m_S^2  + m_{H_d}^2 + m_{H_u}^2 + m_{R_d}^2 + m_{R_u}^2\Big)\lambda_{S}^{2} +m_{R_d}^2 \mbox{Tr}\Big({\lambda_{SE} \lambda_{SE}^{T} }\Big) \nonumber \\ 
 &+m_{R_u}^2 \mbox{Tr}\Big({\lambda_{SE} \lambda_{SE}^{T} }\Big) +2 m_S^2 \mbox{Tr}\Big({\lambda_{SE} \lambda_{SE}^{T} }\Big) +\mbox{Tr}\Big({\lambda_{SE} m_{\hat{\bar{E}}}^2 \lambda_{SE}^{T} }\Big)+\mbox{Tr}\Big({\lambda_{SE} \lambda_{SE}^{T} m_{\hat{E}}^2 }\Big)\Big)\\ 
\beta_{m_{R_d}^2}^{(1)} = &  
2 \lambda_{SR}^{2} \Big(m_{R_d}^2 + m_{R_u}^2 + m_S^2\Big) + 3 \lambda_{TR}^{2} \Big(m_{R_d}^2 + m_{R_u}^2 + m_T^2\Big) + \sqrt{\frac{3}{5}} g_1 \sigma_{1,1} \\ 
\beta_{m_{R_d}^2}^{(2)} = &  
-12 \lambda_{SR}^{4} \Big(m_{R_d}^2 + m_{R_u}^2 + m_S^2\Big)-15 \lambda_{TR}^{4} \Big(m_{R_d}^2 + m_{R_u}^2 + m_T^2\Big)\nonumber \\ 
 &+3 \lambda_{TR}^{2} \Big(- \Big(2 m_T^2  + m_{H_d}^2 + m_{H_u}^2 + m_{R_d}^2 + m_{R_u}^2\Big)\lambda_{T}^{2}  + 4 g_{2}^{2} \Big(m_{R_d}^2 + m_{R_u}^2 + m_T^2\Big)\Big)\nonumber \\ 
 &+6 g_{2}^{4} \sigma_{2,2} +\frac{6}{5} g_{1}^{2} \sigma_{2,1} +4 \sqrt{\frac{3}{5}} g_1 \sigma_{3,1} -2 \lambda_{SR}^{2} \Big(3 \lambda_{TR}^{2} \Big(2 m_{R_d}^2  \nonumber \\ 
 &+ 2 m_{R_u}^2  + m_S^2 + m_T^2\Big)+2 \Big(2 m_S^2  + m_{H_d}^2 + m_{H_u}^2 + m_{R_d}^2 + m_{R_u}^2\Big)\lambda_{S}^{2} +m_{R_d}^2 \mbox{Tr}\Big({\lambda_{SE} \lambda_{SE}^{T} }\Big) \nonumber \\ 
 &+m_{R_u}^2 \mbox{Tr}\Big({\lambda_{SE} \lambda_{SE}^{T} }\Big) +2 m_S^2 \mbox{Tr}\Big({\lambda_{SE} \lambda_{SE}^{T} }\Big) +\mbox{Tr}\Big({\lambda_{SE} m_{\hat{\bar{E}}}^2 \lambda_{SE}^{T} }\Big)+\mbox{Tr}\Big({\lambda_{SE} \lambda_{SE}^{T} m_{\hat{E}}^2 }\Big)\Big)\\ 
\beta_{m_{\hat{E}}^2}^{(1)} = &  
2 \lambda_{SE} m_{\hat{\bar{E}}}^2 \lambda_{SE}^{T}   + 2 m_S^2 \lambda_{SE} \lambda_{SE}^{T}   + 2 \sqrt{\frac{3}{5}} g_1 {\bf 1} \sigma_{1,1}  + \lambda_{SE} \lambda_{SE}^{T} m_{\hat{E}}^2  + m_{\hat{E}}^2 \lambda_{SE} \lambda_{SE}^{T} \\ 
\beta_{m_{\hat{E}}^2}^{(2)} = &  
-2 \lambda_{SR}^{2} \lambda_{SE} \lambda_{SE}^{T} m_{\hat{E}}^2  -2 \lambda_{S}^{2} \lambda_{SE} \lambda_{SE}^{T} m_{\hat{E}}^2  -2 \lambda_{SR}^{2} m_{\hat{E}}^2 \lambda_{SE} \lambda_{SE}^{T}  -2 \lambda_{S}^{2} m_{\hat{E}}^2 \lambda_{SE} \lambda_{SE}^{T}  \nonumber \\ 
 &-4 m_S^2 \lambda_{SE} \lambda_{SE}^{T} \lambda_{SE} \lambda_{SE}^{T}  -2 \lambda_{SE} m_{\hat{\bar{E}}}^2 \lambda_{SE}^{T} \lambda_{SE} \lambda_{SE}^{T}  -2 \lambda_{SE} \lambda_{SE}^{T} \lambda_{SE} m_{\hat{\bar{E}}}^2 \lambda_{SE}^{T}  \nonumber \\ 
 &- \lambda_{SE} \lambda_{SE}^{T} \lambda_{SE} \lambda_{SE}^{T} m_{\hat{E}}^2  -2 \lambda_{SE} \lambda_{SE}^{T} m_{\hat{E}}^2 \lambda_{SE} \lambda_{SE}^{T}  \nonumber \\ 
 &- m_{\hat{E}}^2 \lambda_{SE} \lambda_{SE}^{T} \lambda_{SE} \lambda_{SE}^{T}  +\frac{24}{5} g_{1}^{2} {\bf 1} \sigma_{2,1} +8 \sqrt{\frac{3}{5}} g_1 {\bf 1} \sigma_{3,1} - \lambda_{SE} \lambda_{SE}^{T} m_{\hat{E}}^2  \mbox{Tr}\Big({\lambda_{SE} \lambda_{SE}^{T} }\Big) \nonumber \\ 
 &- m_{\hat{E}}^2 \lambda_{SE} \lambda_{SE}^{T}  \mbox{Tr}\Big({\lambda_{SE} \lambda_{SE}^{T} }\Big) -2 \lambda_{SE} m_{\hat{\bar{E}}}^2 \lambda_{SE}^{T}  \Big(2 \Big(\lambda_{SR}^{2} + \lambda_{S}^{2}\Big) + \mbox{Tr}\Big({\lambda_{SE} \lambda_{SE}^{T} }\Big)\Big)\nonumber \\ 
 &-2 \lambda_{SE} \lambda_{SE}^{T}  \Big(2 \lambda_{SR}^{2} m_{R_d}^2 +2 \lambda_{SR}^{2} m_{R_u}^2 +4 \lambda_{SR}^{2} m_S^2 +2 m_{H_d}^2 \lambda_{S}^{2} +2 m_{H_u}^2 \lambda_{S}^{2} +4 m_S^2 \lambda_{S}^{2} \nonumber \\ 
 &+2 m_S^2 \mbox{Tr}\Big({\lambda_{SE} \lambda_{SE}^{T} }\Big) +\mbox{Tr}\Big({\lambda_{SE} m_{\hat{\bar{E}}}^2 \lambda_{SE}^{T} }\Big)+\mbox{Tr}\Big({\lambda_{SE} \lambda_{SE}^{T} m_{\hat{E}}^2 }\Big)\Big)\\ 
\beta_{m_{\hat{\bar{E}}}^2}^{(1)} = &  
2 \lambda_{SE}^{T} m_{\hat{E}}^2 \lambda_{SE}   + 2 m_S^2 \lambda_{SE}^{T} \lambda_{SE}   -2 \sqrt{\frac{3}{5}} g_1 {\bf 1} \sigma_{1,1}  + \lambda_{SE}^{T} \lambda_{SE} m_{\hat{\bar{E}}}^2  + m_{\hat{\bar{E}}}^2 \lambda_{SE}^{T} \lambda_{SE} \\ 
\beta_{m_{\hat{\bar{E}}}^2}^{(2)} = &  
-2 \lambda_{SR}^{2} \lambda_{SE}^{T} \lambda_{SE} m_{\hat{\bar{E}}}^2  -2 \lambda_{S}^{2} \lambda_{SE}^{T} \lambda_{SE} m_{\hat{\bar{E}}}^2  -4 \lambda_{SR}^{2} \lambda_{SE}^{T} m_{\hat{E}}^2 \lambda_{SE}  -4 \lambda_{S}^{2} \lambda_{SE}^{T} m_{\hat{E}}^2 \lambda_{SE}  \nonumber \\ 
 &-4 m_S^2 \lambda_{SE}^{T} \lambda_{SE} \lambda_{SE}^{T} \lambda_{SE}  - m_{\hat{\bar{E}}}^2 \lambda_{SE}^{T} \lambda_{SE} \lambda_{SE}^{T} \lambda_{SE}  -2 \lambda_{SE}^{T} \lambda_{SE} m_{\hat{\bar{E}}}^2 \lambda_{SE}^{T} \lambda_{SE} \nonumber \\ 
 & - \lambda_{SE}^{T} \lambda_{SE} \lambda_{SE}^{T} \lambda_{SE} m_{\hat{\bar{E}}}^2  -2 \lambda_{SE}^{T} \lambda_{SE} \lambda_{SE}^{T} m_{\hat{E}}^2 \lambda_{SE}  \nonumber \\ 
 &-2 \lambda_{SE}^{T} m_{\hat{E}}^2 \lambda_{SE} \lambda_{SE}^{T} \lambda_{SE}  +\frac{24}{5} g_{1}^{2} {\bf 1} \sigma_{2,1} -8 \sqrt{\frac{3}{5}} g_1 {\bf 1} \sigma_{3,1} - \lambda_{SE}^{T} \lambda_{SE} m_{\hat{\bar{E}}}^2  \mbox{Tr}\Big({\lambda_{SE} \lambda_{SE}^{T} }\Big)\nonumber \\ 
 & -2 \lambda_{SE}^{T} m_{\hat{E}}^2 \lambda_{SE}  \mbox{Tr}\Big({\lambda_{SE} \lambda_{SE}^{T} }\Big) - m_{\hat{\bar{E}}}^2 \lambda_{SE}^{T} \lambda_{SE}  \Big(2 \Big(\lambda_{SR}^{2} + \lambda_{S}^{2}\Big) + \mbox{Tr}\Big({\lambda_{SE} \lambda_{SE}^{T} }\Big)\Big)\nonumber \\ 
 &-2 \lambda_{SE}^{T} \lambda_{SE}  \Big(2 \lambda_{SR}^{2} m_{R_d}^2 +2 \lambda_{SR}^{2} m_{R_u}^2 +4 \lambda_{SR}^{2} m_S^2 +2 m_{H_d}^2 \lambda_{S}^{2} +2 m_{H_u}^2 \lambda_{S}^{2} +4 m_S^2 \lambda_{S}^{2} \nonumber \\ 
 &+2 m_S^2 \mbox{Tr}\Big({\lambda_{SE} \lambda_{SE}^{T} }\Big) +\mbox{Tr}\Big({\lambda_{SE} m_{\hat{\bar{E}}}^2 \lambda_{SE}^{T} }\Big)+\mbox{Tr}\Big({\lambda_{SE} \lambda_{SE}^{T} m_{\hat{E}}^2 }\Big)\Big)
\end{align}}

%% file: BGSP.bbl
\begin{thebibliography}{10}

\bibitem{fayet}
P.~Fayet,
\newblock Phys.Lett. {\bf B78}, 417 (1978).

\bibitem{Polchinski:1982an}
J.~Polchinski and L.~Susskind,
\newblock Phys.Rev. {\bf D26}, 3661 (1982).

\bibitem{Hall:1990hq}
L.~Hall and L.~Randall,
\newblock Nucl.Phys. {\bf B352}, 289 (1991).

\bibitem{fnw}
P.~J. Fox, A.~E. Nelson, and N.~Weiner,
\newblock JHEP {\bf 0208}, 035 (2002), arXiv:hep-ph/0206096.

\bibitem{Nelson:2002ca}
A.~E. Nelson, N.~Rius, V.~Sanz, and M.~Unsal,
\newblock JHEP {\bf 0208}, 039 (2002), arXiv:hep-ph/0206102.

\bibitem{Antoniadis:2005em}
I.~Antoniadis, A.~Delgado, K.~Benakli, M.~Quiros, and M.~Tuckmantel,
\newblock Phys.Lett. {\bf B634}, 302 (2006), arXiv:hep-ph/0507192.

\bibitem{Antoniadis:2006uj}
I.~Antoniadis, K.~Benakli, A.~Delgado, and M.~Quiros,
\newblock Adv.Stud.Theor.Phys. {\bf 2}, 645 (2008), arXiv:hep-ph/0610265.

\bibitem{kpw}
G.~D. Kribs, E.~Poppitz, and N.~Weiner,
\newblock Phys.Rev. {\bf D78}, 055010 (2008), arXiv:0712.2039.

\bibitem{Amigo:2008rc}
S.~D.~L. Amigo, A.~E. Blechman, P.~J. Fox, and E.~Poppitz,
\newblock JHEP {\bf 0901}, 018 (2009), arXiv:0809.1112.

\bibitem{Plehn:2008ae}
T.~Plehn and T.~M. Tait,
\newblock J.Phys. {\bf G36}, 075001 (2009), arXiv:0810.3919.

\bibitem{Benakli:2008pg}
K.~Benakli and M.~Goodsell,
\newblock Nucl.Phys. {\bf B816}, 185 (2009), arXiv:0811.4409.

\bibitem{Belanger:2009wf}
G.~Belanger, K.~Benakli, M.~Goodsell, C.~Moura, and A.~Pukhov,
\newblock JCAP {\bf 0908}, 027 (2009), arXiv:0905.1043.

\bibitem{Benakli:2009mk}
K.~Benakli and M.~Goodsell,
\newblock Nucl.Phys. {\bf B830}, 315 (2010), arXiv:0909.0017.

\bibitem{Choi:2009ue}
S.~Choi, J.~Kalinowski, J.~Kim, and E.~Popenda,
\newblock Acta Phys.Polon. {\bf B40}, 2913 (2009), arXiv:0911.1951.

\bibitem{Benakli:2010gi}
K.~Benakli and M.~Goodsell,
\newblock Nucl.Phys. {\bf B840}, 1 (2010), arXiv:1003.4957.

\bibitem{Choi:2010gc}
S.~Choi {\em et~al.},
\newblock JHEP {\bf 1008}, 025 (2010), arXiv:1005.0818.

\bibitem{Carpenter:2010as}
L.~M. Carpenter,
\newblock JHEP {\bf 1209}, 102 (2012), arXiv:1007.0017.

\bibitem{Kribs:2010md}
G.~D. Kribs, T.~Okui, and T.~S. Roy,
\newblock Phys.Rev. {\bf D82}, 115010 (2010), arXiv:1008.1798.

\bibitem{Abel:2011dc}
S.~Abel and M.~Goodsell,
\newblock JHEP {\bf 1106}, 064 (2011), arXiv:1102.0014.

\bibitem{Davies:2011mp}
R.~Davies, J.~March-Russell, and M.~McCullough,
\newblock JHEP {\bf 1104}, 108 (2011), arXiv:1103.1647.

\bibitem{Benakli:2011vb}
K.~Benakli,
\newblock Fortsch.Phys. {\bf 59}, 1079 (2011), arXiv:1106.1649.

\bibitem{Benakli:2011kz}
K.~Benakli, M.~D. Goodsell, and A.-K. Maier,
\newblock Nucl.Phys. {\bf B851}, 445 (2011), arXiv:1104.2695.

\bibitem{Kalinowski:2011zz}
J.~Kalinowski,
\newblock PoS {\bf EPS-HEP2011}, 265 (2011).

\bibitem{Frugiuele:2011mh}
C.~Frugiuele and T.~Gregoire,
\newblock Phys.Rev. {\bf D85}, 015016 (2012), arXiv:1107.4634.

\bibitem{Itoyama:2011zi}
H.~Itoyama and N.~Maru,
\newblock Int.J.Mod.Phys. {\bf A27}, 1250159 (2012), arXiv:1109.2276.

\bibitem{Rehermann:2011ax}
K.~Rehermann and C.~M. Wells,
\newblock (2011), arXiv:1111.0008.

\bibitem{Bertuzzo:2012su}
E.~Bertuzzo and C.~Frugiuele,
\newblock JHEP {\bf 1205}, 100 (2012), arXiv:1203.5340.

\bibitem{Davies:2012vu}
R.~Davies,
\newblock JHEP {\bf 1210}, 010 (2012), arXiv:1205.1942.

\bibitem{Argurio:2012cd}
R.~Argurio, M.~Bertolini, L.~Di~Pietro, F.~Porri, and D.~Redigolo,
\newblock JHEP {\bf 1208}, 086 (2012), arXiv:1205.4709.

\bibitem{Fok:2012fb}
R.~Fok, G.~D. Kribs, A.~Martin, and Y.~Tsai,
\newblock Phys.Rev. {\bf D87}, 055018 (2013), arXiv:1208.2784.

\bibitem{Argurio:2012bi}
R.~Argurio, M.~Bertolini, L.~Di~Pietro, F.~Porri, and D.~Redigolo,
\newblock JHEP {\bf 1210}, 179 (2012), arXiv:1208.3615.

\bibitem{Frugiuele:2012pe}
C.~Frugiuele, T.~Gregoire, P.~Kumar, and E.~Ponton,
\newblock JHEP {\bf 1303}, 156 (2013), arXiv:1210.0541.

\bibitem{Frugiuele:2012kp}
C.~Frugiuele, T.~Gregoire, P.~Kumar, and E.~Ponton,
\newblock JHEP {\bf 1305}, 012 (2013), arXiv:1210.5257.

\bibitem{Benakli:2012cy}
K.~Benakli, M.~D. Goodsell, and F.~Staub,
\newblock JHEP {\bf 1306}, 073 (2013), arXiv:1211.0552.

\bibitem{Itoyama:2013sn}
H.~Itoyama and N.~Maru,
\newblock Phys.Rev. {\bf D88}, 025012 (2013), arXiv:1301.7548.

\bibitem{Chakraborty:2013gea}
S.~Chakraborty and S.~Roy,
\newblock JHEP {\bf 1401}, 101 (2014), arXiv:1309.6538.

\bibitem{Csaki:2013fla}
C.~Csaki, J.~Goodman, R.~Pavesi, and Y.~Shirman,
\newblock (2013), arXiv:1310.4504.

\bibitem{Itoyama:2013vxa}
H.~Itoyama and N.~Maru,
\newblock (2013), arXiv:1312.4157.

\bibitem{Beauchesne:2014pra}
H.~Beauchesne and T.~Gregoire,
\newblock (2014), arXiv:1402.5403.

\bibitem{Benakli:2014daa}
K.~Benakli,
\newblock (2014), arXiv:1402.4286.

\bibitem{Bertuzzo:2014bwa}
E.~Bertuzzo, C.~Frugiuele, T.~Gregoire, and E.~Ponton,
\newblock (2014), arXiv:1402.5432.

\bibitem{Camargo-Molina:2013sta}
J.~Camargo-Molina, B.~O'Leary, W.~Porod, and F.~Staub,
\newblock JHEP {\bf 1312}, 103 (2013), arXiv:1309.7212.

\bibitem{Blinov:2013fta}
N.~Blinov and D.~E. Morrissey,
\newblock (2013), arXiv:1310.4174.

\bibitem{Chowdhury:2013dka}
D.~Chowdhury, R.~M. Godbole, K.~A. Mohan, and S.~K. Vempati,
\newblock (2013), arXiv:1310.1932.

\bibitem{Heikinheimo:2011fk}
M.~Heikinheimo, M.~Kellerstein, and V.~Sanz,
\newblock JHEP {\bf 1204}, 043 (2012), arXiv:1111.4322.

\bibitem{Kribs:2012gx}
G.~D. Kribs and A.~Martin,
\newblock Phys.Rev. {\bf D85}, 115014 (2012), arXiv:1203.4821.

\bibitem{Alves:2013wra}
D.~S.~M. Alves, J.~Liu, and N.~Weiner,
\newblock (2013), arXiv:1312.4965.

\bibitem{Fok:2012me}
R.~Fok,
\newblock (2012), arXiv:1208.6558.

\bibitem{Dudas:2013gga}
E.~Dudas, M.~Goodsell, L.~Heurtier, and P.~Tziveloglou,
\newblock (2013), arXiv:1312.2011.

\bibitem{Donagi:2008ca}
R.~Donagi and M.~Wijnholt,
\newblock Adv.Theor.Math.Phys. {\bf 15}, 1237 (2011), arXiv:0802.2969.

\bibitem{Beasley:2008dc}
C.~Beasley, J.~J. Heckman, and C.~Vafa,
\newblock JHEP {\bf 0901}, 058 (2009), arXiv:0802.3391.

\bibitem{Beasley:2008kw}
C.~Beasley, J.~J. Heckman, and C.~Vafa,
\newblock JHEP {\bf 0901}, 059 (2009), arXiv:0806.0102.

\bibitem{Blumenhagen:2008aw}
R.~Blumenhagen,
\newblock Phys.Rev.Lett. {\bf 102}, 071601 (2009), arXiv:0812.0248.

\bibitem{Mayrhofer:2013ara}
C.~Mayrhofer, E.~Palti, and T.~Weigand,
\newblock JHEP {\bf 1309}, 082 (2013), arXiv:1303.3589.

\bibitem{Staub:2008uz}
F.~Staub,
\newblock (2008), arXiv:0806.0538.

\bibitem{Staub:2010jh}
F.~Staub,
\newblock Comput.Phys.Commun. {\bf 182}, 808 (2011), arXiv:1002.0840.

\bibitem{Staub:2009bi}
F.~Staub,
\newblock Comput.Phys.Commun. {\bf 181}, 1077 (2010), arXiv:0909.2863.

\bibitem{Staub:2012pb}
F.~Staub,
\newblock Comput.Phys.Commun. {\bf 184}, 1792 (2013), arXiv:1207.0906.

\bibitem{Dreiner:2012dh}
H.~Dreiner, K.~Nickel, W.~Porod, and F.~Staub,
\newblock Comput.Phys.Commun. {\bf 184}, 2604 (2013), arXiv:1212.5074.

\bibitem{Staub:2013tta}
F.~Staub,
\newblock (2013), arXiv:1309.7223.

\bibitem{Porod:2003um}
W.~Porod,
\newblock Comput.Phys.Commun. {\bf 153}, 275 (2003), arXiv:hep-ph/0301101.

\bibitem{Porod:2011nf}
W.~Porod and F.~Staub,
\newblock Comput.Phys.Commun. {\bf 183}, 2458 (2012), arXiv:1104.1573.

\bibitem{Goodsell:2012fm}
M.~D. Goodsell,
\newblock JHEP {\bf 1301}, 066 (2013), arXiv:1206.6697.

\bibitem{Arvanitaki:2013yja}
A.~Arvanitaki, M.~Baryakhtar, X.~Huang, K.~van Tilburg, and G.~Villadoro,
\newblock JHEP {\bf 1403}, 022 (2014), arXiv:1309.3568.

\bibitem{ArkaniHamed:1997ab}
N.~Arkani-Hamed and H.~Murayama,
\newblock Phys.Rev. {\bf D56}, 6733 (1997), arXiv:hep-ph/9703259.

\bibitem{Binetruy:1994bn}
P.~Binetruy and E.~Dudas,
\newblock Nucl.Phys. {\bf B442}, 21 (1995), arXiv:hep-ph/9411413.

\bibitem{Achard:2001qw}
L3 Collaboration, P.~Achard {\em et~al.},
\newblock Phys.Lett. {\bf B517}, 75 (2001), arXiv:hep-ex/0107015.

\bibitem{Carpenter:2010bs}
L.~M. Carpenter, A.~Rajaraman, and D.~Whiteson,
\newblock (2010), arXiv:1010.1011.

\bibitem{Kannike:2011ng}
K.~Kannike, M.~Raidal, D.~M. Straub, and A.~Strumia,
\newblock JHEP {\bf 1202}, 106 (2012), arXiv:1111.2551.

\bibitem{Dermisek:2013gta}
R.~Dermisek and A.~Raval,
\newblock (2013), arXiv:1305.3522.

\bibitem{Ishiwata:2013gma}
K.~Ishiwata and M.~B. Wise,
\newblock (2013), arXiv:1307.1112.

\bibitem{Lavoura:2003xp}
L.~Lavoura,
\newblock Eur.Phys.J. {\bf C29}, 191 (2003), arXiv:hep-ph/0302221.

\bibitem{Baron:2013eja}
ACME Collaboration, J.~Baron {\em et~al.},
\newblock (2013), arXiv:1310.7534.

\bibitem{Barbieri:2006bg}
R.~Barbieri, L.~J. Hall, Y.~Nomura, and V.~S. Rychkov,
\newblock Phys.Rev. {\bf D75}, 035007 (2007), arXiv:hep-ph/0607332.

\bibitem{Hall:2011aa}
L.~J. Hall, D.~Pinner, and J.~T. Ruderman,
\newblock JHEP {\bf 1204}, 131 (2012), arXiv:1112.2703.

\bibitem{Goodsell:2014inprep}
M.~D. Goodsell and P.~Slavich,
\newblock (2014), {\it in preparation}.

\bibitem{Bechtle:2008jh}
P.~Bechtle, O.~Brein, S.~Heinemeyer, G.~Weiglein, and K.~E. Williams,
\newblock Comput.Phys.Commun. {\bf 181}, 138 (2010), arXiv:0811.4169.

\bibitem{Bechtle:2009ic}
P.~Bechtle, O.~Brein, S.~Heinemeyer, G.~Weiglein, and K.~Williams,
\newblock p.~55 (2009), arXiv:0905.2190.

\bibitem{Bechtle:2011sb}
P.~Bechtle, O.~Brein, S.~Heinemeyer, G.~Weiglein, and K.~E. Williams,
\newblock Comput.Phys.Commun. {\bf 182}, 2605 (2011), arXiv:1102.1898.

\bibitem{Bechtle:2013wla}
P.~Bechtle {\em et~al.},
\newblock (2013), arXiv:1311.0055.

\bibitem{Skands:2003cj}
P.~Z. Skands {\em et~al.},
\newblock JHEP {\bf 0407}, 036 (2004), arXiv:hep-ph/0311123.

\bibitem{Allanach:2008qq}
B.~Allanach {\em et~al.},
\newblock Comput.Phys.Commun. {\bf 180}, 8 (2009), arXiv:0801.0045.

\bibitem{Konar:2010bi}
P.~Konar, K.~T. Matchev, M.~Park, and G.~K. Sarangi,
\newblock Phys.Rev.Lett. {\bf 105}, 221801 (2010), arXiv:1008.2483.

\bibitem{Dreiner:2012wm}
H.~Dreiner, F.~Staub, A.~Vicente, and W.~Porod,
\newblock Phys.Rev. {\bf D86}, 035021 (2012), arXiv:1205.0557.

\bibitem{Jack:1999ud}
I.~Jack and D.~Jones,
\newblock Phys.Lett. {\bf B457}, 101 (1999), arXiv:hep-ph/9903365.

\bibitem{Jack:1999fa}
I.~Jack and D.~Jones,
\newblock Phys.Rev. {\bf D61}, 095002 (2000), arXiv:hep-ph/9909570.

\bibitem{Benakli:2013msa}
K.~Benakli, L.~Darm\'e, M.~D. Goodsell, and P.~Slavich,
\newblock (2013), arXiv:1312.5220.

\bibitem{Shafi:1978gg}
Q.~Shafi,
\newblock Phys.Lett. {\bf B79}, 301 (1978).

\bibitem{Dvali:1994vj}
G.~Dvali and Q.~Shafi,
\newblock Phys.Lett. {\bf B339}, 241 (1994), arXiv:hep-ph/9404334.

\bibitem{Dvali:1994wj}
G.~Dvali and Q.~Shafi,
\newblock Phys.Lett. {\bf B326}, 258 (1994), arXiv:hep-ph/9401337.

\end{thebibliography}
